\documentclass[twocolumn,trackchanges]{aastex701}

\usepackage{amsmath}
\begin{document}

\title{SPHEREx 0.75 to 5 $\mu$m Spectra for a Sequence of Nearby Brown Dwarfs}

\shorttitle{SPHEREx Brown Dwarfs}
\shortauthors{Rustamkulov et al.}

\begin{abstract}
    The SPHEREx all-sky survey has now measured the R$\sim$40-100 infrared spectra of thousands of nearby brown dwarfs in the chemically rich 0.75-5 $\mu$m range. The survey's wide spectral coverage and high S/N permits flux measurements that capture several broadband molecular absorption features, and upwards of 80$\%$ of the total bolometric luminosity of most brown dwarfs. Atmospheric models are known to yield systematic disagreements in the inferred temperatures and radii of brown dwarfs, necessitating benchmarking against observations. In this work, we present SPHEREx spectra across a broad sequence of 37 nearby field brown dwarfs, ranging from L0 to Y4 ($\sim$2500-250 K) and compare them to theoretical expectations. We additionally compile spectra for separate low-gravity and low-metallicity objects, and show how they trend with constant spectral type. We fit the measured spectra to the well-known forward model grids Sonora Diamondback, Elf Owl, BT-Settl, ATMO2020 and ATMO2020++ and compare their goodness-of-fit as a function of wavelength, spectral type, and treatment of clouds and chemistry. We find that the models continue to struggle to simultaneously fit the J/H/K peaks and the 4 $\mu$m opacity window, especially in L/T transition objects. The largest deviations appear around the chemistry-sensitive CO$_2$ and CO features. Despite these offsets, the models broadly capture their trends across the L/T transition, with the observed sample of field dwarfs strongly preferring the weak vertical mixing ($k_\mathrm{zz}$ = 10$^4$ cm$^2$s$^{-1}$) Elf Owl models over strong mixing. The spectra shown here along with future SPHEREx data will help guide improvements to models.

    \end{abstract}

% \author[0000-0003-4408-0463]{Zafar Rustamkulov}
% \affiliation{IPAC, California Insitute of Technology, 770 S. Wilson Ave, Pasadena, CA 91125, USA}
% \email{zafar@caltech.edu}

% \author[0000-0003-4269-260X]{J. Davy Kirkpatrick}
% \affiliation{IPAC, California Insitute of Technology, 770 S. Wilson Ave, Pasadena, CA 91125, USA}
% \email{davy@ipac.caltech.edu}

% \author[0000-0001-9674-1564]{Rachel Akeson}
% \affiliation{IPAC, California Insitute of Technology, 770 S. Wilson Ave, Pasadena, CA 91125, USA}
% \email{rla@ipac.caltech.edu}

% \author[0000-0003-4990-189X]{Michael W.\ Werner}
% \affiliation{Jet Propulsion Laboratory, California Institute of Technology, 4800 Oak Grove Drive, Pasadena, CA 91109, USA}
% \email{michael.w.werner@jpl.nasa.gov}

% To use this file: simply \input{author_list_<paper_id>.tex} in your main TeX file where you want the author list to go!
% generated for paper: Brown Dwarfs
% 2026-05-04 12:06:15
\author[0000-0003-4408-0463]{Zafar~Rustamkulov}%
\affiliation{IPAC, California Institute of Technology, MC 100-22, 1200 E California Blvd, Pasadena, CA 91125, USA}%
\email[show]{zafar@caltech.edu}%

\author[0000-0003-4269-260X]{J.~Davy~Kirkpatrick}%
\affiliation{IPAC, California Institute of Technology, MC 100-22, 1200 E California Blvd, Pasadena, CA 91125, USA}%
\email{davy@caltech.edu}%

\author[0000-0001-9674-1564]{Rachel~Akeson}%
\affiliation{IPAC, California Institute of Technology, MC 100-22, 1200 E California Blvd, Pasadena, CA 91125, USA}%
\email{rla@ipac.caltech.edu}%

\author[0000-0003-4990-189X]{Michael~W.~Werner}%
\affiliation{Jet Propulsion Laboratory, California Institute of Technology, 4800 Oak Grove Drive, Pasadena, CA 91109, USA}%
\email{michael.w.werner@jpl.nasa.gov}%

\author[0000-0002-3993-0745]{Matthew~L.~N.~Ashby}%
\affiliation{Center for Astrophysics $|$ Harvard \& Smithsonian, Optical and Infrared Astronomy Division, Cambridge, MA 01238, USA}%
\email{mashby@cfa.harvard.edu}%
\author[0000-0001-5929-4187]{Tzu-Ching~Chang}%
\affiliation{Jet Propulsion Laboratory, California Institute of Technology, 4800 Oak Grove Drive, Pasadena, CA 91109, USA}%
\affiliation{Department of Physics, California Institute of Technology, 1200 E. California Boulevard, Pasadena, CA 91125, USA}%
\email{tzu@caltech.edu}%
\author[0009-0000-3415-2203]{Shuang-Shuang~Chen}%
\affiliation{Department of Physics, California Institute of Technology, 1200 E. California Boulevard, Pasadena, CA 91125, USA}%
\email{schen6@caltech.edu}%
\author[0000-0002-3892-0190]{Asantha~Cooray}%
\affiliation{Department of Physics \& Astronomy, University of California Irvine, Irvine, CA 92697, USA}%
\email{acooray@uci.edu}%
\author[0000-0002-4650-8518]{Brendan~P.~Crill}%
\affiliation{Jet Propulsion Laboratory, California Institute of Technology, 4800 Oak Grove Drive, Pasadena, CA 91109, USA}%
\affiliation{Department of Physics, California Institute of Technology, 1200 E. California Boulevard, Pasadena, CA 91125, USA}%
\email{bcrill@jpl.nasa.gov}%
\author[0000-0001-7432-2932]{Olivier~Dor\'{e}}%
\affiliation{Jet Propulsion Laboratory, California Institute of Technology, 4800 Oak Grove Drive, Pasadena, CA 91109, USA}%
\affiliation{Department of Physics, California Institute of Technology, 1200 E. California Boulevard, Pasadena, CA 91125, USA}%
\email{olivier.dore@caltech.edu }%
\author[0009-0002-0098-6183]{C.~Darren~Dowell}%
\affiliation{Jet Propulsion Laboratory, California Institute of Technology, 4800 Oak Grove Drive, Pasadena, CA 91109, USA}%
\affiliation{Department of Physics, California Institute of Technology, 1200 E. California Boulevard, Pasadena, CA 91125, USA}%
\email{charles.d.dowell@jpl.nasa.gov}%
\author[0000-0002-9382-9832]{Andreas~L.~Faisst}%
\affiliation{IPAC, California Institute of Technology, MC 100-22, 1200 E California Blvd, Pasadena, CA 91125, USA}%
\email{afaisst@caltech.edu}%
\author[0000-0001-5812-1903]{Howard~Hui}%
\affiliation{Department of Physics, California Institute of Technology, 1200 E. California Boulevard, Pasadena, CA 91125, USA}%
\affiliation{Jet Propulsion Laboratory, California Institute of Technology, 4800 Oak Grove Drive, Pasadena, CA 91109, USA}%
\email{hhui@caltech.edu}%
\author[0000-0002-2770-808X]{Woong-Seob~Jeong}%
\affiliation{Korea Astronomy and Space Science Institute (KASI), 776 Daedeok-daero, Yuseong-gu, Daejeon 34055, Republic of Korea}%
\email{jeongws@kasi.re.kr}%
\author[0000-0002-5016-050X]{Miju~Kang}%
\affiliation{Korea Astronomy and Space Science Institute (KASI), 776 Daedeok-daero, Yuseong-gu, Daejeon 34055, Republic of Korea}%
\email{mjkang@kasi.re.kr}%
\author[0009-0003-8869-3365]{Phil~M.~Korngut}%
\affiliation{Department of Physics, California Institute of Technology, 1200 E. California Boulevard, Pasadena, CA 91125, USA}%
\email{pkorngut@caltech.edu}%
\author[0000-0002-9548-1526]{Carey~M.~Lisse}%
\affiliation{Johns Hopkins University, 3400 N Charles St, Baltimore, MD 21218, USA}%
\affiliation{Johns Hopkins University Applied Physics Laboratory, Laurel, MD 20723, USA}%
\email{carey.lisse@jhuapl.edu}%
\author[0000-0001-5382-6138]{Daniel~C.~Masters}%
\affiliation{IPAC, California Institute of Technology, MC 100-22, 1200 E California Blvd, Pasadena, CA 91125, USA}%
\email{dmasters@ipac.caltech.edu}%
\author[0000-0002-6025-0680]{Gary~J.~Melnick}%
\affiliation{Center for Astrophysics $|$ Harvard \& Smithsonian, Optical and Infrared Astronomy Division, Cambridge, MA 01238, USA}%
\email{gmelnick@cfa.harvard.edu}%
\author[0000-0001-9368-3186]{Chi~H.~Nguyen}%
\affiliation{Department of Physics, California Institute of Technology, 1200 E. California Boulevard, Pasadena, CA 91125, USA}%
\email{chnguyen@caltech.edu}%
\author[0000-0002-5158-243X]{Roberta~Paladini}%
\affiliation{IPAC, California Institute of Technology, MC 100-22, 1200 E California Blvd, Pasadena, CA 91125, USA}%
\email{paladini@ipac.caltech.edu}%
\author[0000-0003-1841-2241]{Volker~Tolls}%
\affiliation{Center for Astrophysics $|$ Harvard \& Smithsonian, Optical and Infrared Astronomy Division, Cambridge, MA 01238, USA}%
\email{vtolls@cfa.harvard.edu}%
\author[0000-0003-3078-2763]{Yujin~Yang}%
\affiliation{Korea Astronomy and Space Science Institute (KASI), 776 Daedeok-daero, Yuseong-gu, Daejeon 34055, Republic of Korea}%
\email{yyang@kasi.re.kr}%
\author[0000-0001-8253-1451]{Michael~Zemcov}%
\affiliation{School of Physics and Astronomy, Rochester Institute of Technology, 1 Lomb Memorial Dr., Rochester, NY 14623, USA}%
\affiliation{Jet Propulsion Laboratory, California Institute of Technology, 4800 Oak Grove Drive, Pasadena, CA 91109, USA}%
\email{mbzsps@rit.edu}
% \collaboration{The SPHEREx Science Team}

% \collaboration{all}%{The SPHEREx Science Team}
\keywords{\uat{Brown dwarfs} {185}, \uat{L dwarfs} {894}, \uat{T dwarfs} {1679}, \uat{Y dwarfs} {1827}, \uat{Sky surveys} {1464}, \uat{Infrared spectroscopy} {2285}, \uat{Stellar atmospheres} {1584}}

\section{Introduction}

Brown dwarfs, substellar objects not massive enough to sustain stable hydrogen fusion, are an exceptionally colorful and varied group of celestial bodies. Among the $\sim$10,000 known brown dwarfs \citep{Gagne2026}, effective temperatures span 250 to 2,500 K while intrinsic luminosities range over four orders of magnitude depending on mass and age \citep[e.g.,][]{Sanghi2023}. The vertically thin photosphere of a brown dwarf is the ultimate mediator of its interior heat flux and therefore controls its radius evolution. Unlike stars, brown dwarfs cool, contract, and spin up with age \citep[e.g.,][]{Schneider2018, Vos2020}, resulting in observable population-wide changes in the character of their clouds and chemistry. In this way, their molecule-rich spectra tell stories of the delicate couplings linking their atmosphere, weather, climate, and interior through time. 

Compositional trends in brown dwarf atmospheres resemble those of irradiated gas giant exoplanets---the hottest objects show strong H$_2$O, CO, and alkali opacity, with magnesium silicate clouds appearing, then sinking below the photosphere as they cool below $\sim$1400 K with age \citep[e.g.,][]{Gao2020, Suarez2022, Lothringer2026}. At colder temperatures and older ages, sulfide and chloride cloud species appear as CH$_4$ becomes chemically favored over CO below $\sim$1000 K in T dwarf atmospheres \citep[e.g.,][]{Lodders2002, Tannock2022, Burgasser2025}. The coldest brown dwarfs, the Y dwarfs, begin to resemble Jupiter and Saturn, whose CH$_4$ and NH$_3$-dominated atmospheres have vertically stratified clouds of diverse compositions, and with trace quantities of H$_2$S and PH$_3$ \citep{Beiler2024, Kothari2024}. These objects show diversity in composition and cloud coverage \citep[e.g.,][]{Lueber2026}. Unlike mature giant planets and exoplanets, unbound field brown dwarf atmospheres are heated entirely from within. By virtue of their independent thermochemical evolution, field brown dwarfs represent an idealized glimpse into the interacting processes liberating their primordial heat. Yet despite being internally heated, some brown dwarfs exhibit heat sources of uncertain origin manifesting as mysterious thermal inversions, or stratospheres, and chromospheric activity \citep[e.g.,][]{Faherty2024, Reiners2008}. Oddities aside, the bulk of brown dwarf fundamental parameters can only be estimated from model fits to their spectra. Namely, for an isolated self-luminous field dwarf, the radius, temperature, and mass estimates rely on the information content encoded by theoretical atmosphere models \citep{Burrows2001, Baraffe2003, Saumon2008, Marley2021}. To constrain their ages, these fundamental parameters are compared to evolution models that track the temperature and radius for a brown dwarf of given mass \citep{Filippazzo2015, Dupuy2017}. 

Today, forward models gridded along the axes of temperature and gravity only qualitatively represent the major features seen in spectra, with large, $\gtrsim$20\% systematic differences dependent on the chosen input physics assumptions \citep[e.g.,][]{Suarez2021, Petrus2024, Zhang2021}. While theoretical and lab-measured molecular databases have been improving, model substellar spectra still differ from observations due to differences in their handling of clouds, chemistry, and the thermal profile. Empirical spectral classification continues to be a useful practice to augment these model-dependent determinations by populating a relative scale upon which to compare different objects. %Spectral type, defined by the observable markers, is generally a rough proxy for temperature, but any accurate determination of the radius requires requires a temperature estimate from model fitting.

\subsection{Spectral Classification\label{sec:spectral_classification}}
In this paper, we present sequences of low-mass, low-temperature spectral standards at Solar age/composition, at lower gravities, and at subsolar metallicities. As explained in \cite{kirkpatrick2005}, spectral types for these objects follow the form \{metallicity class\}+\{temperature class\}+\{gravity class\}, where 

\[\text{metallicity class} \in \{\text{d, sd, esd, usd}\}\footnote{For late-M to mid-L dwarfs, these correspond very roughly to values of $[Fe/H] \approx 0.0$, $-1.0 < [Fe/H] < 0.0$, $-1.7 < [Fe/H] < -1.0$, and $-2.3 < [Fe/H] < -1.7$, respectively (Figure 21 of \citealt{zhang2017}). The corresponding metallicity ranges for T and Y subdwarfs have yet to be determined due to the rarity of these objects in current samples.}\]
\[\text{temperature class} \in \{\text{M$n$, L$n$, T$n$, Y$n$}\}\footnote{These correspond roughly to $T_{eff}$ ranges of 3800-2250K, 2250-1250K, 1250-450K, and $<$450K, respectively (Figure 9.19 of \citealt{gray2009} and Figure 22b of \citealt{kirkpatrick2021}.)}\]
\[\text{gravity class} \in \{\alpha, \beta, \gamma, \delta\}\footnote{These are believed to correspond roughly to age ranges of $>$1 Gyr, $\sim$100 Myr, $\sim$10 Myr, and $\sim$1 Myr, respectively (Section 5.2 of \citealt{kirkpatrick2005}). If we consider the range from the M/L dwarf boundary down to the T/Y dwarf boundary, the models of \cite{burrows1997} suggest that these gravity designations correspond to overlapping ranges of $5.4 < log(g) < 4.4$ for $\alpha$, $5.2 < log(g) < 4.0$ for $\beta$, $4.8 < log(g) < 3.8$ for $\gamma$, and $4.3 < log(g) < 3.4$ for $\delta$, with the exact gravity value depending upon the object's temperature class.}\]

For the metallicity classes, d = dwarf, sd = subdwarf, esd = extreme subdwarf, and usd = ultra subdwarf. For the gravity classes, $\alpha$ = field age and field gravity, with $\beta$, $\gamma$, and $\delta$ marking objects with progressively lower gravities and younger ages \citep{Cruz2009}. The $n$ under temperature class is a gradation that generally ranges from 0 to 9.5 except for the Y dwarfs, for which the range currently runs only from 0 to 2. 

For example, a normal field early-L dwarf might have a full spectral type of dL0$\alpha$; however, the d prefix and $\alpha$ suffix are almost never appended for these Solar-composition and Solar-age objects so that the type can be written more simply as L0. A late-M subdwarf, which can be interpreted as an old object with subsolar metallicity, might be denoted by sdM9. An even lower metallicity late-T extreme subdwarf would be denoted as esdT8, and an even lower metallicity early-Y ultra subdwarf denoted as usdY0. Likewise, a solar-composition early-L exhibiting mild effects of lower gravity -- generally interpreted as a sign of youth, as these objects have not yet contracted to their final equilibrium radii -- might be denoted as L2$\beta$. An even lower gravity and possibly even younger late-M-type brown dwarf might be denoted as M8$\gamma$, and an even lower gravity and possibly even younger mid-T denoted as T5$\delta$.

Late-M, L, T, and Y dwarf spectral standards were taken from several sources. Standards with typical field age and field composition were taken from \cite{kirkpatrick2010}, who drew from late-M and L primary and secondary standards first established at optical wavelengths by \cite{boeshaar1985}, \cite{kirkpatrick1991}, and \cite{kirkpatrick1999} and further drew from T dwarf spectral standards established in the near-infrared by \cite{burgasser2006}. The latter reference reached as cold as spectral type T8, but later near-infrared standards at types $\ge$T9 were proposed by \cite{cushing2011} and \cite{kirkpatrick2012}.

Standards at low gravity were taken from the lists of \cite{Cruz2009}, 
%who often listed multiple standards per subtype. 
%Here, we have chosen a single object to serve as a representative of each integral type. 
and those at low metallicity were drawn from the lists of \cite{lepine2007} and \cite{zhang2017}.
%, where again we have chosen a single representative for each integral subtype when multiple ones were given.

\subsection{Brown Dwarf Atmosphere Panorama} 

\begin{figure*}
    \centering
    \includegraphics[width=\textwidth]{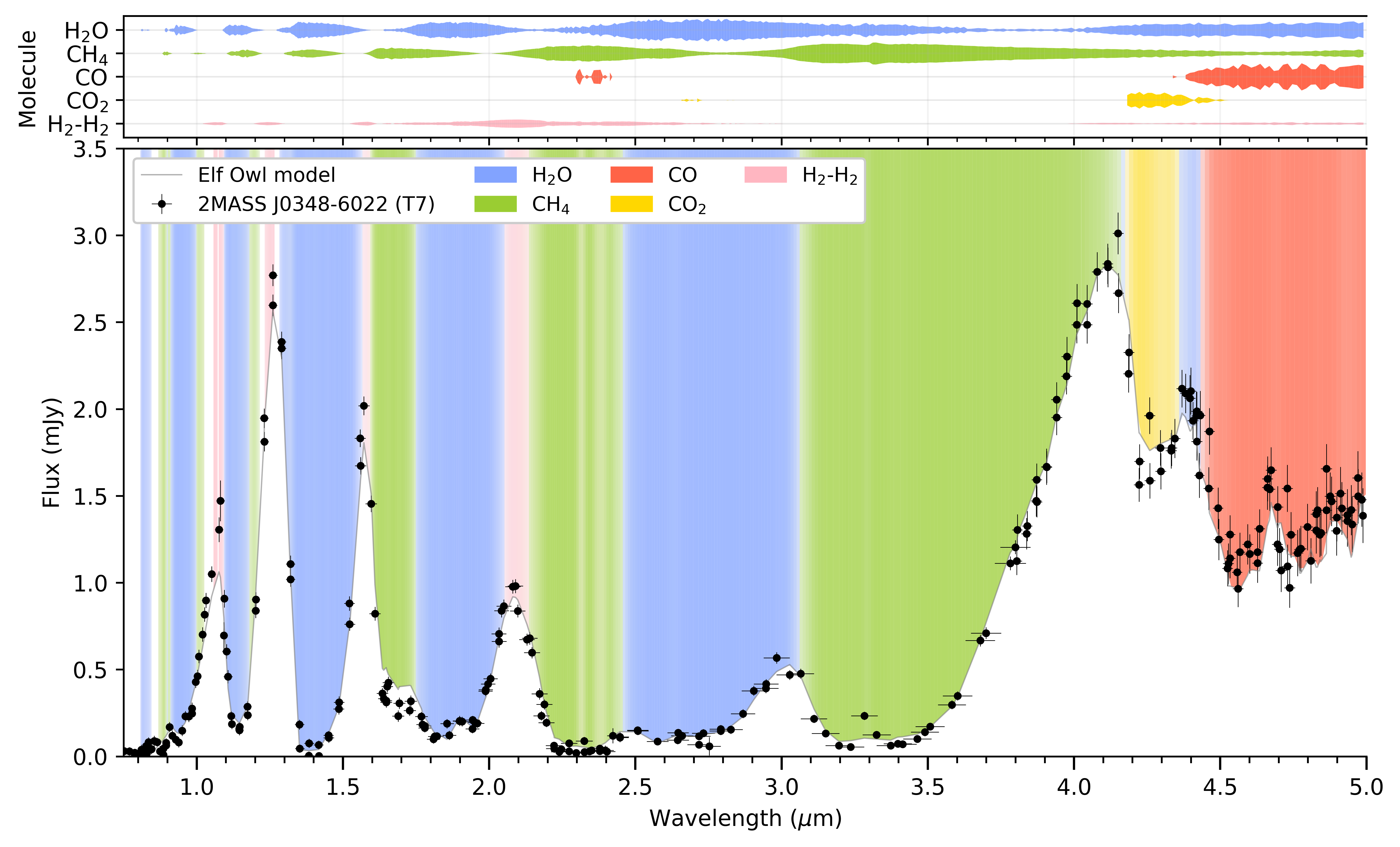}
    \caption{The SPHEREx spectrum of the typical T7 field dwarf 2MASS J03480772-6022270 overplotted with a map of the key molecular opacity contributors and the best-fitting Sonora Elf Owl model. The widths of the shaded regions in the top panel are proportional to the log of the opacity of each molecule. The corresponding shaded regions in the bottom panel highlight the dominant absorber at a given wavelength. The spectrum is dominated by alternating and overlapping bands of water and methane, with carbon dioxide and carbon monoxide appearing at redder wavelengths. There is a newly-apparent weak CH$_4$ emission feature at 3.3 $\mu$m, possibly resulting from a thermal inversion \citep[e.g.,][]{Faherty2024}}
    \label{fig:opacity_map}
\end{figure*}

The 0.75-5.0 $\mu$m SPHEREx spectra of brown dwarfs have many features that can be used to study the physical and chemical conditions of their atmospheres. The H$_2$O and CH$_4$ absorption band strengths are a direct probe of temperature \citep{Lodders2002, kirkpatrick2010}. Carbon tracers -- CH$_4$, CO$_2$, and CO -- are thought to provide clues about vertical mixing and disequilibrium chemistry \citep{Kothari2024}; the C/O ratio itself is believed to hold clues about the brown dwarf's formation pathway \citep{phillips2024}. The main tracer of nitrogen -- NH$_3$ -- can independently be used to study vertical mixing, although its presence is dwarfed by overlying bands of H$_2$O and CH$_4$ \citep{saumon2006, cushing2011}, and is unlikely to be detectable with SPHEREx. Another strong absorber is collision-induced absorption (CIA) by H$_2$ centered at 2.1 $\mu$m \citep{borysow2002}, whose broad pressure-sensitive continuum opacity is a probe of the depth of the photosphere \citep{saumon1994}. Figure \ref{fig:opacity_map} shows a representative SPHEREx spectrum with prominent features shaded by the dominant molecular opacity source at a given wavelength. The opacities are taken from the compiled PICASO opacity database \citep[][and references therein]{Batalha2019}. The spectrum is dominated by H$_2$O, the main oxygen tracer, and CH$_4$ up to the opacity window at 4.2 $\mu$m, beyond which CO$_2$ and CO become prominent. 

\subsection{SPHEREx}

SPHEREx\footnote{\url{https://spherex.caltech.edu}}, the Spectro-Photometer for the History of the Universe, Epoch of Reionization, and Ices Explorer~\citep{Bock2025}, is a NASA medium explorer mission which has been in science operations since 1 May, 2025. SPHEREx is conducting an all-sky near-infrared spectral survey in 102 spectral channels spanning from 0.75 to 5 $\mu$m with spectral resolution of $\sim$40 shortward of 4 $\mu$m and $\sim$120 from 4 to 5 $\mu$m. SPHEREx will survey the whole sky 4 times during its 25 month prime mission.  In addition, SPHEREx observes deep fields in the north and south ecliptic poles, where sources are observed up to hundreds of times per spectral channel. The SPHEREx Science Team goals, from cosmology to ices in the Milky Way, are described in \cite{Bock2025}.  
In the SPHEREx instrument, the focal plane is organized into two sides, both of which include 3$\times$1 H2RG detector arrays \citep{Korngut2026}.  Each detector has been overlaid with a linear variable filter, providing sensitivity to a specific wavelength band, and covers 3.5$\times$3.5 degrees on the sky with 6$\farcs$15 pixels.
The SPHEREx observations are a unique all-sky data set that is available at the NASA/IPAC Infrared Science Archive (IRSA)\footnote{\url{https://irsa.ipac.caltech.edu}}.

\subsection{Motivation}

Before SPHEREx, survey characterization of brown dwarf atmospheres was largely enabled by the broadband near-infrared magnitudes from 2MASS $J,H,K_s$ at 1.1-1.36, 1.5-1.8, and 2-2.4 $\mu{\rm m}$, with 5$\sigma$ depths of $J,H,K_s\simeq17.4,17.2,16.9$ AB mag, and from WISE/AllWISE $W1,W2$ at $2.8$--$3.9~\mu{\rm m}$ and $4$--$5.2~\mu{\rm m}$, with 5$\sigma$ depths of $W1,W2\simeq19.6,19.3$ AB mag \citep{Skrutskie2006, Wright2010}. The combined 2MASS+WISE broadband SED therefore samples brown dwarfs at an effective photometric resolving power of only $R\sim5$--$8$. In comparison, SPHEREx achieves 5$\sigma$ per-exposure depths of $\sim$19.5 in the 0.7-3.8 $\mu{\rm m}$ range with R$\sim$40, and $\sim$17.5-18.5 AB mag depths in the 3.8-5 $\mu{\rm m}$ range with R$\sim$110. 

With their higher sensitivity, increased spectral resolution, and wide wavelength grasp, SPHEREx spectra are uniquely well-suited for detailed atmospheric characterization of nearby brown dwarfs (2-30 pc, see Table \ref{tab:field}). For these reasons, the SPHEREx spectral atlas represents the frontier of brown dwarf survey science. In particular, the poorly studied 3-5 $\mu$m range encapsulates the main opacity features of CH$_4$, CO$_2$, and CO. These species are sensitive diagnostics of thermochemistry and composition, and so far only a handful of brown dwarfs have been observed at these wavelengths with JWST and AKARI. SPHEREx has now gathered the spectra of thousands of brown dwarfs \citep[e.g.,][]{Gagne2026}, with many new discoveries readily possible.

Another important and surprising application of SPHEREx specra is in cosmology, where Y-dwarfs can be mistaken for high-redshift galaxies due to their similar magnitudes and infrared colors. A recent such case of mistaken identity was pointed out in \cite{Bradac2026}, where two high-redshift galaxy candidates turned out to be faint, distant brown dwarfs. Thanks to SPHEREx spectra, many instances of such contamination in galaxy surveys can be cleared up by comparing the object's spectrum to representative templates. Conversely, these spectra may help identify extremely distant brown dwarfs in our galaxy in future surveys.

In this work, we present a curated sample of high-S/N SPHEREx spectra across the full gamut of brown dwarf temperature, gravity, and metallicity to directly compare state-of-the-art theoretical model grids, with the motivation of distilling the most relevant physics to inform the models' continued improvement. Given that field brown dwarfs are in some ways simpler to model and observe than transiting exoplanets, adequately fitting their observables is imperative to improving the characterization of Jovian and lower mass worlds.
 
\section{Observations and Data Analysis}
\subsection{SPHEREx Pipeline}
The SPHEREx science data processing occurs at the SPHEREx Science Data Center (SSDC) at Caltech-IPAC and by the SPHEREx Science Team.  The SSDC performs the first three levels of processing, including astrometric and photometric calibrations, while the Science Team produces higher-level image and catalog products.  The SSDC pipeline infrastructure is described in \cite{Akeson2026} and the individual pipeline modules are documented in the SPHEREx Explanatory Supplement\footnote{\url{https://irsa.ipac.caltech.edu/data/SPHEREx/docs/SPHEREx_Expsupp_QR.pdf}} at IRSA. 

For this work, we use calibrated Level-2 spectral image files from IRSA, with calibration version QR2.  Given the sampling strategy employed by SPHEREx \citep{bryan2025}, the exact spectral sampling is not uniform for each source.  These calibrated images are the input for the Spectrophotometry Tool, one of the SPHEREx tools hosted by IRSA.  This tool performs photometric measurements of selected sources using position and morphology information provided by the user.  Here, we used proper motion corrected positions, specified that all sources are point-like, and used the tool version with the corrected point-spread function (PSF) (as of April 2026).  The tool algorithm is based on the community software package Tractor\footnote{\url{http://thetractor.org}}, which uses generative modeling of astronomical images to measure a maximum likelihood solution (Section 3.3 of the Explanatory Supplement).  A local background is also calculated and subtracted.  If the local region does not contain a sufficient number of unflagged pixels, a global solution is used.  In addition to the source flux and uncertainty, a per point fit quality metric is calculated and flags from the image are propagated to the output. The spectra have spectral flux density units, $F_\nu$, in Jy. At the time of writing, the spectral absolute gain calibration of SPHEREx has a quoted reliability of $\lesssim3\%$ for Bands 1-4, and $\lesssim5\%$ for Bands 5 and 6 in the QR2 data, which are used in this work. The gain calibration is currently being refined to a higher accuracy and reliability in \cite{Ashby2026}.

% \subsection{Spectrophotometry Tool} (RACHEL)  I put all the text in the previous section.  I think it flows better this way, but if not we can pull the tool stuff out into it's own section
\subsection{Target Selection Caveats}

%Some of the brown dwarf spectral standards are faint and have relatively low S/N, while others have only one full-sky pass of spectral coverage. We replace these objects with brighter 20-parsec objects in the same spectral subclass. The low-gravity and low-metallicity sequences are sourced entirely from the list of standards, while 14 of the 24 field dwarfs are selected from the 20-parsec sample. In order to minimize contamination, we exclude known binary systems. Furthermore, we perform visual vetting using the WCS-aligned and stacked SPHEREx images, excluding sources that show extended PSF shapes or obvious contaminators within 3 pixels. Unlike the spectral standards, these 20-parsec stars have not undergone the rigorous M-K Process to qualify their spectral types, and their spectral classification are subject to systematic uncertainties on the order of $\pm$1 SpT. We include the standard spectra separately in the Appendix and the supplemental materials.

The low-gravity and low-metallicity sequences discussed in \S~\ref{sec:spectral_sequences} below (Table \ref{tab:field}) are sourced entirely from the list of standards discussed in \S~\ref{sec:spectral_classification}. Some of the standards for the Solar-age, Solar-metallicity sequence, however, have low S/N compared to other field dwarfs of the same type, have a relatively low number of measurements, or are known binaries, so we have chosen a replacement with the same optical and/or near-infrared spectral type from the 20-pc census of \cite{Kirkpatrick2024} (see Table \ref{tab:field}).  A few other standards were excluded because their PSF shapes were extended in WCS-aligned and stacked SPHEREx images compared to the SPHEREx PSF, they have obvious contaminators within 3 pixels of their centroid position, or they are known to be currently contaminated by a background object based on their proper motion trajectories (as seen in WiseView imagery\footnote{\url{http://byw.tools/wiseview}}; \citealt{caselden2018}). Unlike the spectral standards, these 20-parsec objects should not be used as anchors for the MK Process (\citealt{morgan1943}) but should suffice to show the gross changes in spectral morphology through the sequence.

\subsection{Data Cleaning}

As SPHEREx spectrophotometric measurements can sometimes become affected by cosmic rays, satellite streaks, and Solar particle events, we use the bit-wise pixel quality flags from the Spectrophotometry Tool to apply strict quality cuts to the data. All fluxes with any flag listed in Table 9 of the Explanatory Supplement (other than \texttt{SOURCE}) are excluded to conservatively ensure the cleanliness of the spectra. Visual spot-checks suggest that a poor quality flag does not always reflect an anomalous measurement, and some anomalous measurements do not get flagged, meaning that some objects will show occasional spurious fluxes. There is a known under-flagging of bad pixels in the early data releases, which have since been corrected (as of Dec. 2025). The flagging update was not applied retroactively, so at least half of the observations in this work are subject to underflagging. We remove spurious points by masking fluxes that lie more than 4-$\sigma$ away from the smoothed running median of the data.

%We exclude XX percent of our flux measurements this way. 

\section{Spectral Sequences\label{sec:spectral_sequences}}

In this section, we present our distinct field, low-gravity, and low-metallicity sequences. The field objects are summarized in Table \ref{tab:field}, the low-gravity in Table \ref{tab:lowg}, and the low-metallicity in Table \ref{tab:lowz}. 

\begin{deluxetable*}{llcccc}
\tabletypesize{\scriptsize}
\tablecaption{Objects in the spectral sequence of Fig. \ref{fig:waterfall}\label{tab:field}}
\tablehead{
\colhead{Spec.\ Type} &
\colhead{Name} &
\colhead{Dist.\ (pc)} &
\colhead{Ref.} &
\colhead{Standard?} &
\colhead{Note}
}
\startdata
L0 & 2MASS J17312974+2721233 & $11.95 \pm 0.01$ & 2 & No & A \\
L1 & 2MASS J21304464-0845205 & $26.69 \pm 0.22$ & 1 & Yes & \nodata \\
L2 & 2MASS J21041491-1037369 & $17.24 \pm 0.08$ & 2 & No & B \\
L3 & 2MASS J15065441+1321060 & $11.71 \pm 0.03$ & 2 & Yes & \nodata \\
L4 & 2MASS J21580457-1550098 & $23.20 \pm 0.49$ & 1 & Yes & \nodata \\
L5 & 2MASS J15074769-1627386 & $7.41 \pm 0.01$ & 2 & No & C \\
L6 & 2MASS J10101480-0406499 & $17.33 \pm 1.08$ & 3 & Yes & \nodata \\
L7 & 2MASS J03400942-6724051 & $9.36 \pm 0.04$ & 2 & No & D \\
L8 & 2MASS J16322911+1904407 & $15.09 \pm 0.37$ & 4 & Yes & \nodata \\
L9 & DENIS J025503.3-470049 & $4.87 \pm 0.00$ & 2 & Yes & \nodata \\
T0 & PSO J319.3102-29.6682 & $13.14 \pm 0.60$ & 4 & No & \nodata \\
T1 & WISE J124629.65-313934.2 & $11.52 \pm 0.20$ & 2 & No & E \\
T2 & 2MASS J11220826-3512363 & $13.26 \pm 0.28$ & 2 & No & F \\
T3 & PSO J247.3273+03.5932 & $12.32 \pm 0.45$ & 3 & No & \nodata \\
T4 & 2MASS J22541892+3123498 & $13.89 \pm 0.58$ & 6 & Yes & \nodata \\
T5 & WISE J004542.56+361139.1 & $17.54 \pm 1.14$ & 3 & No & G \\
T6 & DENIS J081730.0-615520 & $5.21 \pm 0.01$ & 2 & No & H \\
T7 & 2MASS J03480772-6022270 & $8.33 \pm 0.12$ & 7 & No & I \\
T8 & 2MASS J04151954-0935066 & $5.71 \pm 0.06$ & 8 & Yes & \nodata \\
T9 & WISE J000517.48+373720.5 & $7.88 \pm 0.13$ & 3 & No & J \\
Y0 & WISE J205628.91+145953.2 & $7.10 \pm 0.10$ & 3 & No & K \\
Y1 & WISE J154151.65-225024.9 & $5.99 \pm 0.07$ & 3 & No & L \\
Y2?\tablenotemark{a} & WISE J182831.08+265037.7 & $9.97 \pm 0.20$ & 3 & Yes & \nodata \\
$\sim$Y4?\tablenotemark{a} & WISE J085510.83-071442.5 & $2.28 \pm 0.01$ & 3 & Yes & \nodata \\
\enddata
\tablenotetext{a}{Standards have not yet been established at types later than Y1, so these classifications should be considered tentative.}
\tablecomments{
Reference codes for astrometry:
(1) Gaia DR3;
(2) Gaia EDR3;
(3) Kirkpatrick et al.\ 2021a;
(4) Dahn et al.\ 2017 + CatWISE 2020;
(5) Best et al.\ 2020;
(6) Manjavacas et al.\ 2013 + CatWISE 2020;
(7) Kirkpatrick et al.\ 2019;
(8) Dupuy and Liu 2012.
Notes:
(A) The L0 standard, 2MASS J03454316+2540233, has much lower S/N than the L0 listed.
(B) The L2 standard, Kelu-1 AB, is a known binary.
(C) The L5 standard, 2MASS J08350622+1953050, has much lower S/N than the L5 listed.
(D) The L7 standard, 2MASS J01033203+1935361, has much lower S/N than the L7 listed.
(E) The T1 standard, WISE J083717.17-000020.0, has much lower S/N than the T1 chosen.
(F) The T2 standard is 2MASS J12545393-0122474.
(G) The T5 standard, 2MASS J15031961+2525196, has poorer SPHEREx spectral coverage than the T5 listed.
(H) The T6 standard, 2MASS J16241436+0029158, has poorer S/N and spectral coverage than the chosen T6.
(I) The T7 standard, 2MASS J07271824+1710012, has poorer spectral coverage than the T7 listed.
(J) The T9 standard, UGPS J072227.51-054031.2, has possible contamination in SPHEREx since it lies deep within the Galactic Plane.
(K) The Y0 standard, WISE J173835.53+273259.0, has poorer S/N than the Y0 listed.
(L) The Y1 standard, WISE J035000.32-565830.2, has poorer S/N than the Y1 listed.
}
\tablecomments{The listed spectral types are all near-infrared classifications except for the L3, L6, L8, and T3, whose classifications were done in the optical.}
\end{deluxetable*}

\begin{figure*}
    \centering
    \includegraphics[width=0.99\linewidth]{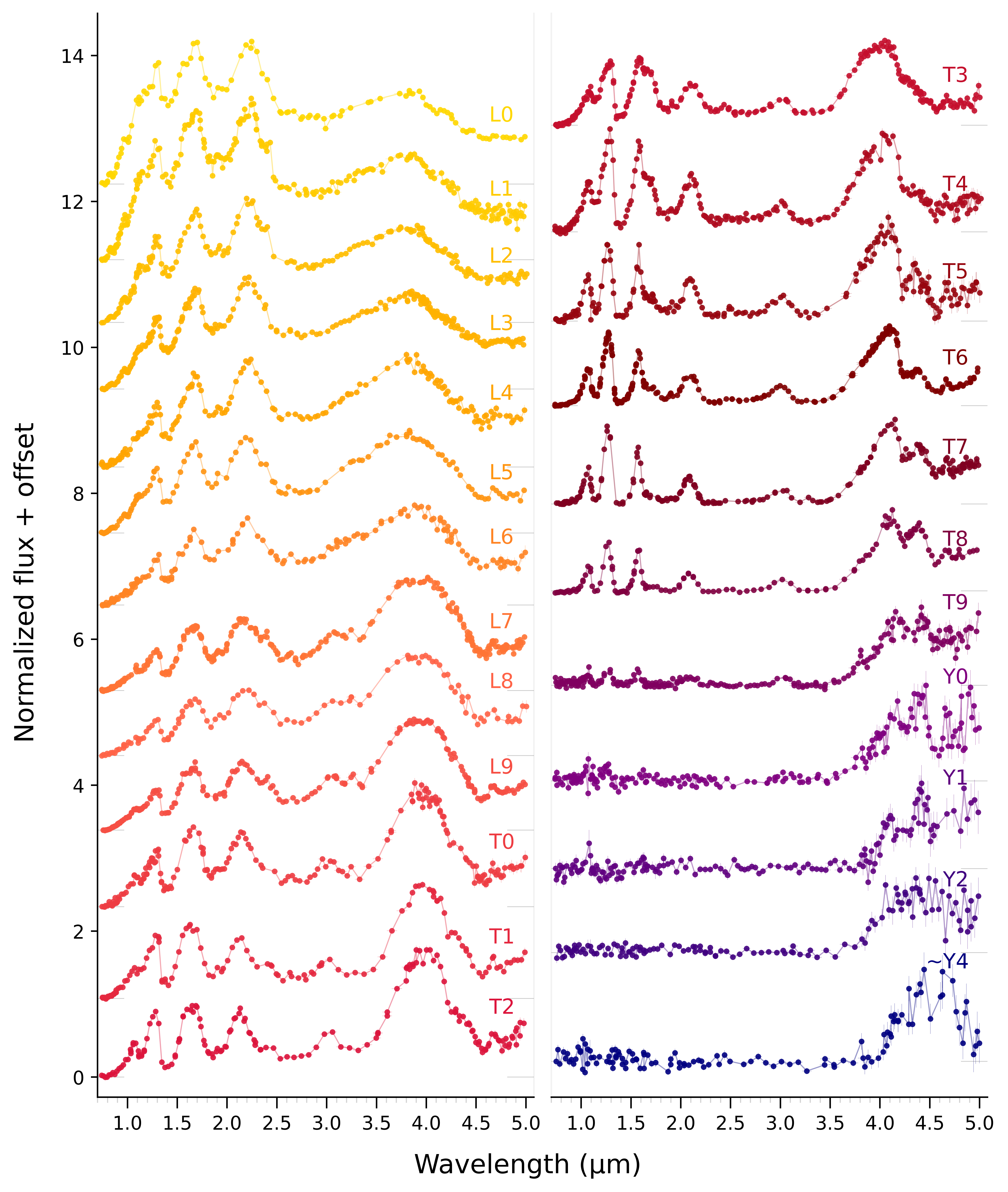}
    \caption{The SPHEREx spectral sequence of the field brown dwarfs listed in Table \ref{tab:field}. Each spectrum is normalized by the mean of the flux between 3.7 and 4.7 $\mu$m, and arbitrarily vertically spaced for visual clarity. The zero-flux level for each spectrum is indicated by the pairs of short grey lines.}
    \label{fig:waterfall}
\end{figure*}

\begin{figure*}
    \centering
    \includegraphics[width=0.8\linewidth]
    {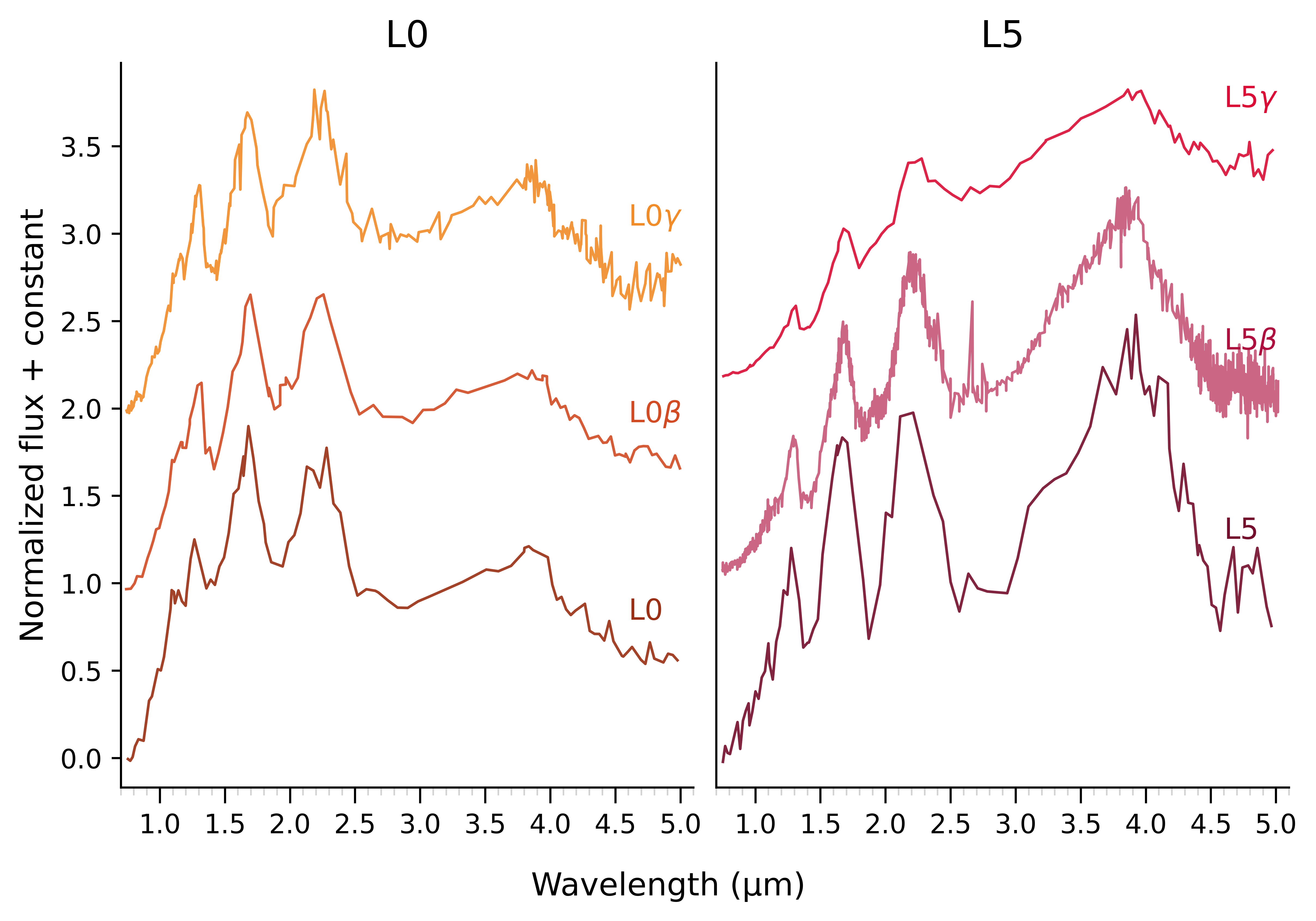}
    \caption{The SPHEREx spectral sequence of low gravity brown dwarfs for L0-types (left) and L5-types (right). The lowest-gravity $\gamma$ types (top) differ from the field-gravity types (bottom) are most distinguished from each other by the relative heights of the NIR peaks, with the K-band peak growing taller relative to the H-band peak with decreasing gravity. For the L5 subtypes, the slope of the 0.7-1.2 $\mu$m range and the amplitudes of the water features get weaker with declining gravity, indicative of cloud scattering and isothermality, respectively.}
    \label{fig:lowg}
\end{figure*}

\begin{deluxetable*}{llcccc}
\label{tab:lowg}
\tabletypesize{\scriptsize}
\tablecaption{Low-gravity comparison sequence at L0 and L5.}
\tablehead{
\colhead{Spec.\ Type} &
\colhead{Name} &
\colhead{Dist.\ (pc)} &
\colhead{Astrometry Ref.} &
\colhead{Standard?} &
\colhead{Note} \\
}
\startdata
L0$\gamma$ & 2MASS J00325584-4405058 & $34.5 \pm 0.5$ & Gaia DR3 & Yes & \nodata \\
L0$\beta$ & 2MASS J15525906+2948485 & $20.5 \pm 0.07$ & Gaia DR3 & Yes & \nodata \\
L0 & 2MASS J03454316+2540233 & $26.4 \pm 0.18$ & Gaia DR3 & Yes & \nodata \\
L5$\gamma$ & 2MASS J03552337+1133437 & $9.16 \pm 0.04$ & Gaia EDR3 & Yes & \nodata \\
L5$\beta$ & 2MASS J04210718-6306022 & $20.0 \pm 1.3$ & Kirkpatrick2021a & Yes & \nodata \\
L5 & 2MASS J08350622+1953050 & $36 \pm 4$ & CatWISE2020 & Yes & \nodata \\
\enddata
\tablecomments{The low-gravity objects above are as defined by \cite{Cruz2009}. The listed spectral types are all optical classifications except for the L5, whose classification was done in the near infrared.}
\end{deluxetable*}

\begin{figure*}
    \centering
    \includegraphics[width=0.8\linewidth]{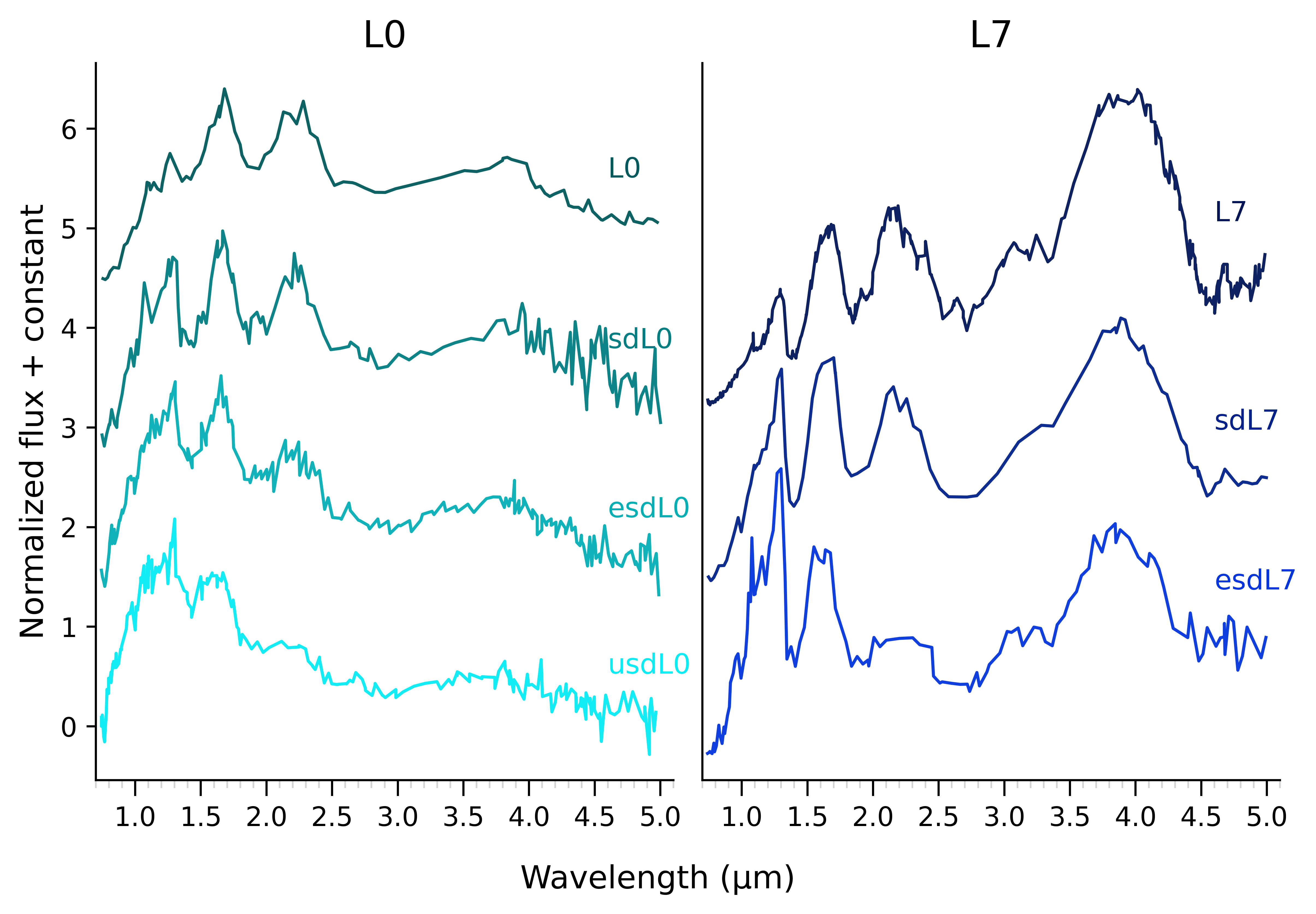}
    \caption{The low-metallicity sequence for the L0 and L7 subtype standards. The objects decrease in metallicity from top to bottom. The designations correspond to [Fe/H] metallicities of $\approx 0.0$, $-1.0 < [Fe/H] < 0.0$, $-1.7 < [Fe/H] < -1.0$, and $-2.3 < [Fe/H] < -1.7$ for the d, sd, esd, and usd L-subtypes, respectively. The decrease in molecular opacity results in a hotter, deeper photosphere, driving an increase in NIR flux, probing deeper where the pressure-broadened H$_2$-H$_2$ CIA opacity dominates from 2-3 $\mu$m. }
    \label{fig:lowz}
\end{figure*}

\begin{deluxetable*}{llccc}
\label{tab:lowz}
\tabletypesize{\scriptsize}
%\tablenum{1}
\tablecaption{Low-metallicity comparison sequence at L0 and L7.}
\tablehead{
\colhead{Spec.\ Type} &
\colhead{Name} &
\colhead{Dist.\ (pc)} &
\colhead{Astrometry Ref.} &
\colhead{Standard?} \\
% \colhead{Note} \\
% \colhead{(1)} &
% \colhead{(2)} &
% \colhead{(3)} &
% \colhead{(4)} &
% \colhead{(5)} &
% \colhead{(6)} \\
}
\startdata
esdL0 & 2MASS J00144919-0838207 & $50.28 \pm 0.50$ & Gaia DR3 & Yes \\
sdL0 & 2MASS J01301256-1047285 & $62.31 \pm 2.62$ & Gaia DR3 & Yes \\
L0 & 2MASS J03454316+2540233 & $26.39 \pm 0.18$ & Gaia DR3 & Yes \\
usdL0 & 2MASS J10130734-1356204 & $56.47 \pm 0.52$ & Gaia DR3 & Yes \\
esdL7 & 2MASS J05325346+8246465 & $24.56 \pm 0.28$ & Gaia DR3 & Yes \\
L7 & 2MASS J03400942-6724051 & $9.36 \pm 0.04$ & Gaia EDR3 & Yes \\
sdL7 & 2MASS J14162408+1348263 & $9.28 \pm 0.02$ & Gaia EDR3 & Yes \\
\enddata
\tablecomments{The low-metallicity objects above are as defined by \cite{zhang2017}. The listed spectral types are all near-infrared classifications except for the L0, whose classification was done in the optical.}
\end{deluxetable*}

\subsection{Qualitative Description}
The brown dwarf spectral sequence is mostly carved by the temperature-sensitive bands of H$_2$O, CH$_4$, CO, CO$_2$, and the opacity windows between them (Fig. \ref{fig:waterfall}). Across the field-type L dwarfs, the onset of CH$_4$ becomes visually apparent for objects later than L4, appearing initially as a downward concavity from 3-3.6 $\mu$m, then completely dominating the blue side of the 4$\mu$m opacity window all the way through the cooler sequence. CH$_4$ and H$_2$O opacity similarly sculpts the shapes of the NIR J/H/K opacity windows, making them sharper at progressively colder temperatures, and introducing a narrow absorption feature at 1.15 $\mu$m. When plotted with flux (Jy) on the vertical axis, the relative heights of the J- and K-band peaks (1.3, 2.2 $\mu$m) begin to reverse at the L/T transition, with the K-band peak falling below the H-band peak (1.6 $\mu$m) at subtype T0. The H-band peak eventually becomes shorter than the J-band peak at subtype T4. In the mid-IR, CO$_2$ and CO become increasingly apparent beyond subtype L3, with growing absorption notches at 4.15-4.3, and 4.4-4.9 $\mu$m, respectively. The fundamental vibration band of the CO band is visible as a subtle peak at 4.67 $\mu$m that appears at subtype L4 and persists through the rest of the sequence. The opacity window near 4 $\mu$m shifts redward for the late-T and Y dwarfs due to the saturation of the CH$_4$ opacity band, and the declining CO abundance with decreasing temperature. The prominence and sharpness of the 4 $\mu$m window peak are maximized at the L/T transition, where CH$_4$ and CO opacity together sandwich the peak of the emergent flux.

The low-gravity sequence for subtypes L0 and L5 is shown in Fig. \ref{fig:lowg}. The most noticeable trend with decreasing gravity is the slope from 0.75-1.2 $\mu$m, which becomes less steep. The L5$\gamma$ object 2MASS J08350622+1953050 shows relatively low-amplitude water features alongside a more clearly defined continuum level, indicative of a more isothermal atmosphere. The field-gravity L5 shows water feature amplitudes roughly 4 times stronger than the $\gamma$ equivalent. Furthermore, with decreasing gravity, the height of the 2.2 $\mu$m flux peak grows by tens of percent relative to the 1.6 $\mu$m peak. 

The low-metallicity sequence for subtypes L0 and L7 is shown in Fig. \ref{fig:lowz}. Decreasing metallicity at a fixed subtype imparts a drastic change in the appearance of the NIR region, showing a reversal in the relative heights of the opacity windows at 1.3 and 2.2 $\mu$m. For both subtypes, the flux ratios of these features range from $\sim$0.5 for the field objects to $\sim$2.5 for the esd objects. This change is leveraged for spectral metallicity classification, and is attributed to the onset of H$_2$-H$_2$ pressure-broadened collisionally-induced absorption opacity (CIA). Namely, as the metallicity and therefore opacity of the atmosphere declines, the photosphere moves deeper to higher pressures where the broad continuum H$_2$-H$_2$ CIA feature becomes a dominant opacity source from 1.7-3 $\mu$m.

\section{Model Grids}

\begin{figure*}
    \centering
    \includegraphics[width=0.95\linewidth]{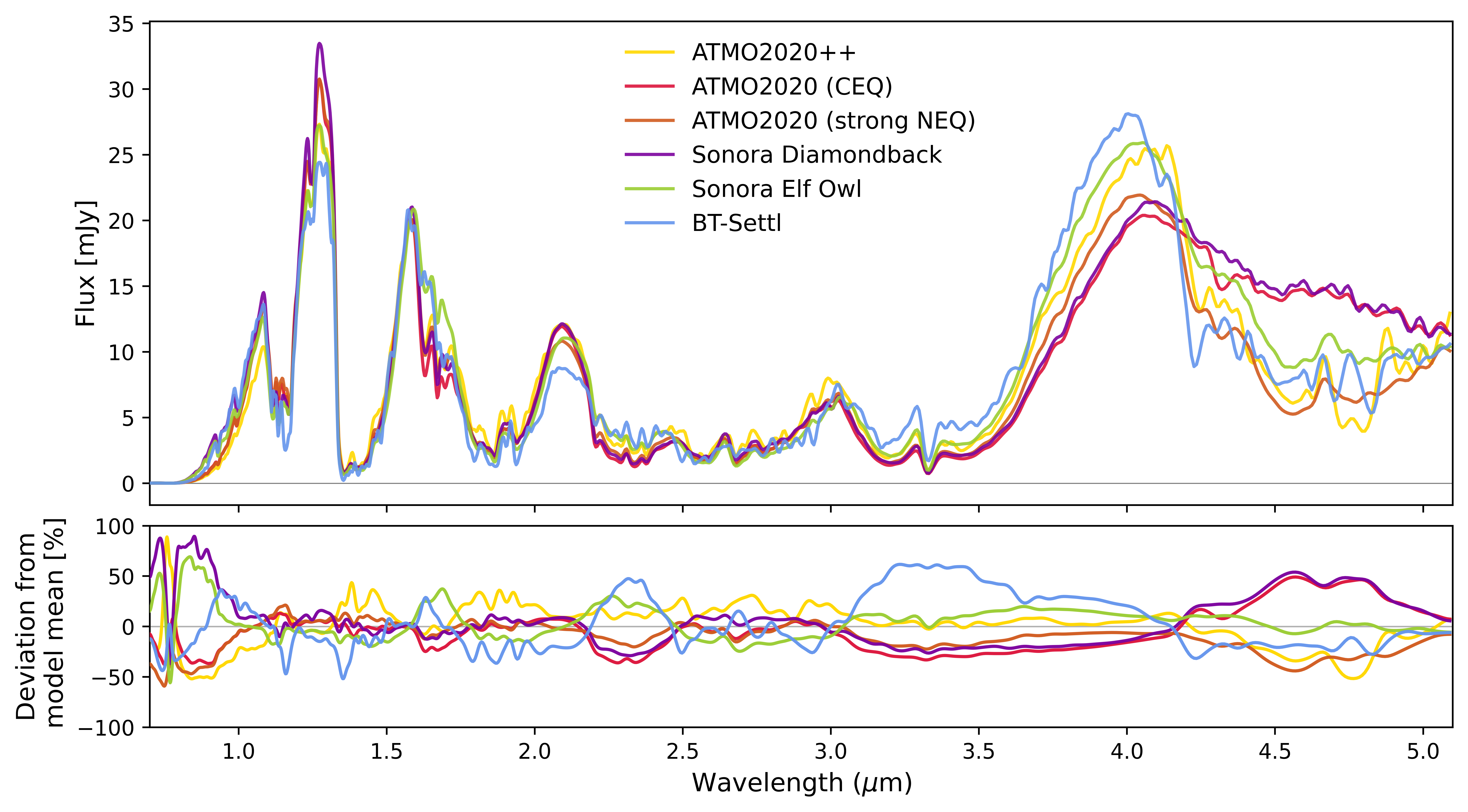}
    \caption{Top: Models of identical gravity (log($g$) = 5.5), metallicity (Solar), and temperature (1200 K) for a 1 $R_J$ brown dwarf at 5 pc across all the grids considered in this work. Bottom: Each model's percent residual relative to the smoothed mean of the models. The Elf Owl model has Solar C/O = 0.5, and log($k_{zz}$) = 2 (weak mixing), and the Diamondback model is cloud-free with Solar C/O. Despite having identical temperature, gravity, and composition, the chemistry and mixing treatment have significant impacts on the spectra. The largest discrepancies (factor of $\sim$2) occur in the optical and the less-explored 3-5 micron region, where differing chemistry assumptions strongly influence the abundance ratios of the carbon species.}
    \label{fig:modcomp}
\end{figure*}

We compare the observed spectra to five separate state-of-the-art model grids (see Table \ref{tab:model_grids}), each of which emphasizes different aspects of the thermochemistry, and includes or omits clouds. The grids are mostly spaced by 100 K in $T_\mathrm{eff}$ and 0.5 in log($g$), with other variables changing by grid. These models emerge from an enormous variety of subjectively chosen input assumptions that govern the coupled composition, thermodynamics, and radiative transfer of each atmosphere. The selected molecular and continuum opacity databases, equations of state, atomic abundances, chemical reaction rates, thermodynamic coefficients, optical constants, convective mixing length scales---and the computational schemes connecting them---vary greatly from model to model. As we show in Fig. \ref{fig:modcomp}, even grid models with the same input temperature, gravity, and metallicity show relative flux differences as high as 2 at some wavelengths due to their different handling of chemistry. The models differ the most in the thermochemistry-sensitive 3-5 $\mu$m region covering the features of CO, CH$_4$, and CO$_2$. Given that they have more differences than similarities, each grid represents a valiant independent attempt at emulating the delightfully complex atmospheres of brown dwarfs. The aim of our comparison is not to rank their relative ``quality" or ``truthfulness" but rather to see which input assumptions and parameters correlate with poor agreements between observations and models.

Each grid reports its spectra in different units. We convert each model to the spectral flux density $F_\lambda\;[\mathrm{W\,m^{-2}\,m^{-1}}]$, which is the luminosity per wavelength emerging from a square meter of the object's photosphere. These flux densities are then rescaled by the scaling factor

\begin{equation}
    F_{\lambda}^{\mathrm{obs}} = \left(\frac{R_\star}{d}\right)^2 F_{\lambda}^{\mathrm{surf}}
    \label{eq:scaling}
\end{equation}

where $R_\star$ is the radius, and $d$ is the measured distance to the object. We finally bring the model units to the Jy units of the SPHEREx observations

\begin{equation}
F_\nu(\mathrm{mJy}) = F_\lambda \frac{\lambda^2}{c} \times 10^{29}
\end{equation}

where $\lambda$ and $c$ are the wavelength and speed of light, respectively.

The following subsections describe each model grid in broad terms. Table \ref{tab:model_grids} summarizes their key attributes, including their treatment of clouds and chemistry, their temperature ranges, native resolving power $R_\lambda$ in the SPHEREx bandpass, and the other parameter dimensions. For more detailed descriptions of each model, the reader is referred to the first reference of each subsection, and the references within. 

% Assuming that each grid is equally trustworthy, this imposes systematic model-dependent uncertainties that are far larger than the observational uncertainties.

% While no grid can be expected to match all 

% There are two main approaches for constraining atmospheric parameters from complex brown dwarf spectra: atmospheric retrievals, and model grid fits. The former approach is sophisticated and involves fitting an individual object with a flexible thermal profile, chemistry, and clouds. 
% The challenge of interpreting the complex spectra of brown dwarfs 
% Forward model grids are 
% theoretical products that can 
% The model grids are binned to the SPHEREx spectral channels for each object, 

\begin{table*}
\centering
\caption{Summary of model grids used in spectral fitting.}
\label{tab:model_grids}
\begin{tabular}{lllllll}
\hline\hline
Grid & Cloud Treatment & Chemistry & $T_{\rm eff}$ (K) & Varied Parameters & $R_\lambda$ & $N$ models \\
\hline
ATMO2020++ & Cloud-free  & Non-eq. & 200--1200 & $T_{\rm eff}$, $\log g$, $[\rm M/H]$ & 3{,}000 & 299 \\
ATMO2020   & Cloud-free  & Eq., non-eq. & 200--3000 & $T_{\rm eff}$, $\log g$, chem.\ mode & 1{,}000 & 510 \\
BT-Settl   & Microphysical (Rossow) & Eq. & 400--2400 & $T_{\rm eff}$, $\log g$ & 8{,}000 & 106 \\
Sonora: Elf Owl & Cloud-free & Non-eq. & 275--2400 & $T_{\rm eff}$, $\log g$, $[\rm M/H]$, C/O, $k_{zz}$ & 60{,}000 & 43{,}252 \\
Sonora: Diamondback & Sedimentation ($f_{\rm sed}$) & Eq. & 900--2400 & $T_{\rm eff}$, $\log g$, $[\rm M/H]$, $f_{\rm sed}$ & 200{,}000 & 1{,}440 \\
\hline
\end{tabular}
\end{table*}

\subsection{ATMO2020}
ATMO2020 \citep{Phillips2020} is a coupled atmosphere-interior evolution grid computed at Solar metallicity using the state-of-the-art ATMO 1D radiative-convective equilibrium code. The models span 200-3000 K in 50-100 K intervals, and log($g$) 2.5-5.5 in steps of 0.5 dex. This grid includes non-equilibrium chemistry, and parametrizes the vertical diffusion coefficient with a power-law gravity dependence ($k_{zz} \propto g^{-2}$) to impose a constant dynamical timescale for objects of different mass. In addition to including chemical equilibrium models where the timescale for chemical equilibration is infinitesimal, they compute strong and weak non-equilibrium chemistry models, corresponding to $k_{zz} = 10^4$ and $10^6$ cm s$^{-2}$ at log($g$) = 4.5, respectively. These values were selected following the finding of \cite{Leggett2017} that this range of vertical mixing best matches the $[4.5]-M'$ (Spitzer Ch. 2 minus M' band) colors of late T- and Y-dwarfs at the same fixed gravity. As seen in Figs. 7 and 8 of that work, none of the several plotted model grids successfully predict the color-magnitude behavior across the full LTY span, with systematic departures on the order of 1-2 magnitudes. The models are downloaded from the ERC Opendata server\footnote{https://noctis.erc-atmo.eu/sharing/zyU96xA6o}.

\subsection{ATMO2020++}
The ATMO2020++ model grid \citep{Leggett2024} targets late T and Y dwarfs (250-1200 K), and includes a disequilibrium chemistry scheme with a fixed vertical mixing diffusion coefficient of $k_{zz} = 10^7$ cm s$^{-2}$. This grid also uniquely computes a diabatic thermal profile in place of the usual adiabatic assumption, to better account for the disruption of convection by the observed rapid rotation of real brown dwarfs. This results in a cooler lower atmosphere, and a warmer upper atmosphere. \cite{Leggett2023} find that this grid better replicates the spectra of late-type brown dwarfs as measured by JWST. The grid is spaced by 25 K from 225 to 350 K, by 50 K from 350 to 450 K, and 100 K onward. The model log-gravity is spaced by 0.5 dex, and ranges from 2.5 to 5.5, which corresponds to a mass range of 0.12 to 120 Jupiter masses, assuming a Jupiter radius. The models are computed with metallicities of [Fe/H] = -0.5, 0.0, and 0.3. We download the full grid from the ERC Opendata server\footnote{https://noctis.erc-atmo.eu/sharing/9puhIZma2}.

\subsection{BT Settl}
The BT-Settl model grid is computed with the PHOENIX code, and includes a sophisticated parameter-free treatment of cloud formation and settling with vertical mixing \citep{Allard2012, Allard2013, Allard2014}. The radiative-convective equilibrium model self-consistently computes the condensation of 55 optically active condensible species that contribute to a vertical grain size distribution profile informed by separate timescales for nucleation, condensation, and settling. These timescales are derived from more sophisticated 3D hydrodynamics \citep{Freytag2010} and cloud microphysics \citep{Rossow1978} models. BT-Settl uses the now-outdated water opacities of \cite{Barber2006}. Despite this, the grid closely reproduces the $J-K_s$ colors of stars, and also matches the bottom of the Main Sequence as it connects to the L dwarfs. We download the AGSS2009 400-2400 K, log($g$) 3-5.5 Solar metallicity models from the SVO Theory Server \footnote{svo2.cab.inta-csic.es/theory/newov2/index.php?models=bt-settl-agss}.

\subsection{Sonora Diamondback}
The Sonora Diamondback coupled atmosphere-interior evolution grid \citep{Morley2024} includes clouds parameterized using the mass-balance approach of \citep{Ackerman2001}. The radiative-convective equilibrium models are computed using the approach of \cite{Saumon2008} (and references therein). The parameter $f_\mathrm{sed}$ is held constant with height and encodes the cloud sedimentation efficiency, with small values resulting in thick clouds. The models incorporate the Mie scattering absorption coefficients of the amorphous form of 4 cloud species: MgSiO$_3$, Mg$_2$SiO$_4$, Fe, and Al$_2$O$_3$. These are identified as the main contributors of aerosol opacity. The gradual contraction and cooling of the model interior is coupled to the thermal structure of the atmosphere: when a model atmosphere cools below $\sim$1300 K, the clouds suddenly clear, resulting in a brief stalling of the brown dwarf's cooling and a brief acceleration in its contraction \citep[e.g. Fig. 14 and 15,][]{Morley2024}. This theoretical prediction is seen as a local occurrence rate maximum in the observed population of brown dwarfs at the L/T transition \citep[e.g., Fig.~17 of][]{Kirkpatrick2024}; \citep[e.g., Fig.~13 of][]{Saumon2008}. Infrared spectroscopy across the transition with Spitzer shows a similar clearing of the clouds, with $f_\mathrm{sed}$ rapidly rising below 1300 K and with fits preferring cloud-free models \citep[e.g.][]{Stephens2009}. The Diamondback grid also includes cloud-free models, which show a lack of the distinctive evolutionary radius ``bump." While the models impose chemical equilibrium, they include the effect of vertical mixing via the vertical eddy diffusion coefficient $k_{zz}$ varies with height and is computed using mixing length theory, without including $k_{zz}$ as a free parameter (like Elf Owl). The Diamondback grid ranges from 900 to 2400 K, 3.5-5.5 in log($g$), with $f_{sed}$ spanning 1-8, and is computed at Solar, +0.5, and -0.5 metallicity. We download Version 2 of the full grid from Zenodo$\footnote{https://zenodo.org/records/12735103}$.

\subsection{Sonora Elf Owl}
The Sonora Elf Owl model grid \citep{Mukherjee2024} is built using the same opacity database and a similar modeling approach to the Diamondback grid, but with an updated python-based implementation \citep{Mukherjee2023}. Unlike Diamondback, Elf Owl includes non-equilibrium chemistry due to vertical mixing, as governed by the fixed vertical mixing coefficient $k_{zz}$. Also unlike Diamondback, these atmosphere models are strictly cloud-free, and do not couple to an interior evolution scheme. The grid spans a wider range of temperatures than Diamondback, 275-2400 K, with a fine, 25 K spacing at the cold end. The strength of non-equilibrium chemistry governs the relative strengths of the carbon-bearing species in the mid-IR, with the CO/CH$_4$ and CO$_2$/CH$_4$ ratios varying by two orders of magnitude across the span of considered log($k_{zz}$) values (2-9) for a 700 K object. The resulting effect is a factor of 2 flux difference in the carbon-rich 4.1-4.9 $\mu$m region. The spread in metallicity has a similarly pronounced effect on the spectra in the same CO/CO$_2$ region, and shows a strong influence on the flux window at 2.2 $\mu$m, corresponding to the broadband continuum opacity of H$_2$-H$_2$ collisional absorption. The relative strengths of the carbon features are not only sensitive to the vertical mixing and metallicity of the atmosphere, but also the carbon-to-oxygen ratio, which spans values of 0.5 (sub-Solar) to 2.5 (greatly enhanced) in the grid. The Elf Owl grid therefore represents an important early foray into disentangling the competing effects of non-equilibrium chemistry and composition. Future Sonora grids will additionally include clouds alongside non-equilibrium chemistry. We download the latest version of the full grid from Zenodo$\footnote{https://zenodo.org/records/15150881}$.

\section{Spectral Fitting}
We compare each measured spectrum to the ensemble of models, binning the models to the SPHEREx wavelength channels. We bin the models by taking the mean of the model flux values bounded by the bandwidth of each spectral measurement. Given the proximity of these objects, we neglect extinction by the interstellar medium. The stellar radius is fitted as a free parameter, while the distance is fixed to the central value of the distance measurement (see Eq. \ref{eq:scaling}). The distance uncertainty is propagated through the radius uncertainty. The SPHEREx science team has examined repeat measurements of stable calibrators in the deep fields at the same wavelengths, which suggest a typical visit-to-visit RMS scatter of 3$\%$ (SPHERE Science Team, private comm.).  This term will be added in later data processing, but we add it here in quadrature with the original flux uncertainties. In addition to estimating goodness-of-fit with $\chi^2_\nu$, we compute the $G_k$ statistic as defined in Eq. 1 of \cite{Cushing2008}, which weighs the information content of a spectral channel in proportion to its wavelength width. Given that the SPHEREx spectra vary from R$\sim$40 in the NIR to R$\sim$110 in the mid-IR, the $G_k$ statistic prevents the narrower and lower-S/N channels from dominating the fits. We do not include the number of fitted grid variables in our goodness-of-fit estimates, since the goal is not to compare the different grids. Also, since the number of fitted points is in the hundreds to thousands, the effect of adding a couple of free parameters to the denominator is negligible.

%  This and the fact that each dimension has a different sampling resolution means that attempts at estimating uniform statistical uncertainties between grids are futile without interpolating and greatly super-sampling them. Sophisticated spectral fitters that account for model systematics \citep[e.g.,][]{Czekala2015, Zhang2021} and forward model retrievals with flexible atmospheric profiles \citep[e.g.,][]{Line2017, Lothringer2024} produce true posteriors and covariances, but are beyond the scope of the present work. Therefore, to avoid undue confusion, we choose to report only the central value of the best-fitting model along each parameter axis.

The coarse nature of model grids makes for jagged, discretized fit posteriors which do not approximate a normal distribution. In this work, we intentionally restrict our model fits to the original parameter grid points rather than generating continuous posteriors through interpolation in $T_{\mathrm{eff}}$, $\log(g)$, [Fe/H], or cloud and chemistry parameters, as is common in the field. As a result, we do not quote formal uncertainties on the best-fitting parameters. The grids are coarsely sampled, with typical spacings of $\Delta T_{\mathrm{eff}}\sim100$ K, $\Delta\log(g)\sim0.5$ dex, and $\Delta\mathrm{[Fe/H]}\sim0.25-0.5$ dex. In practice, brown dwarf spectra can vary nonlinearly over comparable scales, especially near the L/T transition where cloud properties, vertical mixing, and chemistry change rapidly. Even sophisticated Bayesian fitting frameworks that account for interpolation uncertainties and correlated model residuals struggle to reconcile the temperatures, radii, metallicities, and gravities of well-studied benchmark objects, often finding systematic differences that are larger than the grid spacings \citep[e.g.,][]{Zhang2021}. Using their interpolated Bayesian grid fitting approach, they find offsets on the order of 1.3 dex in $\log(g)$ and 0.4 dex in [Fe/H] for well-understood benchmark objects, which is much larger than the spacing of their grids. With their ``traditional" interpolated grid-fitting approach, they report similar offsets, with uncertainties that are questionably 10 or even 100 times smaller than the grid spacing. Alternatively, sophisticated atmospheric retrievals with flexible atmospheric profiles \citep[e.g.,][]{Line2017, Lothringer2024} produce true posteriors and covariances, but are beyond the scope of the present work. Therefore, to avoid constructing posteriors with assumptions of the subgrid behavior, and to avoid imposing additional and unjustified systematics, we elect to instead report the best-fitting 10$\%$ of native models for each grid, and leave a more sophisticated model fitting analysis for a future study. These best-fitting model parameters along with their goodness-of-fit are included in a machine-readable table, which one may interpolate if they wish to derive sub-grid best-fit parameters and confidence intervals.

A handful of selected spectral fits are shown below, in Figures \ref{fig:L1}, \ref{fig:L9}, \ref{fig:T1}, \ref{fig:T6}, \ref{fig:T9}, and \ref{fig:Y4}. The figure legends report the temperature, $\chi^2_\nu$, radius scaling, metallicity, and gravity for each best-fit model. For the Diamondback and Elf Owl models, the legends also report the model's $f_\mathrm{sed}$ and log($k_\mathrm{zz}$), respectively.

The color image in each figure's inset is a median shift-aligned and distortion-corrected stack 2.5' x 2.5' of all available SPHEREx images centered on the target. The blue color channel is assigned to the median of Bands 1 and 2, green assigned to the median of Bands 3 and 4, and red assigned to the median of Bands 5 and 6. The images are aligned to the celestial coordinate grid, with North pointing up and East pointing left.

None of the model grids stand out with exceptional fit quality, with $\chi^2_\nu$ $\geq$ 10 for the majority of spectra with high S/N. In this regime, the best-fit model parameters do not necessarily reflect the true properties of the object. In the next section, we attempt a different approach at comparing the behavior of the models to the observed sequence.

\begin{figure*}
    \centering
    \includegraphics[width=.75\linewidth]{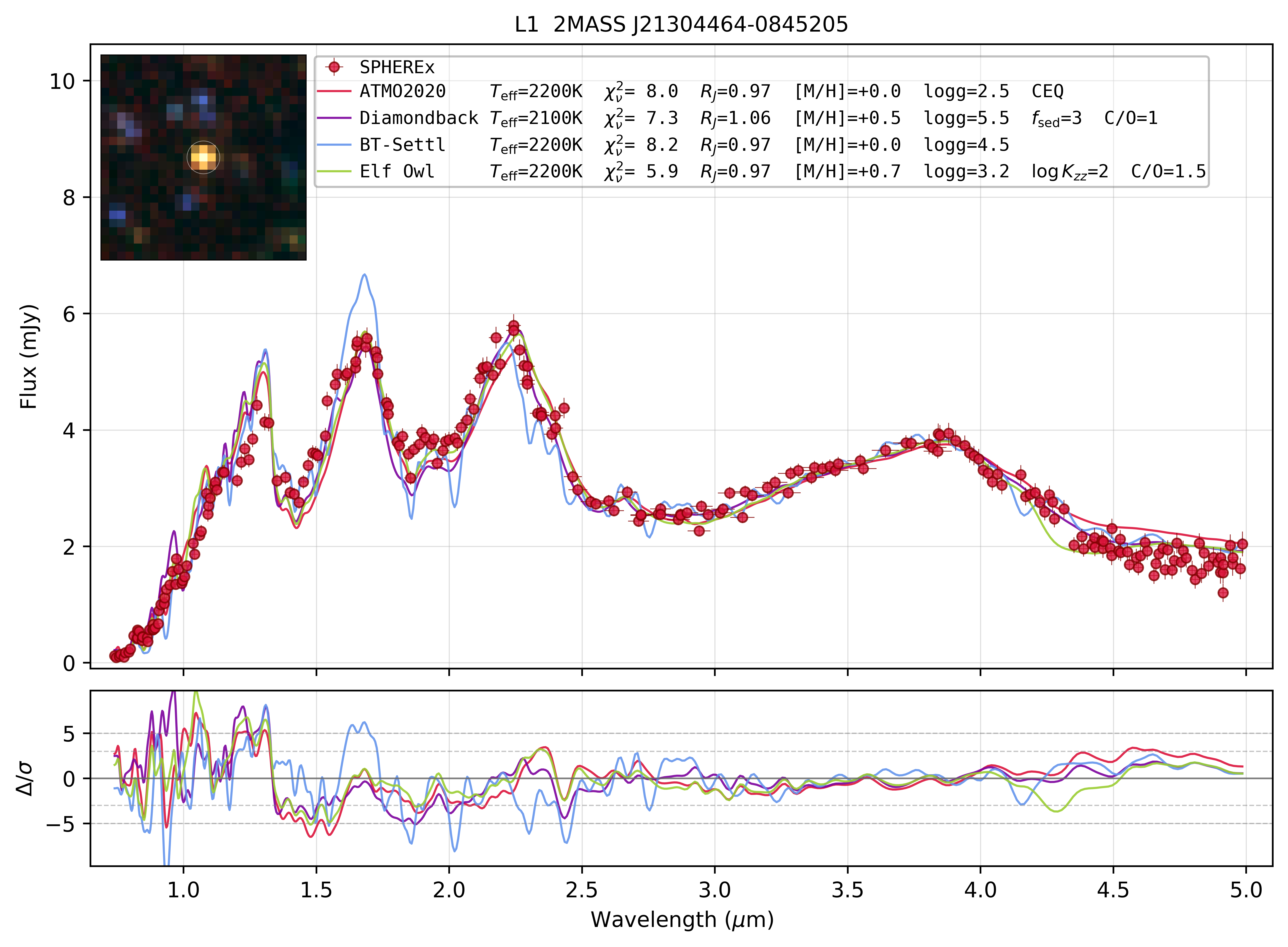}
    \caption{The L1 spectral standard dwarf 2MASS J21304464-0845205. The inset RGB image is the median-stacked SPHEREx image data (see text for full description). The bottom panel shows the flux residuals of the models from the smoothed and interpolated data. The best-fitting models span 100 K in effective temperature, contributing to a $\sim10\%$ systematic range in the best-fit radius. The cloudy Diamondback and BT-Settl models best capture the amplitudes of the NIR opacity windows and the mid-IR water absorption. All models have $\chi^2_\nu$ $\geq$ 5, and span 3 orders-of-magnitude in best-fit gravity.}
    \label{fig:L1}
\end{figure*}

\begin{figure*}
    \centering
    \includegraphics[width=.75\linewidth]{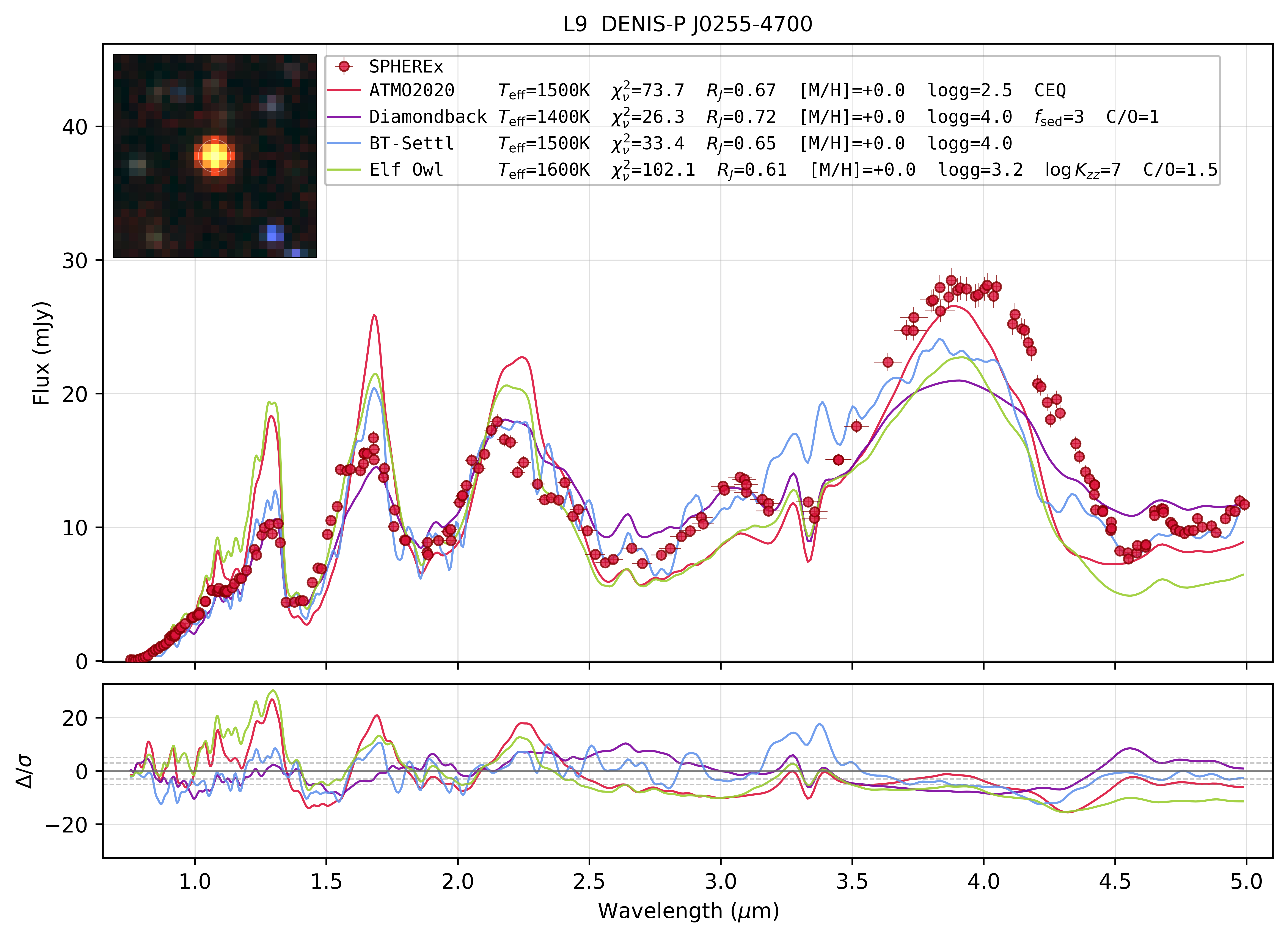}
    \caption{Same as Fig. \ref{fig:L1}, but for the L9 spectral standard dwarf DENIS J025503.3-470049. The Sonora Diamondback fit prefers a cloudy model, and the Elf Owl fit indicates a preference for strong non-equilibrium chemistry, but with $\chi^2_\nu$ $\geq$ 20, these conclusions are not well supported by the data. The 200-K span in recovered temperature, and the unphysically small and disparate radii, further illustrate the negative impact of model systematics on the trustworthiness of recovered physical parameters. The Diamondback model prefers clouds, which help reduce the prominence of the J, H and 4 $\mu$m peaks, helping the fit quality on the blue end of the spectrum.}
    \label{fig:L9}
\end{figure*}

\begin{figure*}
    \centering
    \includegraphics[width=.75\linewidth]{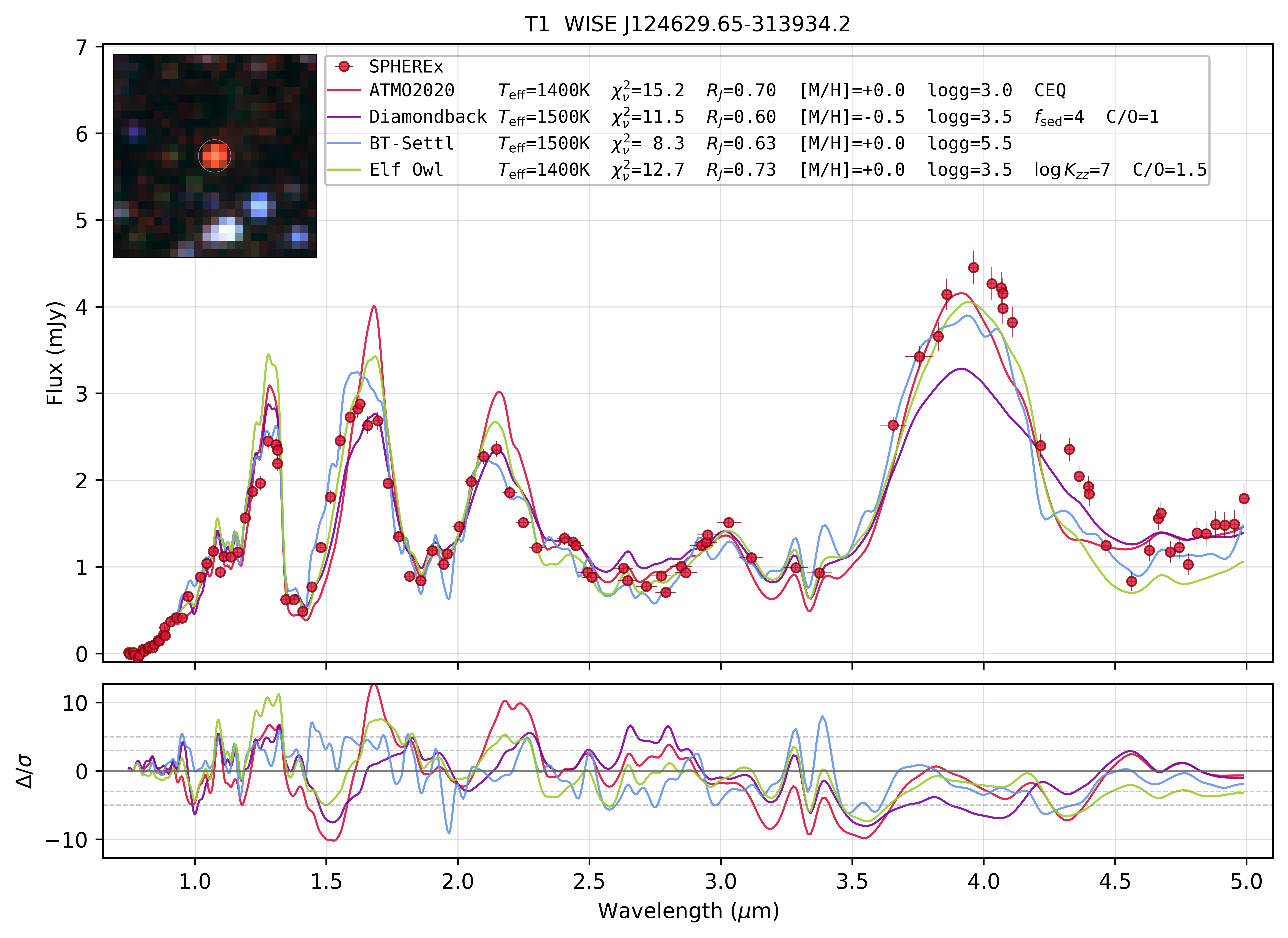}
    \caption{Same as Fig. \ref{fig:L9}, but for the T1 dwarf WISE J124629.65-313934.2. All models fail to capture the true amplitude of the 4 $\mu$m opacity window, while most models overestimate the J/H/K peaks. All models find radii that are considerably smaller than theoretical expectations, with a systematic scatter of $\sim$20$\%$ partially resulting from the 100 K range in effective temperature. The best-fit models are $\sim$200 K warmer than expected for a typical T1 dwarf, possibly due to overfitting of the 4 $\mu$m peak.}
    \label{fig:T1}
\end{figure*}

\begin{figure*}
    \centering
    \includegraphics[width=.75\linewidth]{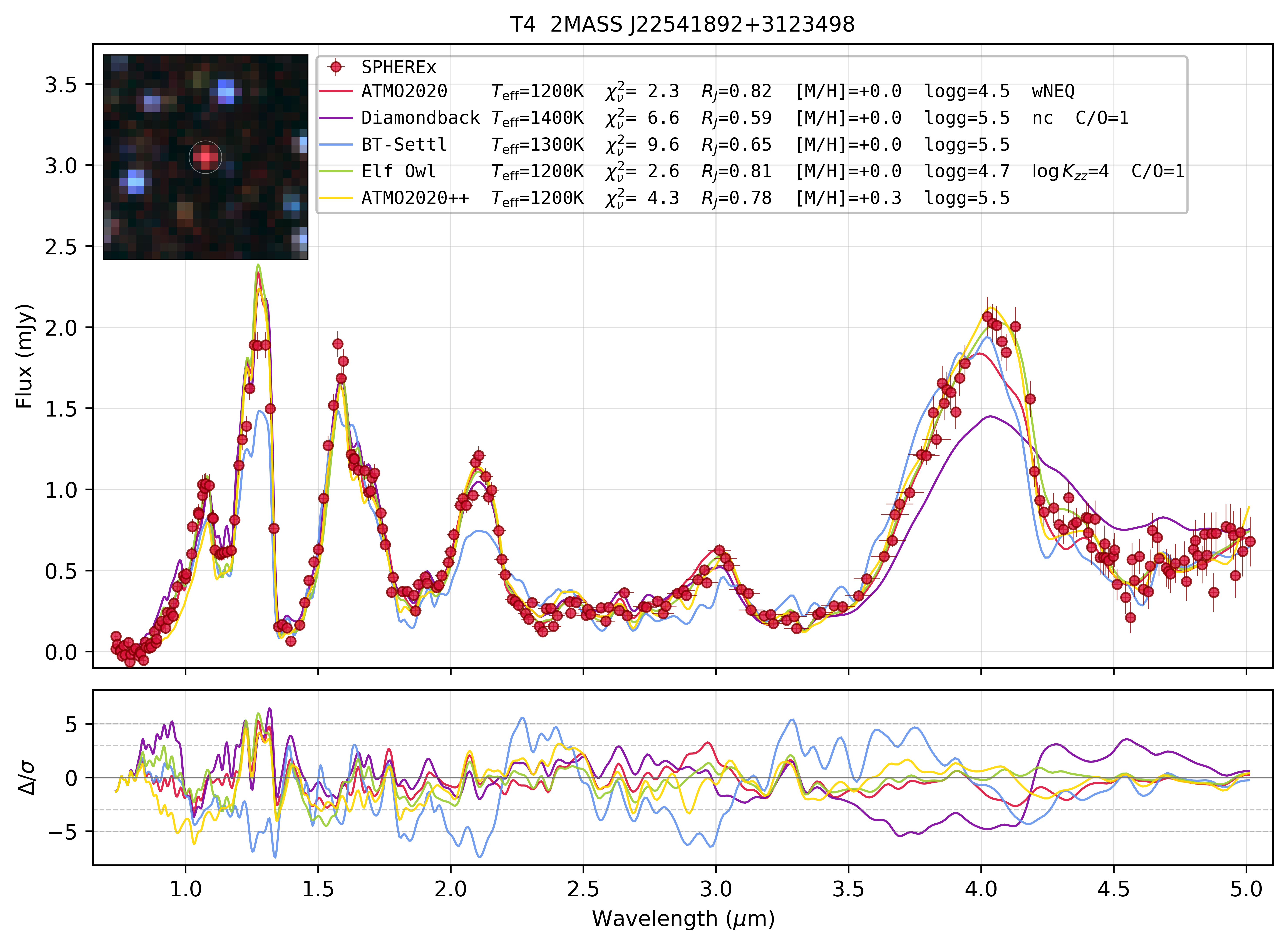}
    \caption{Same as Fig. \ref{fig:L9}, but for the T4 dwarf 2MASS J22541892+3123498.}
    \label{fig:T4}
\end{figure*}

\begin{figure*}
    \centering
    \includegraphics[width=.75\linewidth]{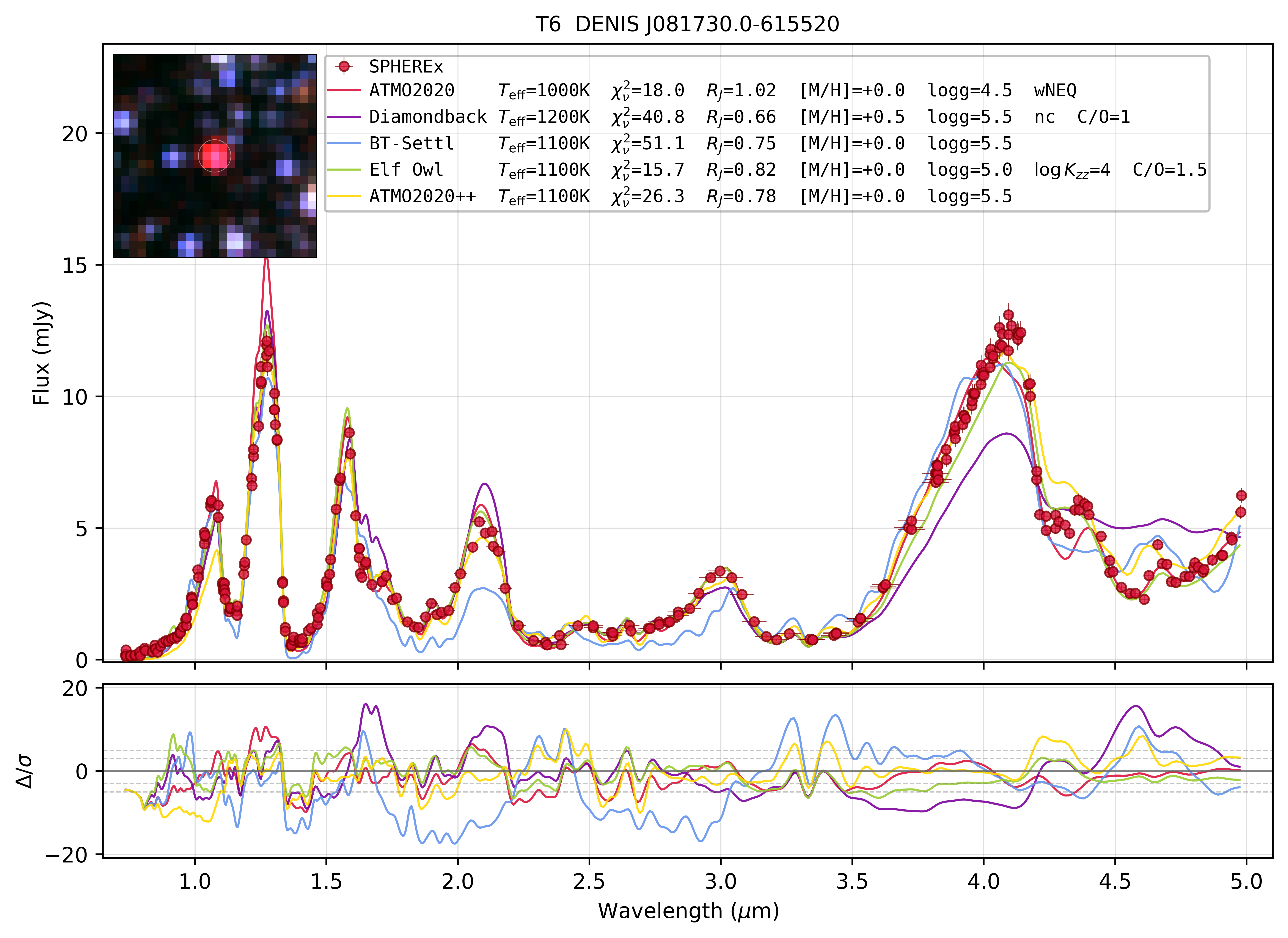}
    \caption{Same as Fig. \ref{fig:L9}, but for the T6 dwarf DENIS J081730.0-615520. The strong absorption seen at the red end of the spectrum indicate a relatively high abundance of CO and CO$_2$. }
    \label{fig:T6}
\end{figure*}

\begin{figure*}
    \centering
    \includegraphics[width=.75\linewidth]{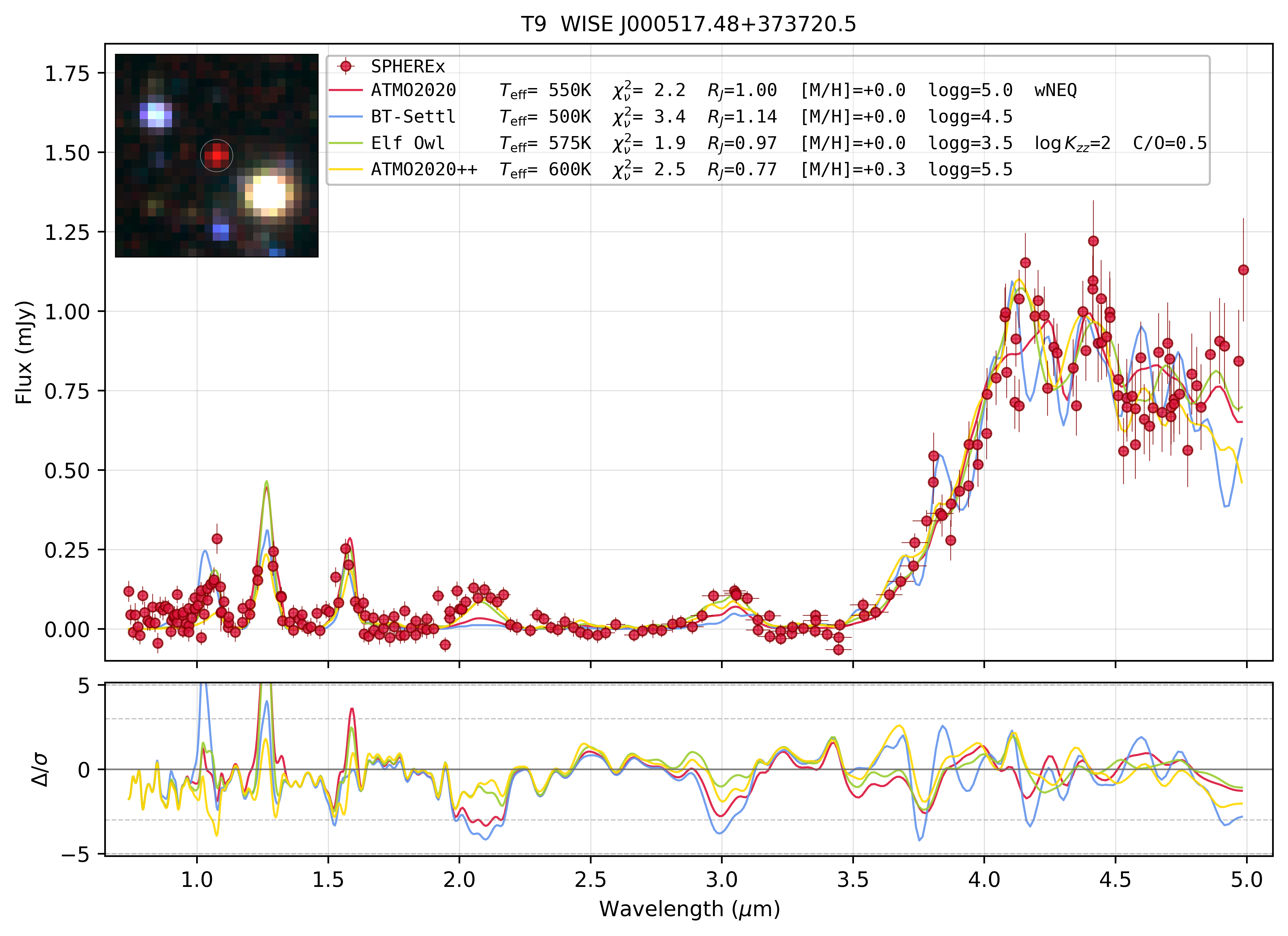}
    \caption{Same as Fig. \ref{fig:L9}, but for the T9 dwarf WISE J000517.48+373720.5. The best-fitting models range from 550-650 K, and yield radii that systematically differ by $\sim50\%$. The recovered metallicities similarly vary strongly between the models, ranging from strongly sub-Solar to super-Solar. Unlike the hotter T6 dwarf, this T9 shows relatively weak CO and CO$_2$ absorption.}
    \label{fig:T9}
\end{figure*}

\begin{figure*}
    \centering
    \includegraphics[width=.75\linewidth]{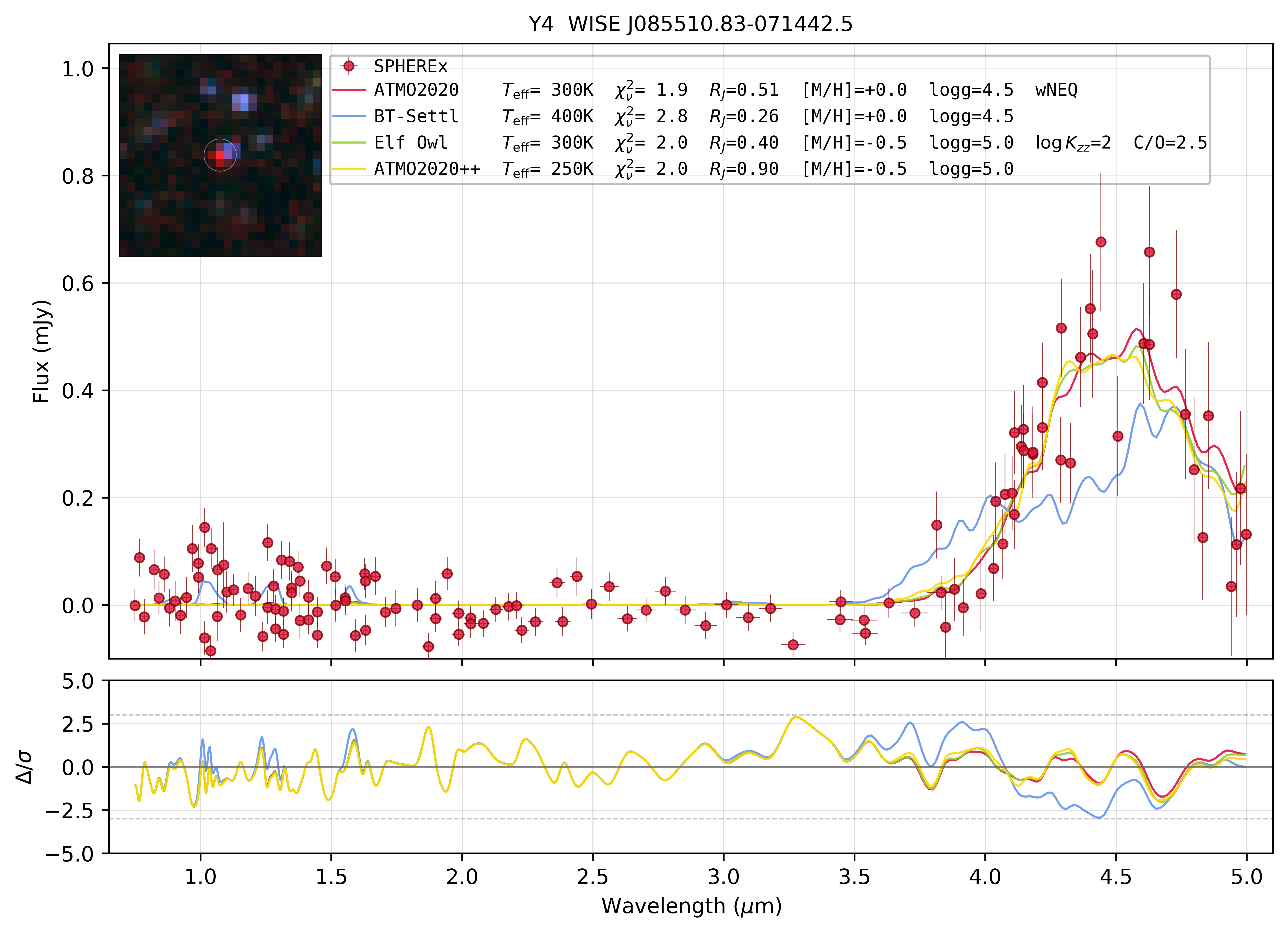}
    \caption{Same as Fig. \ref{fig:L9}, but for the $\gtrsim$Y4 dwarf WISE J085510.83-071442.5, the coldest brown dwarf known. Only flux redward of 3.9 $\mu$m is detected reliably. The factor of 3-4 span of recovered radii demonstrates the enormous systematic uncertainties associated with the characterization of Y dwarfs with atmosphere models.}
    \label{fig:Y4}
\end{figure*}

\subsection{Model Intercomparison}

\begin{figure*}
    \centering
    \includegraphics[width=.8\linewidth]{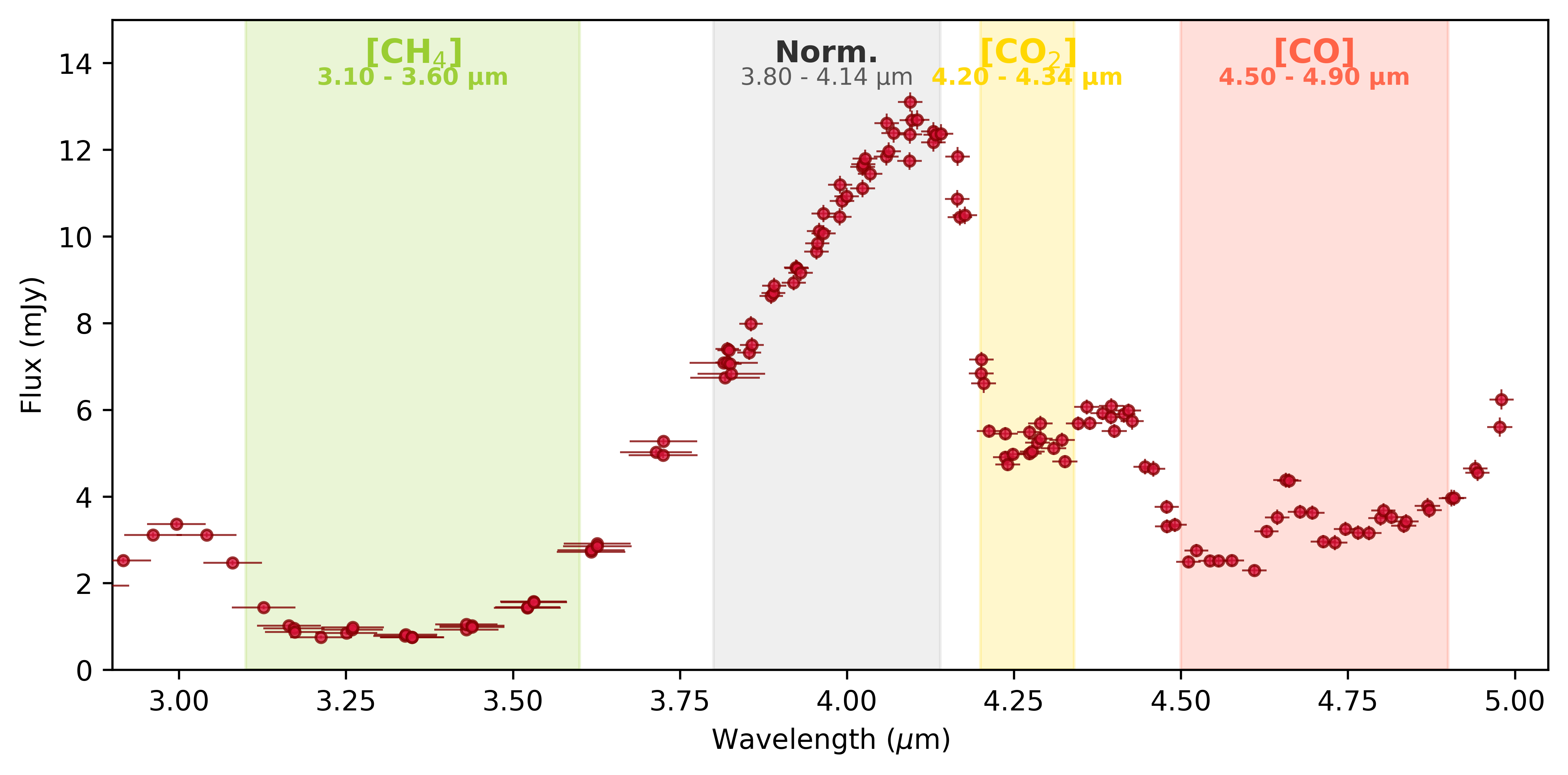}
    \caption{A schematic showing the ranges of the spectral indices defined in this work overlaid with the representative spectrum of the T6 dwarf DENIS J081730.0-615520. The 4 $\mu$m opacity window is the closest any spectrum gets to the ``continuum" flux level, and it stands out across the full sequence of brown dwarfs, making it a natural anchor point.}
    \label{fig:indices}
\end{figure*}

Population studies of brown dwarfs and exoplanets using the 2MASS/WISE filters, and post-cryo Spitzer/IRAC observations have been conducted \citep[e.g.,][]{Patten2006, Baxter2021, Sanghi2023}, but these broadband filters do not have the resolution or placement to identify individual chemical features, and therefore fail to distill the chemistry information. To leverage the wavelength and spectral type coverage of SPHEREx, and to better identify specific model deficiencies, we define a set of custom spectral indices corresponding to the thermochemically sensitive carbon species in the 3-5 $\mu$m range. We define three wavelength ranges centered on the most prominent absorption features of CH$_4$, CO$_2$, and CO, and compute their magnitudes normalized relative to the universally prominent 4 $\mu$m opacity window, which lands between fundamental water absorption bands and encapsulates a region of declining CH$_4$ opacity. The magnitude of a given index species M$_i$ is given by

\begin{equation}
M_i = -2.5 \log_{10}\left(\frac{F_i}{F_{\rm norm}}\right)
\end{equation}

where $F_i$ is the mean of the in-band flux, and $F_\mathrm{norm}$ is the mean of the reference or normalization region. The magnitude uncertainties $\sigma_I$ are given by

\begin{equation}
\sigma_M = \frac{2.5}{\ln 10}
\left[
\left(\frac{\sigma_F}{F}\right)^2 +
\left(\frac{\sigma_{F_{\rm norm}}}{F_{\rm norm}}\right)^2
\right]^{1/2}
\end{equation}

where $\sigma_F$ and $\sigma_{F_{\rm norm}}$ are the quadrature-summed flux uncertainties within the respective bands. We select the spectral indices visually, using the high-S/N spectrum of the T6 DENIS J081730.0-615520, and the opacity ranges in Fig. \ref{fig:opacity_map} to place the region boundaries (see Table \ref{tab:spec_regions} and Fig. \ref{fig:indices}). We find that using narrower ranges does not necessarily improve the color contrast or the strengths of the trends, but does hurt the SNR of the measurements. These regions are somewhat contaminated by overlapping molecules, making them imperfect chemical indicators. Indeed, the lack of a clean continuum baseline flux level in any wavelength region makes brown dwarf spectra a challenge to interpret.

\begin{deluxetable}{lcc}
\tablecaption{Wavelength ranges of our spectral indices.\label{tab:spec_regions}}
\tablewidth{0pt}
\tablehead{
\colhead{Index} &
\colhead{Feature region ($\mu$m)} &
\colhead{Norm.\ region ($\mu$m)}
}
\startdata
CH$_4$  & 3.20--3.65 & 3.70--4.14 \\
CO$_2$  & 4.18--4.35 & 3.70--4.14 \\
CO      & 4.55--4.80 & 3.70--4.14 \\
\enddata
\end{deluxetable}

\begin{figure*}
    \centering
    \includegraphics[width=.9\linewidth]{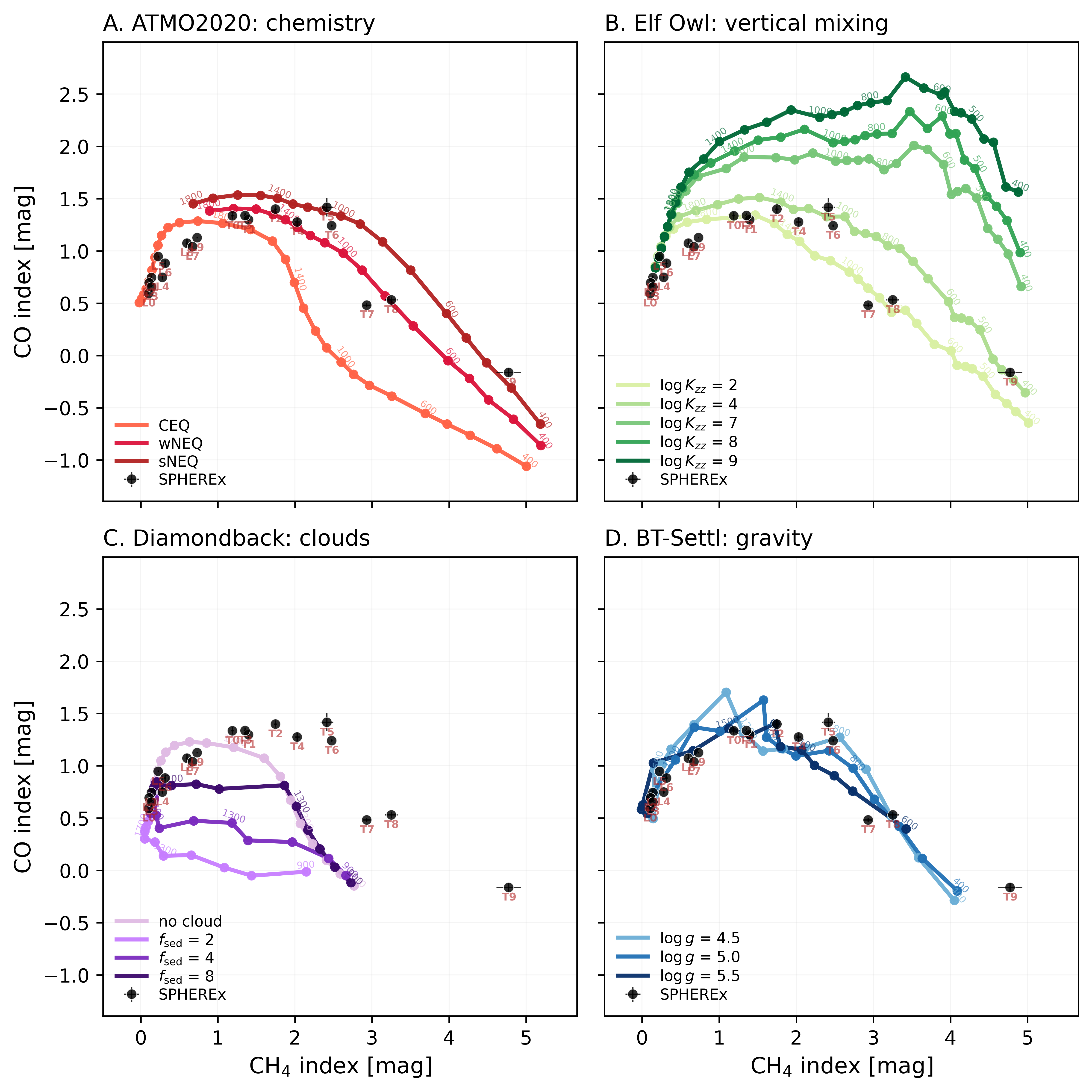}
    \caption{The relative strengths of the CH$_4$ and CO magnitude indices for the SPHEREx field brown dwarf sequence (black points) (Table \ref{tab:field}) compared to four model grids. The models all share Solar metallicity ([Fe/H] = 0) and C/O ratio, and have log($g$) = 5, except for BT-Settl, which spans a range of log($g$). Stronger CH$_4$ and CO absorption results in a larger index value and corresponds to higher abundances of those gases. The CO index drops precipitously from type T6 to type T7, spanning $\sim$1000 to 700 K. Meanwhile, the CH$_4$ index varies smoothly across this chemical transition.}
    \label{fig:co_ch4}
\end{figure*}

\begin{figure*}
    \centering
    \includegraphics[width=.9\linewidth]{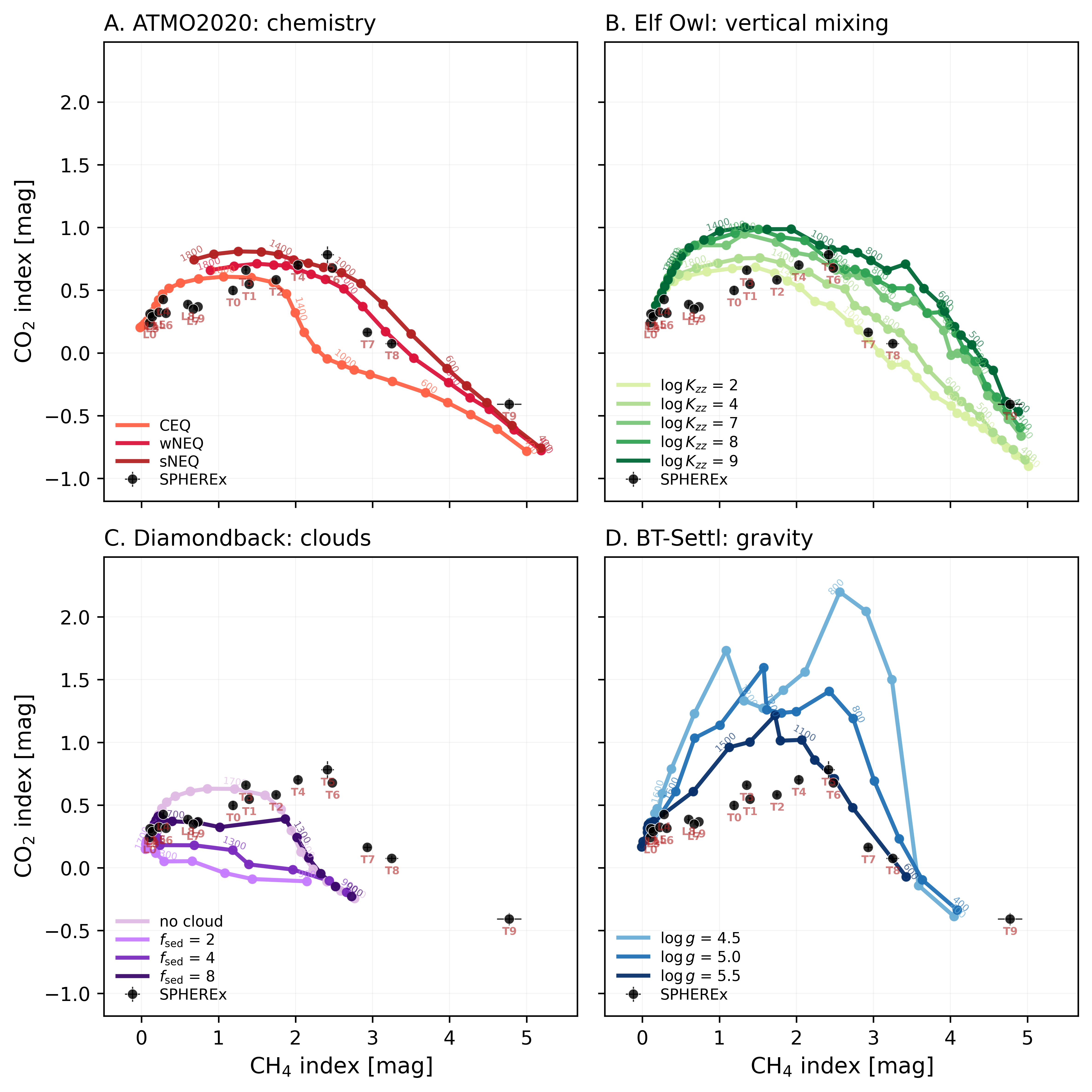}
    \caption{The relative strengths of the CH$_4$ and CO$_2$ magnitude indices for the SPHEREx field brown dwarf sequence (black points) (Table \ref{tab:field}) compared to models. The models all share Solar metallicity ([Fe/H] = 0) and C/O ratio, and have log($g$) = 5, except for BT-Settl, which spans a range of log($g$). Stronger CH$_4$ and CO$_2$ absorption results in a larger index value and corresponds to higher abundances of those gases. Similarly to the CO index, the CO$_2$ index decays significantly from spectral type T6 to T7, indicative of a chemical transition between $\sim$1000 and 700 K.}
    \label{fig:co2_ch4}
\end{figure*}

We next compute the same chemical indices for each model grid as a function of temperature, at a fixed log($g$) of 5, and compare the grids to the models in Figures \ref{fig:co_ch4} and \ref{fig:co2_ch4}, showing CO and CO$_2$ as a function of CH$_4$ strength, respectively. Higher values of the indices generally correspond to stronger absorption by that molecule, and this is made apparent by the trend of increasing CH$_4$ index with later spectral subtypes. Each spectral subtype is labeled, and the grid model temperatures are sparsely labeled to minimize visual crowding. The Y-dwarfs are excluded from these figures due to their low SNR in the CH$_4$ band, which is undetected by SPHEREx.

The sequence of observed field brown dwarfs shows fairly continuous and smooth behavior, with the spectral types increasing nearly monotonically from left to right. The Elf Owl models are shown with weak and strong vertical mixing (dashed and solid green lines, respectively), and the Diamondback models are similarly shown with and without clouds (solid and dashed purple lines, respectively). Two chemistry variants of ATMO 2020, chemical equilibrium and strong non-equilibrium, are shown as light and dark orange lines, respectively.

\section{Discussion}

While no grid matches spectra across the sequence, all of them qualitatively capture the general trend of CO and CO$_2$ with CH$_4$ abundance, showing a gradual rise from subtype L0 through T3 ($\sim$2300-1200 K), and then rapidly descending by 1.5 magnitudes for the cooler, CH$_4$-dominated objects (Fig. \ref{fig:co_ch4}, Fig. \ref{fig:co2_ch4}). The CO$_2$ index trend is similar in shape, with a maximum at subtype T5 ($\sim$1000 K), though the models tend to under-predict the strength of the CO$_2$ feature for the L dwarfs (Fig. \ref{fig:co2_ch4}). The Elf Owl models with weaker vertical mixing (log $k_\mathrm{zz}$ = 4) much more closely match the observations than the models with stronger mixing (log $k_\mathrm{zz}$ = 8). The same is seen for the ATMO2020 models, with the data aligning more with the stronger non-equilibrium chemistry model line. The T dwarfs show a CO dropoff of a similar magnitude to the Elf Owl low $k_\mathrm{zz}$ and ATMO2020 strong non-equilibrium (NEQ) models, while the earlier T and L dwarfs veer more toward the equilibrium models. The addition of clouds to the Diamondback models results in worse agreement in the CO index, and none of the models are able to match the strength of the CH$_4$ feature for objects later than T7. The CO magnitudes of the mid-T dwarfs appear well-represented by the Elf Owl (low $k_\mathrm{zz}$), ATMO2020 (weak NEQ), BT-Settl, and ATMO2020++ model grids alike, and this is corroborated by the relatively good fit statistics for these objects. 

The observed CO/CH$_4$ sequence appears to qualitatively lie along a smooth, curved trend line, with deviations on the order of 0.2 mag. The relatively monotonic progression of spectral type with CH$_4$ index, and the fairly regular spacing along this axis, indicates that CH$_4$ opacity varies smoothly with temperature, or more tautologically that it dominates the spectral classification. The lack of stochasticity may be an indication that either our sample size is too small or well-curated to see much astrophysical scatter, or that the chemical trends are relatively robust against individuality, following a well-traveled path in chemistry/cooling space. More objects with well-defined spectral subtypes are needed to comment on this empirical observation. The recent works of \cite{Gagne2026}, \cite{Brooks2026}, and \cite{Tu2026} present other sequences of SPHEREx brown dwarf spectra. The studies of \cite{Brooks2026} and \cite{Gagne2026} show similar broad H$_2$O, CO, CO$_2$, and CH$_4$ morphology across ultracool spectral types, and the sample of \cite{Tu2026} also quantitatively recovers smooth molecular-index trends and a CO/CO$_2$ turnover near $\sim$1000 K.

Lastly, the wide range of parameter space explored by Elf Owl gives a unique opportunity to test their observability and crosstalk. In Fig. \ref{fig:EO_co_co2_ch4} of the Appendix, we show the CH$_4$ index against CO$-$CO$_2$ for one-parameter slices through the Elf Owl grid to test the coupling of the carbon chemistry in the grid. CO$-$CO$_2$ captures the relative behavior of the two carbon-oxygen species, which are chemically linked to CH$_4$ through the temperature-, metallicity-, and mixing-dependent partitioning of carbon among CO, CO$_2$, and CH$_4$ \citep[e.g.,][]{Fortney2020, Sing2024}. The figure shows the same spectral sequence across varying vertical mixing, gravity, metallicity, and C/O ratio. Changing the vertical mixing and metallicity parameters results in the most CO$-$CO$_2$ variance at a fixed CH$_4$ index. Namely, increasing the vertical mixing from weak (log($k_{zz}$) = 2) to Jupiter-like (log($k_{zz}$) = 7) at SOolar metallicity results in 0-, 0.5-, and 1-dex differences in CO$-$CO$_2$ for L's, mid-T's and late-T's, respectively. A population-wide study of T dwarf carbon chemistry shows that they have log($k_{zz}$) = 5 - 8, which is higher than the log($k_{zz}$) = 2-4 values inferred from the plot. At Solar metallicity, the CO$-$CO$_2$ remains roughly constant, with just 0.5-dex of variation across the sequence. However, at a wider range of modeled [M/H] values, the models overlap at constant CH$_4$ index, making the CO$-$CO$_2$ index degenerate in the interpretation of atmospheric metallicity. C/O ratio and log($g$) contribute a further 0.5-dex of scatter, each. The fact that none of the Elf Owl cross-cuts seem to reproduce the sequence suggests that either our sample of objects spans a gradient in composition and thermochemistry, or that the imperfect spectral fits are due to some inherent model deficiency causing the indices to be poor representations of the chemical trends. Above all, the Fig. \ref{fig:EO_co_co2_ch4} shows that the chemistry-sensitive carbon species map a complex and degenerate space, and that future population-level inferences will require other thermochemical signatures, such as the NIR water features, to break these degeneracies.

\section{Conclusion}

In this work, we have presented 37 SPHEREx spectra of the nearest and highest-S/N L0-Y4 brown dwarfs spanning the full range of spectral subtypes (Fig. \ref{fig:waterfall}). As SPHEREx continues mapping the sky, these spectra will grow in S/N with the growing spectral sampling density, enabling more detailed investigations of their chemistry and variability. However, even the first-epoch SPHEREx spectra analyzed here reveal that state-of-the-art brown dwarf model grids generally fail to trace their complex features, meaning that our understanding of their nature is limited by model systematics rather than observational uncertainty. By examining the sequence through the lens of the thermochemically active species, which are now readily accessible with SPHEREx, we show that the models generally approximate the smoothly coherent abundance trends of CO and CO$_2$ as they are converted to CH$_4$ with declining temperature and increasing age. We also show that the addition of clouds does not improve Diamondback model fits (see Figs. \ref{fig:co_ch4} and \ref{fig:co2_ch4}), especially at the cloudy L/T transition (see Fig. \ref{fig:L9}), implying the need for more theoretical work connecting clouds with chemistry. Optimistically, the models seem to qualitatively represent the low-gravity and low-metallicity L dwarfs (see Fig. \ref{fig:L0gamma}, \ref{fig:L5beta}, \ref{fig:esdL0}, and \ref{fig:usdL0}, which probe a more extreme regime of parameter space. Future observational studies with expanded statistical samples will closely explore empirical trends with luminosity and age, while other works will attempt to decipher the competing effects of vertical mixing, metallicity, and bulk composition. 

\begin{acknowledgements}
    % \section{ACKNOWLEDGEMENTS}
    We thank Federico Marocco and Rocio Kiman for their fruitful discussions about spectral fitting. We additionally thank Jonathan Fortney and Caroline Morley for their suggestions on which models to use. The SPHEREx spectral image data used in this work were obtained from the SPHEREx Quick Release Spectral Images, QR2 \citep{spherex_qr2}, served by IRSA/IPAC. We acknowledge support from the SPHEREx project under a contract from the NASA/Goddard Space Flight Center to the California Institute of Technology. Part of the research described in this paper was carried out at the Jet Propulsion Laboratory, California Institute of Technology, under a contract with the National Aeronautics and Space Administration (80NM0018D0004). The authors acknowledge the Texas Advanced Computing Center (TACC) at The University of Texas at Austin for providing computational resources that have contributed to the research results reported within this paper.
    
\end{acknowledgements}

\appendix

\begin{figure*}
    \centering
    \includegraphics[width=.9\linewidth]{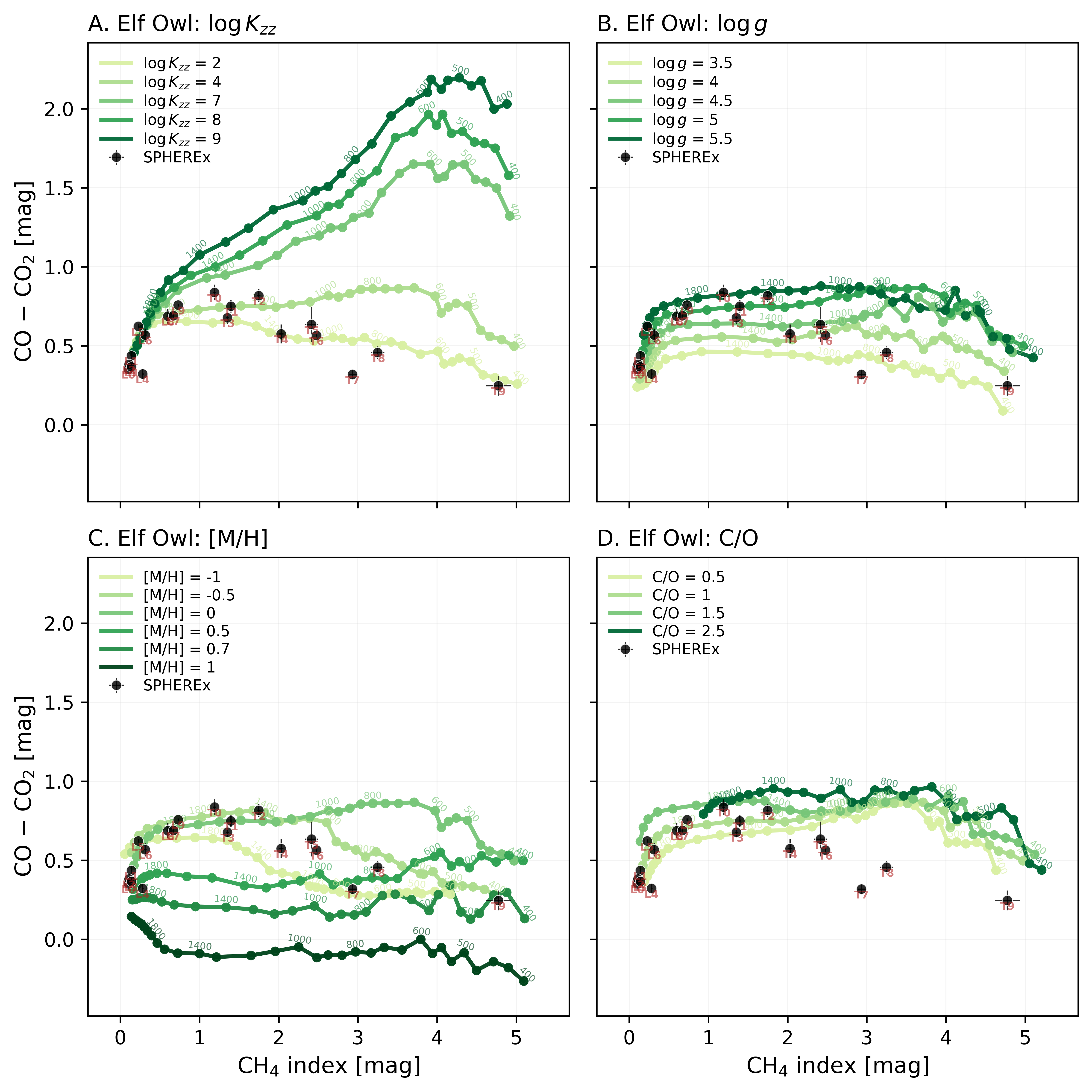}
    \caption{The relative strength of the CH$_4$ index versus the difference of the CO and CO$_2$ indices for the SPHEREx field brown dwarf sequence (black points) (Table \ref{tab:field}) compared to the various Elf Owl free parameters. Unless otherwise indicated by the legend, the models have Solar metallicity and C/O ratio, with log($g$) = 5.0, and log($k_{zz}$) = 4.0. }
    \label{fig:EO_co_co2_ch4}
\end{figure*}

\begin{figure*}
    \centering
    \includegraphics[width=.75\linewidth]{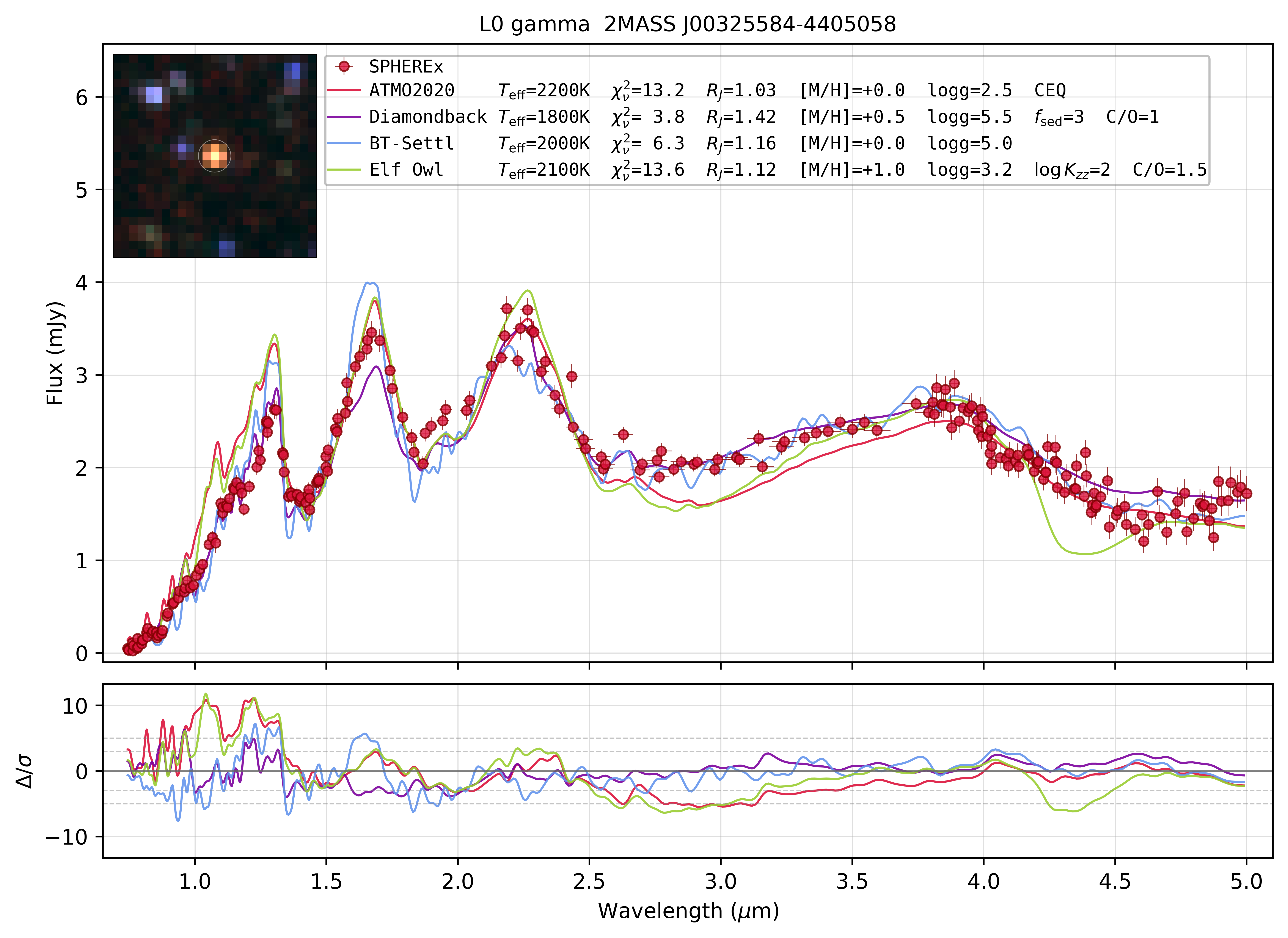}
    \caption{Same as Fig. \ref{fig:L9}, but for the low-gravity L0$\gamma$ dwarf 2MASS J00325584-4405058. }
    \label{fig:L0gamma}
\end{figure*}

\begin{figure*}
    \centering
    \includegraphics[width=.75\linewidth]{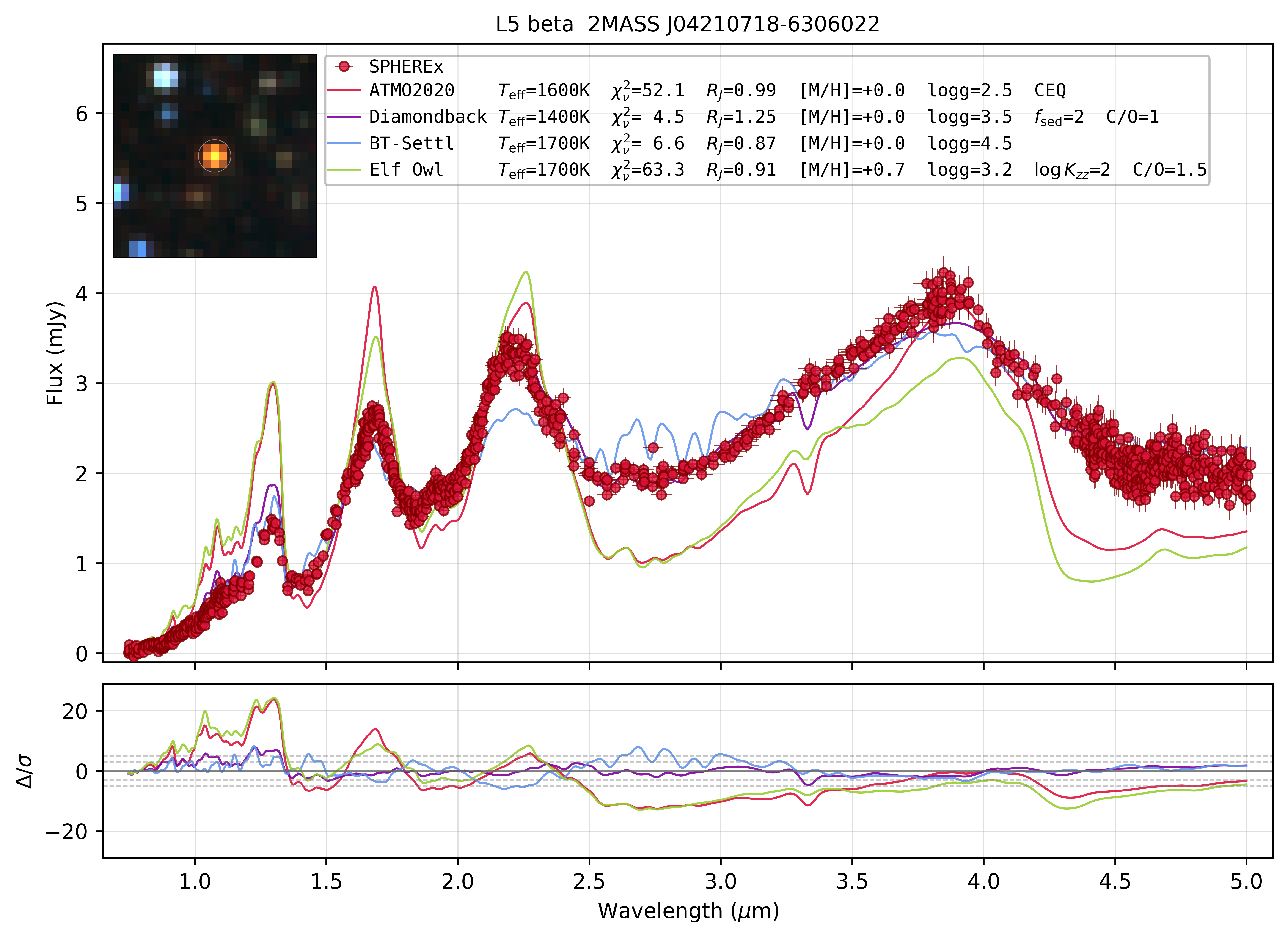}
    \caption{Same as Fig. \ref{fig:L9}, but for the low-gravity L5$\beta$ dwarf 2MASS J04210718-6306022. }
    \label{fig:L5beta}
\end{figure*}

\begin{figure*}
    \centering
    \includegraphics[width=.75\linewidth]{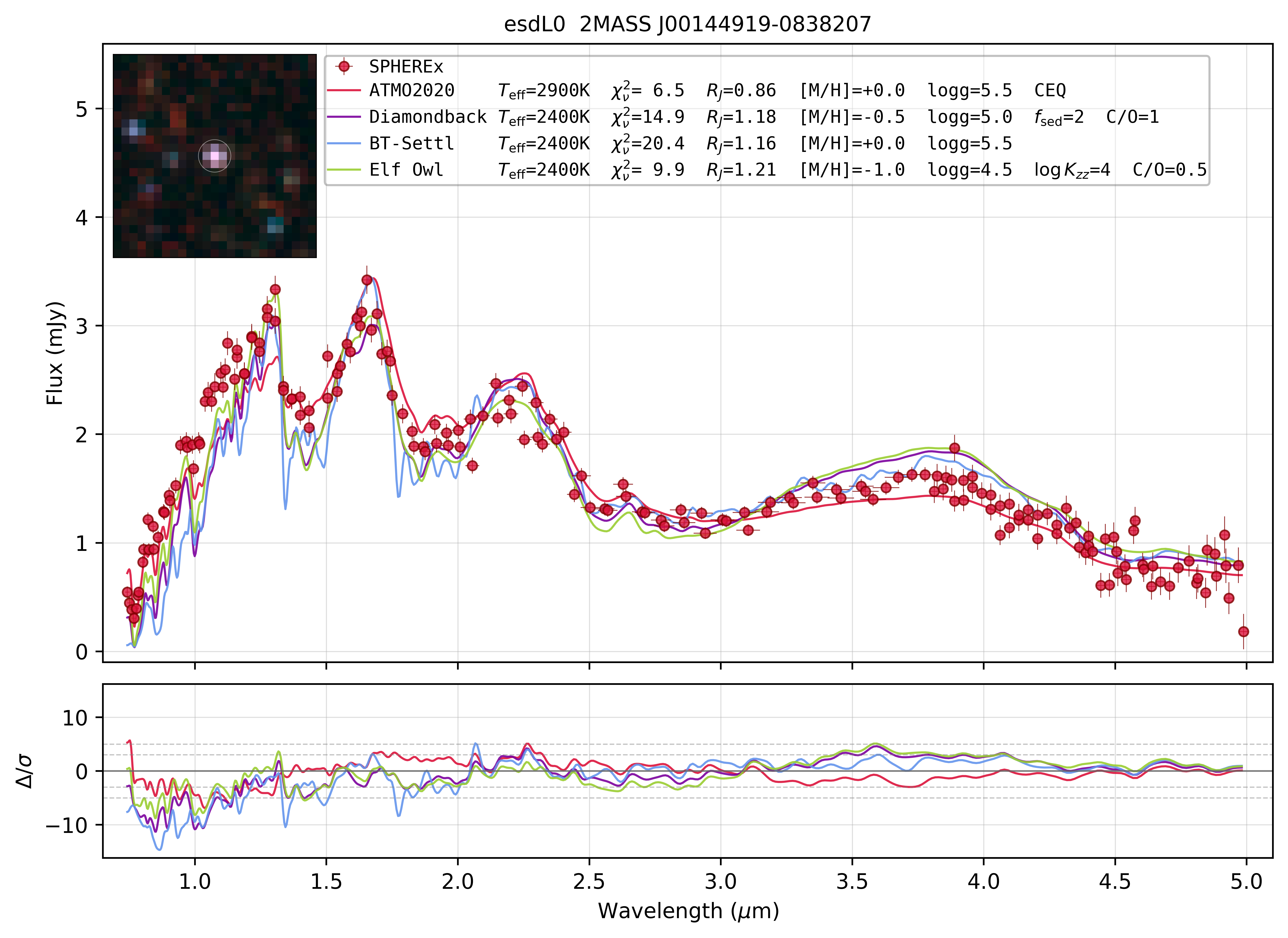}
    \caption{Same as Fig. \ref{fig:L9}, but for the low metallicity esdL0 dwarf 2MASS J00144919-0838207.}
    \label{fig:esdL0}
\end{figure*}

\begin{figure*}
    \centering
    \includegraphics[width=.75\linewidth]{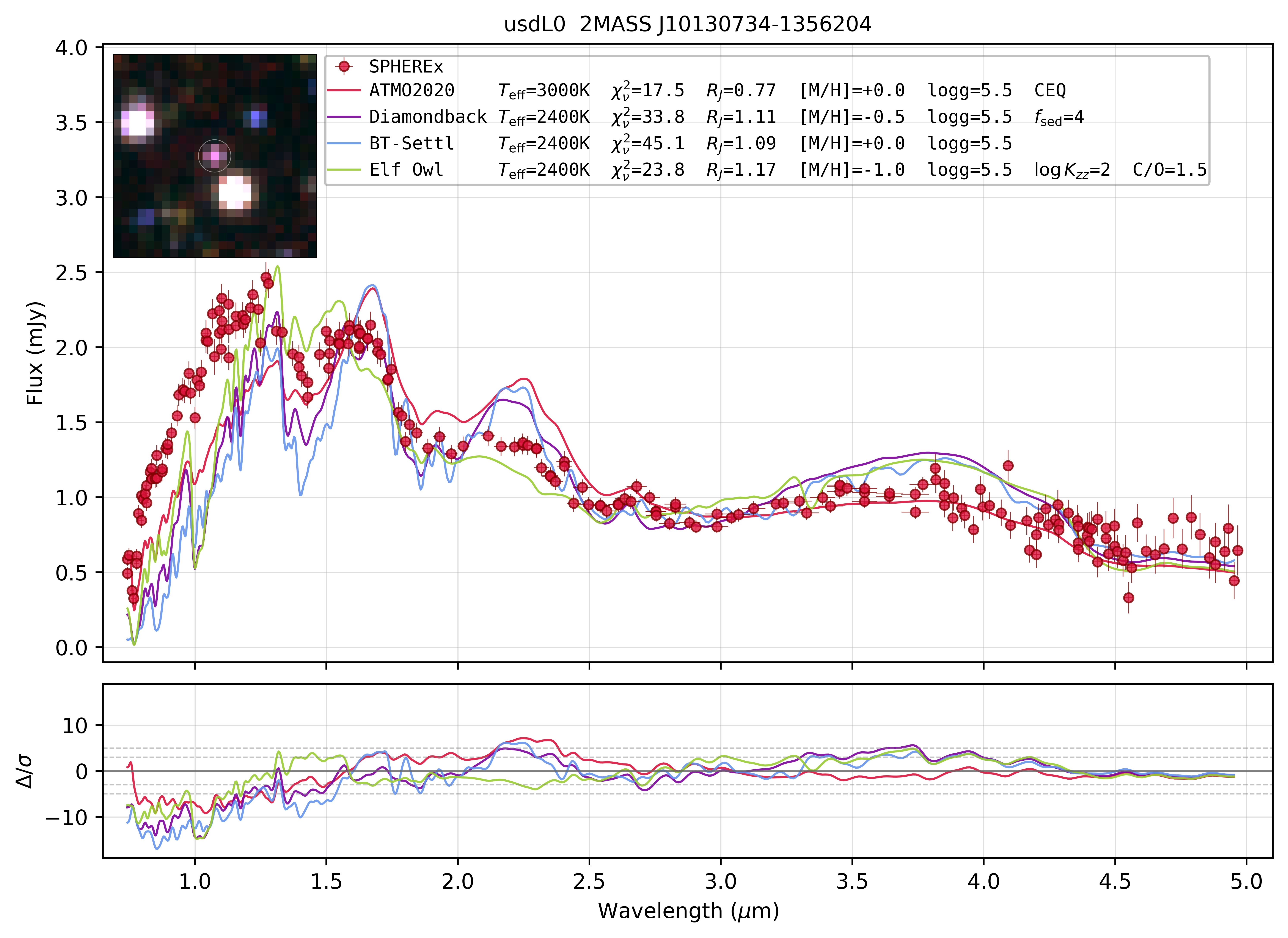}
    \caption{Same as Fig. \ref{fig:L9}, but for the ultra low metallicity usdL0 dwarf 2MASS J10130734-1356204. }
    \label{fig:usdL0}
\end{figure*}

\bibliography{sample701}{}

@ARTICLE{Sing2024,
       author = {{Sing}, David K. and {Rustamkulov}, Zafar and {Thorngren}, Daniel P. and {Barstow}, Joanna K. and {Tremblin}, Pascal and {Alves de Oliveira}, Catarina and {Beck}, Tracy L. and {Birkmann}, Stephan M. and {Challener}, Ryan C. and {Crouzet}, Nicolas and {Espinoza}, N{\'e}stor and {Ferruit}, Pierre and {Giardino}, Giovanna and {Gressier}, Am{\'e}lie and {Lee}, Elspeth K.~H. and {Lewis}, Nikole K. and {Maiolino}, Roberto and {Manjavacas}, Elena and {Rauscher}, Bernard J. and {Sirianni}, Marco and {Valenti}, Jeff A.},
        title = "{A warm Neptune's methane reveals core mass and vigorous atmospheric mixing}",
      journal = {\nat},
         year = 2024,
        month = jun,
       volume = {630},
       number = {8018},
        pages = {831-835},
          doi = {10.1038/s41586-024-07395-z},
archivePrefix = {arXiv},
       eprint = {2405.11027},
 primaryClass = {astro-ph.EP},
       adsurl = {https://ui.adsabs.harvard.edu/abs/2024Natur.630..831S}
}

@ARTICLE{Fortney2020,
       author = {{Fortney}, Jonathan J. and {Visscher}, Channon and {Marley}, Mark S. and {Hood}, Callie E. and {Line}, Michael R. and {Thorngren}, Daniel P. and {Freedman}, Richard S. and {Lupu}, Roxana},
        title = "{Beyond Equilibrium Temperature: How the Atmosphere/Interior Connection Affects the Onset of Methane, Ammonia, and Clouds in Warm Transiting Giant Planets}",
      journal = {\aj},
         year = 2020,
        month = dec,
       volume = {160},
       number = {6},
          eid = {288},
        pages = {288},
          doi = {10.3847/1538-3881/abc5bd},
archivePrefix = {arXiv},
       eprint = {2010.00146},
 primaryClass = {astro-ph.EP},
       adsurl = {https://ui.adsabs.harvard.edu/abs/2020AJ....160..288F}
}

@ARTICLE{Batalha2019,
       author = {{Batalha}, Natasha E. and {Marley}, Mark S. and {Lewis}, Nikole K. and {Fortney}, Jonathan J.},
        title = "{Exoplanet Reflected-light Spectroscopy with PICASO}",
      journal = {\apj},
         year = 2019,
        month = jun,
       volume = {878},
       number = {1},
          eid = {70},
        pages = {70},
          doi = {10.3847/1538-4357/ab1b51},
archivePrefix = {arXiv},
       eprint = {1904.09355},
 primaryClass = {astro-ph.EP},
       adsurl = {https://ui.adsabs.harvard.edu/abs/2019ApJ...878...70B}
}

@article{Ashby2026,
       author = {{Ashby, M., et al.}},
       title = "{Absolute Flux Calibration of the Spectro-Photometer for the History of the Universe, Epoch of Reionization, and Ices Explorer (SPHEREx) Mission}",
       journal = {in prep.},
         year = 2026
}

@ARTICLE{Wright2010,
       author = {{Wright}, Edward L. and {Eisenhardt}, Peter R.~M. and {Mainzer}, Amy K. and {Ressler}, Michael E. and {Cutri}, Roc M. and {Jarrett}, Thomas and {Kirkpatrick}, J. Davy and {Padgett}, Deborah and {McMillan}, Robert S. and {Skrutskie}, Michael and {Stanford}, S.~A. and {Cohen}, Martin and {Walker}, Russell G. and {Mather}, John C. and {Leisawitz}, David and {Gautier}, III, Thomas N. and {McLean}, Ian and {Benford}, Dominic and {Lonsdale}, Carol J. and {Blain}, Andrew and {Mendez}, Bryan and {Irace}, William R. and {Duval}, Valerie and {Liu}, Fengchuan and {Royer}, Don and {Heinrichsen}, Ingolf and {Howard}, Joan and {Shannon}, Mark and {Kendall}, Martha and {Walsh}, Amy L. and {Larsen}, Mark and {Cardon}, Joel G. and {Schick}, Scott and {Schwalm}, Mark and {Abid}, Mohamed and {Fabinsky}, Beth and {Naes}, Larry and {Tsai}, Chao-Wei},
        title = "{The Wide-field Infrared Survey Explorer (WISE): Mission Description and Initial On-orbit Performance}",
      journal = {\aj},
         year = 2010,
        month = dec,
       volume = {140},
       number = {6},
        pages = {1868-1881},
          doi = {10.1088/0004-6256/140/6/1868},
archivePrefix = {arXiv},
       eprint = {1008.0031},
 primaryClass = {astro-ph.IM},
       adsurl = {https://ui.adsabs.harvard.edu/abs/2010AJ....140.1868W}
}

@ARTICLE{Skrutskie2006,
       author = {{Skrutskie}, M.~F. and {Cutri}, R.~M. and {Stiening}, R. and {Weinberg}, M.~D. and {Schneider}, S. and {Carpenter}, J.~M. and {Beichman}, C. and {Capps}, R. and {Chester}, T. and {Elias}, J. and {Huchra}, J. and {Liebert}, J. and {Lonsdale}, C. and {Monet}, D.~G. and {Price}, S. and {Seitzer}, P. and {Jarrett}, T. and {Kirkpatrick}, J.~D. and {Gizis}, J.~E. and {Howard}, E. and {Evans}, T. and {Fowler}, J. and {Fullmer}, L. and {Hurt}, R. and {Light}, R. and {Kopan}, E.~L. and {Marsh}, K.~A. and {McCallon}, H.~L. and {Tam}, R. and {Van Dyk}, S. and {Wheelock}, S.},
        title = "{The Two Micron All Sky Survey (2MASS)}",
      journal = {\aj},
         year = 2006,
        month = feb,
       volume = {131},
       number = {2},
        pages = {1163-1183},
          doi = {10.1086/498708},
       adsurl = {https://ui.adsabs.harvard.edu/abs/2006AJ....131.1163S}
}

@ARTICLE{Cruz2009,
       author = {{Cruz}, Kelle L. and {Kirkpatrick}, J. Davy and {Burgasser}, Adam J.},
        title = "{Young L Dwarfs Identified in the Field: A Preliminary Low-Gravity, Optical Spectral Sequence from L0 to L5}",
      journal = {\aj},
         year = 2009,
        month = feb,
       volume = {137},
       number = {2},
        pages = {3345-3357},
          doi = {10.1088/0004-6256/137/2/3345},
archivePrefix = {arXiv},
       eprint = {0812.0364},
 primaryClass = {astro-ph},
       adsurl = {https://ui.adsabs.harvard.edu/abs/2009AJ....137.3345C}
}

@ARTICLE{Suarez2021,
       author = {{Su{\'a}rez}, Genaro and {Metchev}, Stanimir and {Leggett}, Sandy K. and {Saumon}, Didier and {Marley}, Mark S.},
        title = "{Ultracool Dwarfs Observed with the Spitzer Infrared Spectrograph. I. An Accurate Look at the L-to-T Transition at  300 Myr from Optical Through Mid-infrared Spectrophotometry}",
      journal = {\apj},
         year = 2021,
        month = oct,
       volume = {920},
       number = {2},
          eid = {99},
        pages = {99},
          doi = {10.3847/1538-4357/ac1418},
archivePrefix = {arXiv},
       eprint = {2107.14282},
 primaryClass = {astro-ph.SR},
       adsurl = {https://ui.adsabs.harvard.edu/abs/2021ApJ...920...99S}
}

@ARTICLE{Petrus2024,
       author = {{Petrus}, Simon and {Whiteford}, Niall and {Patapis}, Polychronis and {Biller}, Beth A. and {Skemer}, Andrew and {Hinkley}, Sasha and {Su{\'a}rez}, Genaro and {Palma-Bifani}, Paulina and {Morley}, Caroline V. and {Tremblin}, Pascal and {Charnay}, Benjamin and {Vos}, Johanna M. and {Wang}, Jason J. and {Stone}, Jordan M. and {Bonnefoy}, Micka{\"e}l and {Chauvin}, Ga{\"e}l and {Miles}, Brittany E. and {Carter}, Aarynn L. and {Lueber}, Anna and {Helling}, Christiane and {Sutlieff}, Ben J. and {Janson}, Markus and {Gonzales}, Eileen C. and {Hoch}, Kielan K.~W. and {Absil}, Olivier and {Balmer}, William O. and {Boccaletti}, Anthony and {Bonavita}, Mariangela and {Booth}, Mark and {Bowler}, Brendan P. and {Briesemeister}, Zackery W. and {Bryan}, Marta L. and {Calissendorff}, Per and {Cantalloube}, Faustine and {Chen}, Christine H. and {Choquet}, Elodie and {Christiaens}, Valentin and {Cugno}, Gabriele and {Currie}, Thayne and {Danielski}, Camilla and {De Furio}, Matthew and {Dupuy}, Trent J. and {Factor}, Samuel M. and {Faherty}, Jacqueline K. and {Fitzgerald}, Michael P. and {Fortney}, Jonathan J. and {Franson}, Kyle and {Girard}, Julien H. and {Grady}, Carol A. and {Henning}, Thomas and {Hines}, Dean C. and {Hood}, Callie E. and {Howe}, Alex R. and {Kalas}, Paul and {Kammerer}, Jens and {Kennedy}, Grant M. and {Kenworthy}, Matthew A. and {Kervella}, Pierre and {Kim}, Minjae and {Kitzmann}, Daniel and {Kraus}, Adam L. and {Kuzuhara}, Masayuki and {Lagage}, Pierre-Olivier and {Lagrange}, Anne-Marie and {Lawson}, Kellen and {Lazzoni}, Cecilia and {Leisenring}, Jarron M. and {Lew}, Ben W.~P. and {Liu}, Michael C. and {Liu}, Pengyu and {Llop-Sayson}, Jorge and {Lloyd}, James P. and {Macintosh}, Bruce and {M{\^a}lin}, Mathilde and {Manjavacas}, Elena and {Marino}, Sebasti{\'a}n and {Marley}, Mark S. and {Marois}, Christian and {Martinez}, Raquel A. and {Matthews}, Elisabeth C. and {Matthews}, Brenda C. and {Mawet}, Dimitri and {Mazoyer}, Johan and {McElwain}, Michael W. and {Metchev}, Stanimir and {Meyer}, Michael R. and {Millar-Blanchaer}, Maxwell A. and {Molli{\`e}re}, Paul and {Moran}, Sarah E. and {Mukherjee}, Sagnick and {Pantin}, Eric and {Perrin}, Marshall D. and {Pueyo}, Laurent and {Quanz}, Sascha P. and {Quirrenbach}, Andreas and {Ray}, Shrishmoy and {Rebollido}, Isabel and {Adams Redai}, Jea and {Ren}, Bin B. and {Rickman}, Emily and {Sallum}, Steph and {Samland}, Matthias and {Sargent}, Benjamin and {Schlieder}, Joshua E. and {Stapelfeldt}, Karl R. and {Tamura}, Motohide and {Tan}, Xianyu and {Theissen}, Christopher A. and {Uyama}, Taichi and {Vasist}, Malavika and {Vigan}, Arthur and {Wagner}, Kevin and {Ward-Duong}, Kimberly and {Wolff}, Schuyler G. and {Worthen}, Kadin and {Wyatt}, Mark C. and {Ygouf}, Marie and {Zurlo}, Alice and {Zhang}, Xi and {Zhang}, Keming and {Zhang}, Zhoujian and {Zhou}, Yifan},
        title = "{The JWST Early Release Science Program for Direct Observations of Exoplanetary Systems. V. Do Self-consistent Atmospheric Models Represent JWST Spectra? A Showcase with VHS 1256─1257 b}",
      journal = {\apjl},
         year = 2024,
        month = may,
       volume = {966},
       number = {1},
          eid = {L11},
        pages = {L11},
          doi = {10.3847/2041-8213/ad3e7c},
archivePrefix = {arXiv},
       eprint = {2312.03852},
 primaryClass = {astro-ph.EP},
       adsurl = {https://ui.adsabs.harvard.edu/abs/2024ApJ...966L..11P}
}

@ARTICLE{Dupuy2017,
       author = {{Dupuy}, Trent J. and {Liu}, Michael C.},
        title = "{Individual Dynamical Masses of Ultracool Dwarfs}",
      journal = {\apjs},
         year = 2017,
        month = aug,
       volume = {231},
       number = {2},
          eid = {15},
        pages = {15},
          doi = {10.3847/1538-4365/aa5e4c},
archivePrefix = {arXiv},
       eprint = {1703.05775},
 primaryClass = {astro-ph.SR},
       adsurl = {https://ui.adsabs.harvard.edu/abs/2017ApJS..231...15D}
}

@ARTICLE{Filippazzo2015,
       author = {{Filippazzo}, Joseph C. and {Rice}, Emily L. and {Faherty}, Jacqueline and {Cruz}, Kelle L. and {Van Gordon}, Mollie M. and {Looper}, Dagny L.},
        title = "{Fundamental Parameters and Spectral Energy Distributions of Young and Field Age Objects with Masses Spanning the Stellar to Planetary Regime}",
      journal = {\apj},
         year = 2015,
        month = sep,
       volume = {810},
       number = {2},
          eid = {158},
        pages = {158},
          doi = {10.1088/0004-637X/810/2/158},
archivePrefix = {arXiv},
       eprint = {1508.01767},
 primaryClass = {astro-ph.SR},
       adsurl = {https://ui.adsabs.harvard.edu/abs/2015ApJ...810..158F}
}

@ARTICLE{Marley2021,
       author = {{Marley}, Mark S. and {Saumon}, Didier and {Visscher}, Channon and {Lupu}, Roxana and {Freedman}, Richard and {Morley}, Caroline and {Fortney}, Jonathan J. and {Seay}, Christopher and {Smith}, Adam J.~R.~W. and {Teal}, D.~J. and {Wang}, Ruoyan},
        title = "{The Sonora Brown Dwarf Atmosphere and Evolution Models. I. Model Description and Application to Cloudless Atmospheres in Rainout Chemical Equilibrium}",
      journal = {\apj},
         year = 2021,
        month = oct,
       volume = {920},
       number = {2},
          eid = {85},
        pages = {85},
          doi = {10.3847/1538-4357/ac141d},
archivePrefix = {arXiv},
       eprint = {2107.07434},
 primaryClass = {astro-ph.SR},
       adsurl = {https://ui.adsabs.harvard.edu/abs/2021ApJ...920...85M}
}

@ARTICLE{Saumon2008,
       author = {{Saumon}, D. and {Marley}, Mark S.},
        title = "{The Evolution of L and T Dwarfs in Color-Magnitude Diagrams}",
      journal = {\apj},
         year = 2008,
        month = dec,
       volume = {689},
       number = {2},
        pages = {1327-1344},
          doi = {10.1086/592734},
archivePrefix = {arXiv},
       eprint = {0808.2611},
 primaryClass = {astro-ph},
       adsurl = {https://ui.adsabs.harvard.edu/abs/2008ApJ...689.1327S}
}

@ARTICLE{Baraffe2003,
       author = {{Baraffe}, I. and {Chabrier}, G. and {Barman}, T.~S. and {Allard}, F. and {Hauschildt}, P.~H.},
        title = "{Evolutionary models for cool brown dwarfs and extrasolar giant planets. The case of HD 209458}",
      journal = {\aap},
         year = 2003,
        month = may,
       volume = {402},
        pages = {701-712},
          doi = {10.1051/0004-6361:20030252},
archivePrefix = {arXiv},
       eprint = {astro-ph/0302293},
 primaryClass = {astro-ph},
       adsurl = {https://ui.adsabs.harvard.edu/abs/2003A&A...402..701B}
}

@ARTICLE{Burrows2001,
       author = {{Burrows}, Adam and {Hubbard}, W.~B. and {Lunine}, J.~I. and {Liebert}, James},
        title = "{The theory of brown dwarfs and extrasolar giant planets}",
      journal = {Reviews of Modern Physics},
         year = 2001,
        month = jul,
       volume = {73},
       number = {3},
        pages = {719-765},
          doi = {10.1103/RevModPhys.73.719},
archivePrefix = {arXiv},
       eprint = {astro-ph/0103383},
 primaryClass = {astro-ph},
       adsurl = {https://ui.adsabs.harvard.edu/abs/2001RvMP...73..719B}
}

@ARTICLE{Lodders2002,
       author = {{Lodders}, Katharina and {Fegley}, Bruce},
        title = "{Atmospheric Chemistry in Giant Planets, Brown Dwarfs, and Low-Mass Dwarf Stars. I. Carbon, Nitrogen, and Oxygen}",
      journal = {\icarus},
         year = 2002,
        month = feb,
       volume = {155},
       number = {2},
        pages = {393-424},
          doi = {10.1006/icar.2001.6740},
       adsurl = {https://ui.adsabs.harvard.edu/abs/2002Icar..155..393L}
}

@ARTICLE{Patten2006,
       author = {{Patten}, Brian M. and {Stauffer}, John R. and {Burrows}, Adam and {Marengo}, Massimo and {Hora}, Joseph L. and {Luhman}, Kevin L. and {Sonnett}, Sarah M. and {Henry}, Todd J. and {Raghavan}, Deepak and {Megeath}, S. Thomas and {Liebert}, James and {Fazio}, Giovanni G.},
        title = "{Spitzer IRAC Photometry of M, L, and T Dwarfs}",
      journal = {\apj},
         year = 2006,
        month = nov,
       volume = {651},
       number = {1},
        pages = {502-516},
          doi = {10.1086/507264},
archivePrefix = {arXiv},
       eprint = {astro-ph/0606432},
 primaryClass = {astro-ph},
       adsurl = {https://ui.adsabs.harvard.edu/abs/2006ApJ...651..502P}
}

@article{Brooks2026,
doi = {10.3847/2515-5172/ae6257},
url = {https://doi.org/10.3847/2515-5172/ae6257},
year = {2026},
month = {apr},
publisher = {The American Astronomical Society},
volume = {10},
number = {4},
pages = {94},
author = {Brooks, Hunter and Cushing, Michael C. and Kothari, Harshil},
title = {An Ultracool Dwarf Spectral Sequence Using SPHEREx},
journal = {Research Notes of the AAS}
}

@misc{Gagne2026,
      title={A SPHEREx Pipeline and Spectral Library for Ultracool Dwarfs}, 
      author={Jonathan Gagné and Jacqueline K. Faherty and Azul Ruiz Diaz and Louis-Philippe Coulombe and Thomas P. Bickle and Adam C. Schneider and J. Davy Kirkpatrick and Marc J. Kuchner and Aaron M. Meisner and Dan Caselden and Adam J. Burgasser and Sarah Casewell and Easton J. Honaker and Frank Kiwy and Federico Marocco and Nikolaj Stevnbak Andersen and Lizzeth Ruiz Arroyo and Bruce Baller and Paul Beaulieu and John Bell and Martin Bilsing and Troy K. Bohling and Guillaume Colin and Giovanni Colombo and Sam Deen and Alexandru Dereveanco and Kevin Dixon and Hugo A. Durantini Luca and Deiby Flores and Christoph Franck and Christopher Fulvi and Michael Gallmann and Jean Marc Gantier and Konstantin Glebov and Léopold Gramaize and Leslie K. Hamlet and Ken Hinckley and Kevin Jablonski and Peter A. Jałowiczor and Martin Kabatnik and Peter Kasprowitz and K Ly and David W. Martin and Naoufel Marzak and Alexander McColgan and Neil J. McEwan and Marianne N. Michaels and William Pendrill and Stéphane Perlin and Ben Pumphrey and James Rabe and Henry Raway and Walter Ruben Robledo and David Roser and Animesh Roy and Arttu Sainio and Vincent Schindler and Manfred Schonau and Jö rg Schümann and Karl Selg-Mann and Andrea Serio and Patrick Smith and Andres Stenner and Christopher Tanner and Melina Thévenot and Vinod Thakur and Mayahuel Torres Guerrero and Maurizio Ventura and Nikita V. Voloshin and Jim Walla and Zbigniew Wȩdracki and Bailey Weyandt and Breck Wilhite and Spartacus Zitouni},
      year={2026},
      eprint={2604.22012},
      archivePrefix={arXiv},
      primaryClass={astro-ph.SR},
      url={https://arxiv.org/abs/2604.22012}, 
}

@misc{Tu2026,
      title={SPHEREx Ultracool Dwarf spectral Atlas (SUDA): Atmospheric and Fundamental Parameters of Ultracool Dwarfs}, 
      author={Zhijun Tu and Shu Wang and Haomiao Huang and Xiaodian Chen and Jifeng Liu},
      year={2026},
      eprint={2604.26771},
      archivePrefix={arXiv},
      primaryClass={astro-ph.SR},
      url={https://arxiv.org/abs/2604.26771}, 
}

@misc{spherex_qr2,
  author    = {{SPHEREx Team}},
  title     = {{SPHEREx Quick Release Spectral Images - QR2}},
  year      = {2025},
  publisher = {IPAC},
  doi       = {10.26131/IRSA652},
  url       = {https://doi.org/10.26131/IRSA652}
}

@ARTICLE{Beiler2024,
       author = {{Beiler}, Samuel A. and {Cushing}, Michael C. and {Kirkpatrick}, J. Davy and {Schneider}, Adam C. and {Mukherjee}, Sagnick and {Marley}, Mark S. and {Marocco}, Federico and {Smart}, Richard L.},
        title = "{Precise Bolometric Luminosities and Effective Temperatures of 23 Late-T and Y Dwarfs Obtained with JWST}",
      journal = {\apj},
         year = 2024,
        month = oct,
       volume = {973},
       number = {2},
          eid = {107},
        pages = {107},
          doi = {10.3847/1538-4357/ad6301},
archivePrefix = {arXiv},
       eprint = {2407.08518},
 primaryClass = {astro-ph.SR},
       adsurl = {https://ui.adsabs.harvard.edu/abs/2024ApJ...973..107B}
}

@ARTICLE{Lueber2026,
       author = {{Lueber}, A. and {Kitzmann}, D. and {Heng}, K.},
        title = "{Clouds and chemistry across the brown dwarf T-Y sequence: Insights from JWST atmospheric retrievals}",
      journal = {\aap},
         year = 2026,
        month = mar,
       volume = {707},
          eid = {A92},
        pages = {A92},
          doi = {10.1051/0004-6361/202556207},
archivePrefix = {arXiv},
       eprint = {2601.12575},
 primaryClass = {astro-ph.SR},
       adsurl = {https://ui.adsabs.harvard.edu/abs/2026A&A...707A..92L}
}

@ARTICLE{Reiners2008,
       author = {{Reiners}, A. and {Basri}, G.},
        title = "{Chromospheric Activity, Rotation, and Rotational Braking in M and L Dwarfs}",
      journal = {\apj},
         year = 2008,
        month = sep,
       volume = {684},
       number = {2},
        pages = {1390-1403},
          doi = {10.1086/590073},
archivePrefix = {arXiv},
       eprint = {0805.1059},
 primaryClass = {astro-ph},
       adsurl = {https://ui.adsabs.harvard.edu/abs/2008ApJ...684.1390R}
}

@ARTICLE{Faherty2024,
       author = {{Faherty}, Jacqueline K. and {Burningham}, Ben and {Gagn{\'e}}, Jonathan and {Su{\'a}rez}, Genaro and {Vos}, Johanna M. and {Alejandro Merchan}, Sherelyn and {Morley}, Caroline V. and {Rowland}, Melanie and {Lacy}, Brianna and {Kiman}, Rocio and et al.},
        title = "{Methane emission from a cool brown dwarf}",
      journal = {\nat},
         year = 2024,
        month = apr,
       volume = {628},
       number = {8008},
        pages = {511-514},
          doi = {10.1038/s41586-024-07190-w},
archivePrefix = {arXiv},
       eprint = {2404.10977},
 primaryClass = {astro-ph.SR},
       adsurl = {https://ui.adsabs.harvard.edu/abs/2024Natur.628..511F}
}

@misc{Bradac2026,
      title={Two Exciting High-redshift Galaxy Candidates Turn Out to Be Two Exciting Ultra-cool Brown Dwarfs}, 
      author={Maruša Bradač and Chris Willott and Yoshihisa Asada and Loïc Albert and Gregor Rihtaršič and Anishya Harshan and Jon Judež and Nicholas S. Martis and Andrea Ferrara and Abdurro'uf and Joseph F. V. Allingham and Volker Bromm and John Chisholm and Dan Coe and Guillaume Desprez and Jose M. Diego and Andreas L. Faisst and Seiji Fujimoto and Tiger Yu-Yang Hsiao and Kohei Inayoshi and Anton M. Koekemoer and Vasily Kokorev and Brian C. Lemaux and Paulo A. A. Lopes and Danilo Marchesini and Vladan Markov and Gaël Noirot and Richard Pan and Scott W. Randall and Johan Richard and Luke Robbins and Ghassan T. E. Sarrouh and Marcin Sawicki and Tim Schrabback and Roberta Tripodi and Eros Vanzella and Rogier A. Windhorst},
      year={2026},
      eprint={2604.23668},
      archivePrefix={arXiv},
      primaryClass={astro-ph.GA},
      url={https://arxiv.org/abs/2604.23668}, 
}

@ARTICLE{Tannock2022,
       author = {{Tannock}, Megan E. and {Metchev}, Stanimir and {Hood}, Callie E. and {Mace}, Gregory N. and {Fortney}, Jonathan J. and {Morley}, Caroline V. and {Jaffe}, Daniel T. and {Lupu}, Roxana},
        title = "{A 1.46-2.48 {\ensuremath{\mu}}m spectroscopic atlas of a T6 dwarf (1060 K) atmosphere with IGRINS: first detections of H$_{2}$S and H$_{2}$, and verification of H$_{2}$O, CH$_{4}$, and NH$_{3}$ line lists}",
      journal = {\mnras},
         year = 2022,
        month = aug,
       volume = {514},
       number = {3},
        pages = {3160-3178},
          doi = {10.1093/mnras/stac1412},
archivePrefix = {arXiv},
       eprint = {2206.03519},
 primaryClass = {astro-ph.SR},
       adsurl = {https://ui.adsabs.harvard.edu/abs/2022MNRAS.514.3160T}
}

@ARTICLE{Burgasser2025,
       author = {{Burgasser}, Adam J. and {Gonzales}, Eileen C. and {Beiler}, Samuel A. and {Visscher}, Channon and {Burningham}, Ben and {Mace}, Gregory N. and {Faherty}, Jacqueline K. and {Zhang}, Zenghua and {Sousa-Silva}, Clara and {Lodieu}, Nicolas and {Metchev}, Stanimir A. and {Meisner}, Aaron and {Cushing}, Michael and {Schneider}, Adam C. and {Suarez}, Genaro and {Hsu}, Chih-Chun and {Gerasimov}, Roman and {Aganze}, Christian and {Theissen}, Christopher A.},
        title = "{Observation of undepleted phosphine in the atmosphere of a low-temperature brown dwarf}",
      journal = {Science},
         year = 2025,
        month = nov,
       volume = {390},
       number = {6774},
        pages = {697-701},
          doi = {10.1126/science.adu0401},
archivePrefix = {arXiv},
       eprint = {2510.03916},
 primaryClass = {astro-ph.SR},
       adsurl = {https://ui.adsabs.harvard.edu/abs/2025Sci...390..697B}
}

@ARTICLE{Sanghi2023,
       author = {{Sanghi}, Aniket and {Liu}, Michael C. and {Best}, William M.~J. and {Dupuy}, Trent J. and {Siverd}, Robert J. and {Zhang}, Zhoujian and {Hurt}, Spencer A. and {Magnier}, Eugene A. and {Aller}, Kimberly M. and {Deacon}, Niall R.},
        title = "{The Hawaii Infrared Parallax Program. VI. The Fundamental Properties of 1000+ Ultracool Dwarfs and Planetary-mass Objects Using Optical to Mid-infrared Spectral Energy Distributions and Comparison to BT-Settl and ATMO 2020 Model Atmospheres}",
      journal = {\apj},
         year = 2023,
        month = dec,
       volume = {959},
       number = {1},
          eid = {63},
        pages = {63},
          doi = {10.3847/1538-4357/acff66},
archivePrefix = {arXiv},
       eprint = {2309.03082},
 primaryClass = {astro-ph.SR},
       adsurl = {https://ui.adsabs.harvard.edu/abs/2023ApJ...959...63S}
}

@ARTICLE{Lothringer2026,
       author = {{Lothringer}, Joshua D. and {Lowson}, Nataliea and {Fu}, Guangwei},
        title = "{The Library of Exoplanet Atmospheric Composition Measurements: Population-level Trends in Exoplanet Composition with ExoComp}",
      journal = {\aj},
         year = 2026,
        month = jan,
       volume = {171},
       number = {1},
          eid = {31},
        pages = {31},
          doi = {10.3847/1538-3881/ae1b8e},
archivePrefix = {arXiv},
       eprint = {2510.26785},
 primaryClass = {astro-ph.EP},
       adsurl = {https://ui.adsabs.harvard.edu/abs/2026AJ....171...31L}
}

@ARTICLE{Gao2020,
       author = {{Gao}, Peter and {Thorngren}, Daniel P. and {Lee}, Elspeth K.~H. and {Fortney}, Jonathan J. and {Morley}, Caroline V. and {Wakeford}, Hannah R. and {Powell}, Diana K. and {Stevenson}, Kevin B. and {Zhang}, Xi},
        title = "{Aerosol composition of hot giant exoplanets dominated by silicates and hydrocarbon hazes}",
      journal = {Nature Astronomy},
         year = 2020,
        month = may,
       volume = {4},
        pages = {951-956},
          doi = {10.1038/s41550-020-1114-3},
archivePrefix = {arXiv},
       eprint = {2005.11939},
 primaryClass = {astro-ph.EP},
       adsurl = {https://ui.adsabs.harvard.edu/abs/2020NatAs...4..951G}
}

@ARTICLE{Suarez2022,
       author = {{Su{\'a}rez}, Genaro and {Metchev}, Stanimir},
        title = "{Ultracool dwarfs observed with the Spitzer infrared spectrograph - II. Emergence and sedimentation of silicate clouds in L dwarfs, and analysis of the full M5-T9 field dwarf spectroscopic sample}",
      journal = {\mnras},
         year = 2022,
        month = jul,
       volume = {513},
       number = {4},
        pages = {5701-5726},
          doi = {10.1093/mnras/stac1205},
archivePrefix = {arXiv},
       eprint = {2205.00168},
 primaryClass = {astro-ph.SR},
       adsurl = {https://ui.adsabs.harvard.edu/abs/2022MNRAS.513.5701S}
}

@ARTICLE{Schneider2018,
       author = {{Schneider}, Adam C. and {Hardegree-Ullman}, Kevin K. and {Cushing}, Michael C. and {Kirkpatrick}, J. Davy and {Shkolnik}, Evgenya L.},
        title = "{Spitzer Light Curves of the Young, Planetary-mass TW Hya Members 2MASS J11193254-1137466AB and WISEA J114724.10-204021.3}",
      journal = {\aj},
         year = 2018,
        month = jun,
       volume = {155},
       number = {6},
          eid = {238},
        pages = {238},
          doi = {10.3847/1538-3881/aabfc2},
archivePrefix = {arXiv},
       eprint = {1804.06917},
 primaryClass = {astro-ph.SR},
       adsurl = {https://ui.adsabs.harvard.edu/abs/2018AJ....155..238S}
}

@ARTICLE{Vos2020,
       author = {{Vos}, Johanna M. and {Biller}, Beth A. and {Allers}, Katelyn N. and {Faherty}, Jacqueline K. and {Liu}, Michael C. and {Metchev}, Stanimir and {Eriksson}, Simon and {Manjavacas}, Elena and {Dupuy}, Trent J. and {Janson}, Markus and {Radigan-Hoffman}, Jacqueline and {Crossfield}, Ian and {Bonnefoy}, Micka{\"e}l and {Best}, William M.~J. and {Homeier}, Derek and {Schlieder}, Joshua E. and {Brandner}, Wolfgang and {Henning}, Thomas and {Bonavita}, Mariangela and {Buenzli}, Esther},
        title = "{Spitzer Variability Properties of Low-gravity L Dwarfs}",
      journal = {\aj},
         year = 2020,
        month = jul,
       volume = {160},
       number = {1},
          eid = {38},
        pages = {38},
          doi = {10.3847/1538-3881/ab9642},
archivePrefix = {arXiv},
       eprint = {2005.12854},
 primaryClass = {astro-ph.SR},
       adsurl = {https://ui.adsabs.harvard.edu/abs/2020AJ....160...38V}
}

@ARTICLE{Baxter2021,
       author = {{Baxter}, Claire and {D{\'e}sert}, Jean-Michel and {Tsai}, Shang-Min and {Todorov}, Kamen O. and {Bean}, Jacob L. and {Deming}, Drake and {Parmentier}, Vivien and {Fortney}, Jonathan J. and {Line}, Michael and {Thorngren}, Daniel and {Pierrehumbert}, Raymond T. and {Burrows}, Adam and {Showman}, Adam P.},
        title = "{Evidence for disequilibrium chemistry from vertical mixing in hot Jupiter atmospheres. A comprehensive survey of transiting close-in gas giant exoplanets with warm-Spitzer/IRAC}",
      journal = {\aap},
         year = 2021,
        month = apr,
       volume = {648},
          eid = {A127},
        pages = {A127},
          doi = {10.1051/0004-6361/202039708},
archivePrefix = {arXiv},
       eprint = {2103.07185},
 primaryClass = {astro-ph.EP},
       adsurl = {https://ui.adsabs.harvard.edu/abs/2021A&A...648A.127B}
}

@article{bryan2025,
      title={Optimized Observation Sequencing in Low-Earth Orbit with the SPHEREx Survey Planning Software}, 
      author={Sean Bryan and James Bock and Thomas Burk and Tzu-Ching Chang and Brendan P. Crill and Ari Cukierman and Olivier Dore and C. Darren Dowell and Gregory Dubois-Felsmann and Beth Fabinsky and Sergi Hildebrandt-Rafels and Howard Hui and Kyle Hughes and Phillip Korngut and Philip Mauskopf and Julian Mena and Chi Nguyen and Milad Pourrahmani and Dustin Putnam and Keshav Ramanathan and Flora Ridenhour and Cody Roberson and Amy Trangsrud and Stephen Unwin and Pao-Yu Wang and the SPHEREx Team},
      year={2025},
      eprint={2508.20332},
      archivePrefix={arXiv},
      primaryClass={astro-ph.IM},
      url={https://arxiv.org/abs/2508.20332}, 
}

@ARTICLE{Lothringer2024,
       author = {{Lothringer}, Joshua D. and {Zhou}, Yifan and {Apai}, D{\'a}niel and {Tan}, Xianyu and {Parmentier}, Vivien and {Casewell}, Sarah L.},
        title = "{Atmospheric Retrievals of the Phase-resolved Spectra of Irradiated Brown Dwarfs WD-0137B and EPIC-2122B}",
      journal = {\apj},
         year = 2024,
        month = jun,
       volume = {968},
       number = {2},
          eid = {126},
        pages = {126},
          doi = {10.3847/1538-4357/ad43da},
archivePrefix = {arXiv},
       eprint = {2404.16813},
 primaryClass = {astro-ph.SR},
       adsurl = {https://ui.adsabs.harvard.edu/abs/2024ApJ...968..126L}
}

@ARTICLE{Line2017,
       author = {{Line}, Michael R. and {Marley}, Mark S. and {Liu}, Michael C. and {Burningham}, Ben and {Morley}, Caroline V. and {Hinkel}, Natalie R. and {Teske}, Johanna and {Fortney}, Jonathan J. and {Freedman}, Richard and {Lupu}, Roxana},
        title = "{Uniform Atmospheric Retrieval Analysis of Ultracool Dwarfs. II. Properties of 11 T dwarfs}",
      journal = {\apj},
         year = 2017,
        month = oct,
       volume = {848},
       number = {2},
          eid = {83},
        pages = {83},
          doi = {10.3847/1538-4357/aa7ff0},
archivePrefix = {arXiv},
       eprint = {1612.02809},
 primaryClass = {astro-ph.SR},
       adsurl = {https://ui.adsabs.harvard.edu/abs/2017ApJ...848...83L}
}

@ARTICLE{Zhang2021,
       author = {{Zhang}, Zhoujian and {Liu}, Michael C. and {Marley}, Mark S. and {Line}, Michael R. and {Best}, William M.~J.},
        title = "{Uniform Forward-modeling Analysis of Ultracool Dwarfs. I. Methodology and Benchmarking}",
      journal = {\apj},
         year = 2021,
        month = jul,
       volume = {916},
       number = {1},
          eid = {53},
        pages = {53},
          doi = {10.3847/1538-4357/abf8b2},
archivePrefix = {arXiv},
       eprint = {2011.12294},
 primaryClass = {astro-ph.SR},
       adsurl = {https://ui.adsabs.harvard.edu/abs/2021ApJ...916...53Z}
}

@ARTICLE{burrows1997,
       author = {{Burrows}, A. and {Marley}, M. and {Hubbard}, W.~B. and {Lunine}, J.~I. and {Guillot}, T. and {Saumon}, D. and {Freedman}, R. and {Sudarsky}, D. and {Sharp}, C.},
        title = "{A Nongray Theory of Extrasolar Giant Planets and Brown Dwarfs}",
      journal = {\apj},
         year = 1997,
        month = dec,
       volume = {491},
       number = {2},
        pages = {856-875},
          doi = {10.1086/305002},
archivePrefix = {arXiv},
       eprint = {astro-ph/9705201},
 primaryClass = {astro-ph},
       adsurl = {https://ui.adsabs.harvard.edu/abs/1997ApJ...491..856B}
}

@BOOK{gray2009,
       author = {{Gray}, Richard O. and {Corbally}, J., Christopher},
        title = "{Stellar Spectral Classification}",
         year = 2009,
       adsurl = {https://ui.adsabs.harvard.edu/abs/2009ssc..book.....G}
}

@ARTICLE{Mukherjee2023,
       author = {{Mukherjee}, Sagnick and {Batalha}, Natasha E. and {Fortney}, Jonathan J. and {Marley}, Mark S.},
        title = "{PICASO 3.0: A One-dimensional Climate Model for Giant Planets and Brown Dwarfs}",
      journal = {\apj},
         year = 2023,
        month = jan,
       volume = {942},
       number = {2},
          eid = {71},
        pages = {71},
          doi = {10.3847/1538-4357/ac9f48},
archivePrefix = {arXiv},
       eprint = {2208.07836},
 primaryClass = {astro-ph.EP},
       adsurl = {https://ui.adsabs.harvard.edu/abs/2023ApJ...942...71M}
}

@ARTICLE{Mukherjee2024,
       author = {{Mukherjee}, Sagnick and {Fortney}, Jonathan J. and {Morley}, Caroline V. and {Batalha}, Natasha E. and {Marley}, Mark S. and {Karalidi}, Theodora and {Visscher}, Channon and {Lupu}, Roxana and {Freedman}, Richard and {Gharib-Nezhad}, Ehsan},
        title = "{The Sonora Substellar Atmosphere Models. IV. Elf Owl: Atmospheric Mixing and Chemical Disequilibrium with Varying Metallicity and C/O Ratios}",
      journal = {\apj},
         year = 2024,
        month = mar,
       volume = {963},
       number = {1},
          eid = {73},
        pages = {73},
          doi = {10.3847/1538-4357/ad18c2},
archivePrefix = {arXiv},
       eprint = {2402.00756},
 primaryClass = {astro-ph.EP},
       adsurl = {https://ui.adsabs.harvard.edu/abs/2024ApJ...963...73M}
}

@ARTICLE{Stephens2009,
       author = {{Stephens}, D.~C. and {Leggett}, S.~K. and {Cushing}, Michael C. and {Marley}, Mark S. and {Saumon}, D. and {Geballe}, T.~R. and {Golimowski}, David A. and {Fan}, Xiaohui and {Noll}, K.~S.},
        title = "{The 0.8-14.5 {\ensuremath{\mu}}m Spectra of Mid-L to Mid-T Dwarfs: Diagnostics of Effective Temperature, Grain Sedimentation, Gas Transport, and Surface Gravity}",
      journal = {\apj},
         year = 2009,
        month = sep,
       volume = {702},
       number = {1},
        pages = {154-170},
          doi = {10.1088/0004-637X/702/1/154},
archivePrefix = {arXiv},
       eprint = {0906.2991},
 primaryClass = {astro-ph.SR},
       adsurl = {https://ui.adsabs.harvard.edu/abs/2009ApJ...702..154S}
}

@software{caselden2018,
       author = {{Caselden}, Dan and {Westin}, III, Paul and {Meisner}, Aaron and {Kuchner}, Marc and {Colin}, Guillaume},
        title = "{WiseView: Visualizing motion and variability of faint WISE sources}",
 howpublished = {Astrophysics Source Code Library, record ascl:1806.004},
         year = 2018,
        month = jun,
          eid = {ascl:1806.004},
archivePrefix = {ascl},
       eprint = {1806.004},
       adsurl = {https://ui.adsabs.harvard.edu/abs/2018ascl.soft06004C}
}

@BOOK{morgan1943,
       author = {{Morgan}, William Wilson and {Keenan}, Philip Childs and {Kellman}, Edith},
        title = "{An atlas of stellar spectra, with an outline of spectral classification}",
         year = 1943,
       adsurl = {https://ui.adsabs.harvard.edu/abs/1943assw.book.....M}
}

@ARTICLE{Kirkpatrick2024,
       author = {{Kirkpatrick}, J. Davy and {Marocco}, Federico and {Gelino}, Christopher R. and {Raghu}, Yadukrishna and {Faherty}, Jacqueline K. and {Bardalez Gagliuffi}, Daniella C. and {Schurr}, Steven D. and {Apps}, Kevin and {Schneider}, Adam C. and {Meisner}, Aaron M. and {Kuchner}, Marc J. and {Caselden}, Dan and {Smart}, R.~L. and {Casewell}, S.~L. and {Raddi}, Roberto and {Kesseli}, Aurora and {Stevnbak Andersen}, Nikolaj and {Antonini}, Edoardo and {Beaulieu}, Paul and {Bickle}, Thomas P. and {Bilsing}, Martin and {Chieng}, Raymond and {Colin}, Guillaume and {Deen}, Sam and {Dereveanco}, Alexandru and {Doll}, Katharina and {Durantini Luca}, Hugo A. and {Frazer}, Anya and {Gantier}, Jean Marc and {Gramaize}, L{\'e}opold and {Grant}, Kristin and {Hamlet}, Leslie K. and {Higashimura}, Hiro and {Hyogo}, Michiharu and {Ja{\l}owiczor}, Peter A. and {Jonkeren}, Alexander and {Kabatnik}, Martin and {Kiwy}, Frank and {Martin}, David W. and {Michaels}, Marianne N. and {Pendrill}, William and {Pessanha Machado}, Celso and {Pumphrey}, Benjamin and {Rothermich}, Austin and {Russwurm}, Rebekah and {Sainio}, Arttu and {Sanchez}, John and {Sapelkin-Tambling}, Fyodor Theo and {Sch{\"u}mann}, J{\"o}rg and {Selg-Mann}, Karl and {Singh}, Harshdeep and {Stenner}, Andres and {Sun}, Guoyou and {Tanner}, Christopher and {Th{\'e}venot}, Melina and {Ventura}, Maurizio and {Voloshin}, Nikita V. and {Walla}, Jim and {W{\k{e}}dracki}, Zbigniew and {Adorno}, Jose I. and {Aganze}, Christian and {Allers}, Katelyn N. and {Brooks}, Hunter and {Burgasser}, Adam J. and {Calamari}, Emily and {Connor}, Thomas and {Costa}, Edgardo and {Eisenhardt}, Peter R. and {Gagn{\'e}}, Jonathan and {Gerasimov}, Roman and {Gonzales}, Eileen C. and {Hsu}, Chih-Chun and {Kiman}, Rocio and {Li}, Guodong and {Low}, Ryan and {Mamajek}, Eric and {Pantoja}, Blake M. and {Popinchalk}, Mark and {Rees}, Jon M. and {Stern}, Daniel and {Su{\'a}rez}, Genaro and {Theissen}, Christopher and {Tsai}, Chao-Wei and {Vos}, Johanna M. and {Zurek}, David and {The Backyard Worlds: Planet 9 Collaboration}},
        title = "{The Initial Mass Function Based on the Full-sky 20 pc Census of {\ensuremath{\sim}}3600 Stars and Brown Dwarfs}",
      journal = {\apjs},
         year = 2024,
        month = apr,
       volume = {271},
       number = {2},
          eid = {55},
        pages = {55},
          doi = {10.3847/1538-4365/ad24e2},
archivePrefix = {arXiv},
       eprint = {2312.03639},
 primaryClass = {astro-ph.SR},
       adsurl = {https://ui.adsabs.harvard.edu/abs/2024ApJS..271...55K}
}

@ARTICLE{kirkpatrick2021,
       author = {{Kirkpatrick}, J. Davy and {Gelino}, Christopher R. and {Faherty}, Jacqueline K. and {Meisner}, Aaron M. and {Caselden}, Dan and {Schneider}, Adam C. and {Marocco}, Federico and {Cayago}, Alfred J. and {Smart}, R.~L. and {Eisenhardt}, Peter R. and {Kuchner}, Marc J. and {Wright}, Edward L. and {Cushing}, Michael C. and {Allers}, Katelyn N. and {Bardalez Gagliuffi}, Daniella C. and {Burgasser}, Adam J. and {Gagn{\'e}}, Jonathan and {Logsdon}, Sarah E. and {Martin}, Emily C. and {Ingalls}, James G. and {Lowrance}, Patrick J. and {Abrahams}, Ellianna S. and {Aganze}, Christian and {Gerasimov}, Roman and {Gonzales}, Eileen C. and {Hsu}, Chih-Chun and {Kamraj}, Nikita and {Kiman}, Rocio and {Rees}, Jon and {Theissen}, Christopher and {Ammar}, Kareem and {Andersen}, Nikolaj Stevnbak and {Beaulieu}, Paul and {Colin}, Guillaume and {Elachi}, Charles A. and {Goodman}, Samuel J. and {Gramaize}, L{\'e}opold and {Hamlet}, Leslie K. and {Hong}, Justin and {Jonkeren}, Alexander and {Khalil}, Mohammed and {Martin}, David W. and {Pendrill}, William and {Pumphrey}, Benjamin and {Rothermich}, Austin and {Sainio}, Arttu and {Stenner}, Andres and {Tanner}, Christopher and {Th{\'e}venot}, Melina and {Voloshin}, Nikita V. and {Walla}, Jim and {W{\k{e}}dracki}, Zbigniew and {Backyard Worlds: Planet 9 Collaboration}},
        title = "{The Field Substellar Mass Function Based on the Full-sky 20 pc Census of 525 L, T, and Y Dwarfs}",
      journal = {\apjs},
         year = 2021,
        month = mar,
       volume = {253},
       number = {1},
          eid = {7},
        pages = {7},
          doi = {10.3847/1538-4365/abd107},
archivePrefix = {arXiv},
       eprint = {2011.11616},
 primaryClass = {astro-ph.SR},
       adsurl = {https://ui.adsabs.harvard.edu/abs/2021ApJS..253....7K}
}

@ARTICLE{Ackerman2001,
       author = {{Ackerman}, Andrew S. and {Marley}, Mark S.},
        title = "{Precipitating Condensation Clouds in Substellar Atmospheres}",
      journal = {\apj},
         year = 2001,
        month = aug,
       volume = {556},
       number = {2},
        pages = {872-884},
          doi = {10.1086/321540},
archivePrefix = {arXiv},
       eprint = {astro-ph/0103423},
 primaryClass = {astro-ph},
       adsurl = {https://ui.adsabs.harvard.edu/abs/2001ApJ...556..872A}
}

@ARTICLE{Morley2024,
       author = {{Morley}, Caroline V. and {Mukherjee}, Sagnick and {Marley}, Mark S. and {Fortney}, Jonathan J. and {Visscher}, Channon and {Lupu}, Roxana and {Gharib-Nezhad}, Ehsan and {Thorngren}, Daniel and {Freedman}, Richard and {Batalha}, Natasha},
        title = "{The Sonora Substellar Atmosphere Models. III. Diamondback: Atmospheric Properties, Spectra, and Evolution for Warm Cloudy Substellar Objects}",
      journal = {\apj},
         year = 2024,
        month = nov,
       volume = {975},
       number = {1},
          eid = {59},
        pages = {59},
          doi = {10.3847/1538-4357/ad71d5},
archivePrefix = {arXiv},
       eprint = {2402.00758},
 primaryClass = {astro-ph.SR},
       adsurl = {https://ui.adsabs.harvard.edu/abs/2024ApJ...975...59M}
}

@ARTICLE{Freytag2010,
       author = {{Freytag}, B. and {Allard}, F. and {Ludwig}, H.-G. and {Homeier}, D. and {Steffen}, M.},
        title = "{The role of convection, overshoot, and gravity waves for the transport of dust in M dwarf and brown dwarf atmospheres}",
      journal = {\aap},
         year = 2010,
        month = apr,
       volume = {513},
          eid = {A19},
        pages = {A19},
          doi = {10.1051/0004-6361/200913354},
archivePrefix = {arXiv},
       eprint = {1002.3437},
 primaryClass = {astro-ph.SR},
       adsurl = {https://ui.adsabs.harvard.edu/abs/2010A&A...513A..19F}
}

@ARTICLE{Rossow1978,
       author = {{Rossow}, W.~B.},
        title = "{Cloud Microphysics: Analysis of the Clouds of Earth, Venus, Mars, and Jupiter}",
      journal = {\icarus},
         year = 1978,
        month = oct,
       volume = {36},
       number = {1},
        pages = {1-50},
          doi = {10.1016/0019-1035(78)90072-6},
       adsurl = {https://ui.adsabs.harvard.edu/abs/1978Icar...36....1R}
}

@ARTICLE{Allard2012,
       author = {{Allard}, F. and {Homeier}, D. and {Freytag}, B.},
        title = "{Models of very-low-mass stars, brown dwarfs and exoplanets}",
      journal = {Philosophical Transactions of the Royal Society of London Series A},
         year = 2012,
        month = jun,
       volume = {370},
       number = {1968},
        pages = {2765-2777},
          doi = {10.1098/rsta.2011.0269},
archivePrefix = {arXiv},
       eprint = {1112.3591},
 primaryClass = {astro-ph.SR},
       adsurl = {https://ui.adsabs.harvard.edu/abs/2012RSPTA.370.2765A}
}

@ARTICLE{Allard2013,
       author = {{Allard}, F. and {Homeier}, D. and {Freytag}, B. and {Schaffenberger}, W. and {Rajpurohit}, A.~S.},
        title = "{Progress in modeling very low mass stars, brown dwarfs, and planetary mass objects.}",
      journal = {Memorie della Societa Astronomica Italiana Supplementi},
         year = 2013,
        month = jan,
       volume = {24},
        pages = {128},
          doi = {10.48550/arXiv.1302.6559},
archivePrefix = {arXiv},
       eprint = {1302.6559},
 primaryClass = {astro-ph.SR},
       adsurl = {https://ui.adsabs.harvard.edu/abs/2013MSAIS..24..128A}
}

@ARTICLE{Barber2006,
       author = {{Barber}, R.~J. and {Tennyson}, J. and {Harris}, G.~J. and {Tolchenov}, R.~N.},
        title = "{A high-accuracy computed water line list}",
      journal = {\mnras},
         year = 2006,
        month = may,
       volume = {368},
       number = {3},
        pages = {1087-1094},
          doi = {10.1111/j.1365-2966.2006.10184.x},
archivePrefix = {arXiv},
       eprint = {astro-ph/0601236},
 primaryClass = {astro-ph},
       adsurl = {https://ui.adsabs.harvard.edu/abs/2006MNRAS.368.1087B}
}

@INPROCEEDINGS{Allard2014,
       author = {{Allard}, F.},
        title = "{The BT-Settl Model Atmospheres for Stars, Brown Dwarfs and Planets}",
    booktitle = {Exploring the Formation and Evolution of Planetary Systems},
         year = 2014,
       editor = {{Booth}, Mark and {Matthews}, Brenda C. and {Graham}, James R.},
       series = {IAU Symposium},
       volume = {299},
        month = jan,
        pages = {271-272},
          doi = {10.1017/S1743921313008545},
       adsurl = {https://ui.adsabs.harvard.edu/abs/2014IAUS..299..271A}
}

@ARTICLE{Leggett2017,
       author = {{Leggett}, S.~K. and {Tremblin}, P. and {Esplin}, T.~L. and {Luhman}, K.~L. and {Morley}, Caroline V.},
        title = "{The Y-type Brown Dwarfs: Estimates of Mass and Age from New Astrometry, Homogenized Photometry, and Near-infrared Spectroscopy}",
      journal = {\apj},
         year = 2017,
        month = jun,
       volume = {842},
       number = {2},
          eid = {118},
        pages = {118},
          doi = {10.3847/1538-4357/aa6fb5},
archivePrefix = {arXiv},
       eprint = {1704.03573},
 primaryClass = {astro-ph.SR},
       adsurl = {https://ui.adsabs.harvard.edu/abs/2017ApJ...842..118L}
}

@ARTICLE{Phillips2020,
       author = {{Phillips}, M.~W. and {Tremblin}, P. and {Baraffe}, I. and {Chabrier}, G. and {Allard}, N.~F. and {Spiegelman}, F. and {Goyal}, J.~M. and {Drummond}, B. and {H{\'e}brard}, E.},
        title = "{A new set of atmosphere and evolution models for cool T-Y brown dwarfs and giant exoplanets}",
      journal = {\aap},
         year = 2020,
        month = may,
       volume = {637},
          eid = {A38},
        pages = {A38},
          doi = {10.1051/0004-6361/201937381},
archivePrefix = {arXiv},
       eprint = {2003.13717},
 primaryClass = {astro-ph.SR},
       adsurl = {https://ui.adsabs.harvard.edu/abs/2020A&A...637A..38P}
}

@ARTICLE{Leggett2023,
       author = {{Leggett}, S.~K. and {Tremblin}, Pascal},
        title = "{The First Y Dwarf Data from JWST Show that Dynamic and Diabatic Processes Regulate Cold Brown Dwarf Atmospheres}",
      journal = {\apj},
         year = 2023,
        month = dec,
       volume = {959},
       number = {2},
          eid = {86},
        pages = {86},
          doi = {10.3847/1538-4357/acfdad},
archivePrefix = {arXiv},
       eprint = {2309.14567},
 primaryClass = {astro-ph.SR},
       adsurl = {https://ui.adsabs.harvard.edu/abs/2023ApJ...959...86L}
}

@ARTICLE{Cushing2008,
       author = {{Cushing}, Michael C. and {Marley}, Mark S. and {Saumon}, D. and {Kelly}, Brandon C. and {Vacca}, William D. and {Rayner}, John T. and {Freedman}, Richard S. and {Lodders}, Katharina and {Roellig}, Thomas L.},
        title = "{Atmospheric Parameters of Field L and T Dwarfs}",
      journal = {\apj},
         year = 2008,
        month = may,
       volume = {678},
       number = {2},
        pages = {1372-1395},
          doi = {10.1086/526489},
archivePrefix = {arXiv},
       eprint = {0711.0801},
 primaryClass = {astro-ph},
       adsurl = {https://ui.adsabs.harvard.edu/abs/2008ApJ...678.1372C}
}

@ARTICLE{Leggett2024,
       author = {{Leggett}, S.~K. and {Tremblin}, Pascal},
        title = "{James Webb Space Telescope Spectra of Cold Brown Dwarfs are Well-reproduced by Phosphine-free, Diabatic, ATMO2020++ Models}",
      journal = {Research Notes of the American Astronomical Society},
         year = 2024,
        month = jan,
       volume = {8},
       number = {1},
          eid = {13},
        pages = {13},
          doi = {10.3847/2515-5172/ad1b61},
       adsurl = {https://ui.adsabs.harvard.edu/abs/2024RNAAS...8...13L}
}

@ARTICLE{Akeson2026,
       author = {{Akeson}, Rachel and {Dubois-Felsmann}, Gregory P. and {Crill}, Brendan P. and {Faisst}, Andreas L. and {Fatahi}, Tamim and {Fazar}, Candice M. and {Goldina}, Tatiana and {Masters}, Daniel C. and {Nelson}, Christina and {Paladini}, Roberta and {Teplitz}, Harry I. and {Torrini}, Gabriela and {Velicheti}, Phani and {Ashby}, Matthew L.~N. and {Avner}, Dan and {Bach}, Yoonsoo P. and {Bock}, James J. and {Bruton}, Sean and {Bryan}, Sean A. and {Chang}, Tzu-Ching and {Chen}, Shuang-Shuang and {Cukierman}, Ari J. and {Dore}, O. and {Dowell}, C. Darren and {Everett}, Spencer and {Feder}, Richard M. and {Huai}, Zhaoyu and {Hui}, Howard and {Jeong}, Woong-Seob and {Jo}, Young-Soo and {Korngut}, Phil M. and {Kwon}, Yuna G. and {Lee}, Bomee and {Melnick}, Gary J. and {Murgia}, Giulia and {Nguyen}, Chi H. and {Pourrahmani}, Milad and {Rustamkulov}, Zafar and {Tolls}, Volker and {Wang}, Pao-Yu and {Yang}, Yujin and {Zemcov}, Michael},
        title = "{The SPHEREx Image and Spectrophotometry Processing Pipeline}",
      journal = {arXiv e-prints},
         year = 2025,
        month = nov,
          eid = {arXiv:2511.15823},
        pages = {arXiv:2511.15823},
          doi = {10.48550/arXiv.2511.15823},
archivePrefix = {arXiv},
       eprint = {2511.15823},
 primaryClass = {astro-ph.IM},
       adsurl = {https://ui.adsabs.harvard.edu/abs/2025arXiv251115823A}
}

@ARTICLE{Bock2025,
       author = {{Bock}, James J. and {Aboobaker}, Asad M. and {Adamo}, Joseph and {Akeson}, Rachel and {Alred}, John M. and {Alibay}, Farah and {Ashby}, Matthew L.~N. and {Bach}, Yoonsoo P. and {Bleem}, Lindsey E. and {Bolton}, Douglas and {Braun}, David F. and {Bruton}, Sean and {Bryan}, Sean A. and {Chang}, Tzu-Ching and {Chen}, Shuang-Shuang and {Cheng}, Yun-Ting and {Cheshire}, IV, James R. and {Chiang}, Yi-Kuan and {Choppin de Janvry}, Jean and {Condon}, Samuel and {Cook}, Walter R. and {Cooray}, Asantha and {Crill}, Brendan P. and {Cukierman}, Ari J. and {Dore}, Olivier and {Dowell}, C. Darren and {Dubois-Felsmann}, Gregory P. and {Eifler}, Tim and {Everett}, Spencer and {Fabinsky}, Beth E. and {Faisst}, Andreas L. and {Fanson}, James L. and {Farrington}, Allen H. and {Fatahi}, Tamim and {Fazar}, Candice M. and {Feder}, Richard M. and {Frater}, Eric H. and {Grasshorn Gebhardt}, Henry S. and {Giri}, Utkarsh and {Goldina}, Tatiana and {Gorjian}, Varoujan and {Habib}, Salman and {Hart}, William G. and {Heinrich}, Chen and {Hora}, Joseph L. and {Huai}, Zhaoyu and {Hui}, Howard and {Jo}, Young-Soo and {Jeong}, Woong-Seob and {Kang}, Jae Hwan and {Kang}, Miju and {Kecman}, Branislav and {Kim}, Chul-Hwan and {Kim}, Jaeyeong and {Kim}, Minjin and {Kim}, Young-Jun and {Kim}, Yongjung and {Kirkpatrick}, J. Davy and {Kobayashi}, Yosuke and {Korngut}, Phil M. and {Krause}, Elisabeth and {Lee}, Bomee and {Lee}, Ho-Gyu and {Lee}, Jae-Joon and {Lee}, Jeong-Eun and {Lisse}, Carey M. and {Mariani}, Giacomo and {Masters}, Daniel C. and {Mauskopf}, Philip D. and {Melnick}, Gary J. and {Minasyan}, Mary H. and {Mirocha}, Jordan and {Miyasaka}, Hiromasa and {Moore}, Anne and {Moore}, Bradley D. and {Murgia}, Giulia and {Naylor}, Bret J. and {Nelson}, Christina and {Nguyen}, Chi H. and {Nguyen}, Hien T. and {Noh}, Jinyoung K. and {Padin}, Stephen and {Paladini}, Roberta and {Park}, Sung-Joon and {Penanen}, Konstantin I. and {Putnam}, Dustin S. and {Pyo}, Jeonghyun and {Ramachandra}, Nesar and {Ramanathan}, Keshav and {Rustamkulov}, Zafar and {Reiley}, Daniel J. and {Rice}, Eric B. and {Rocca}, Jennifer M. and {Seok}, Ji Yeon and {Smith}, Roger and {Stober}, Jeremy and {Susca}, Sara and {Teplitz}, Harry I. and {Thelen}, Michael P. and {Tolls}, Volker and {Torrini}, Gabriela and {Trangsrud}, Amy R. and {Unwin}, Stephen and {Velicheti}, Phani and {Wang}, Pao-Yu and {Wen}, Robin Y. and {-Werner}, Michael-W. and {Williams}, Abby E. and {Williamson}, Ross and {Wincentsen}, James and {Windhorst}, Rogier A. and {Yang}, Soung-Chul and {Yang}, Yujin and {Zemcov}, Michael},
        title = "{The SPHEREx Satellite Mission}",
      journal = {arXiv e-prints},
         year = 2025,
        month = nov,
          eid = {arXiv:2511.02985},
        pages = {arXiv:2511.02985},
          doi = {10.48550/arXiv.2511.02985},
archivePrefix = {arXiv},
       eprint = {2511.02985},
 primaryClass = {astro-ph.IM},
       adsurl = {https://ui.adsabs.harvard.edu/abs/2025arXiv251102985B}
}

@ARTICLE{boeshaar1985,
       author = {{Boeshaar}, P.~C. and {Tyson}, J.~A.},
        title = "{New limits on the surface density of M dwarfs. I. Photographic survey and preliminary CCD data.}",
      journal = {\aj},
         year = 1985,
        month = may,
       volume = {90},
        pages = {817-822},
          doi = {10.1086/113791},
       adsurl = {https://ui.adsabs.harvard.edu/abs/1985AJ.....90..817B}
}

@ARTICLE{borysow2002,
       author = {{Borysow}, A.},
        title = "{Collision-induced absorption coefficients of H$_{2}$ pairs at temperatures from 60 K to 1000 K}",
      journal = {\aap},
         year = 2002,
        month = aug,
       volume = {390},
        pages = {779-782},
          doi = {10.1051/0004-6361:20020555},
       adsurl = {https://ui.adsabs.harvard.edu/abs/2002A&A...390..779B}
}

@ARTICLE{burgasser2006,
       author = {{Burgasser}, Adam J. and {Geballe}, T.~R. and {Leggett}, S.~K. and {Kirkpatrick}, J. Davy and {Golimowski}, David A.},
        title = "{A Unified Near-Infrared Spectral Classification Scheme for T Dwarfs}",
      journal = {\apj},
         year = 2006,
        month = feb,
       volume = {637},
       number = {2},
        pages = {1067-1093},
          doi = {10.1086/498563},
archivePrefix = {arXiv},
       eprint = {astro-ph/0510090},
 primaryClass = {astro-ph},
       adsurl = {https://ui.adsabs.harvard.edu/abs/2006ApJ...637.1067B}
}

@ARTICLE{cushing2011,
       author = {{Cushing}, Michael C. and {Kirkpatrick}, J. Davy and {Gelino}, Christopher R. and {Griffith}, Roger L. and {Skrutskie}, Michael F. and {Mainzer}, A. and {Marsh}, Kenneth A. and {Beichman}, Charles A. and {Burgasser}, Adam J. and {Prato}, Lisa A. and {Simcoe}, Robert A. and {Marley}, Mark S. and {Saumon}, D. and {Freedman}, Richard S. and {Eisenhardt}, Peter R. and {Wright}, Edward L.},
        title = "{The Discovery of Y Dwarfs using Data from the Wide-field Infrared Survey Explorer (WISE)}",
      journal = {\apj},
         year = 2011,
        month = dec,
       volume = {743},
       number = {1},
          eid = {50},
        pages = {50},
          doi = {10.1088/0004-637X/743/1/50},
archivePrefix = {arXiv},
       eprint = {1108.4678},
 primaryClass = {astro-ph.SR},
       adsurl = {https://ui.adsabs.harvard.edu/abs/2011ApJ...743...50C}
}

@article{Korngut2026,
       author = {{Korngut, P. et al.}},
      journal = {in prep.},
         year = 2026
}

@ARTICLE{kirkpatrick2012,
       author = {{Kirkpatrick}, J. Davy and {Gelino}, Christopher R. and {Cushing}, Michael C. and {Mace}, Gregory N. and {Griffith}, Roger L. and {Skrutskie}, Michael F. and {Marsh}, Kenneth A. and {Wright}, Edward L. and {Eisenhardt}, Peter R. and {McLean}, Ian S. and {Mainzer}, Amanda K. and {Burgasser}, Adam J. and {Tinney}, C.~G. and {Parker}, Stephen and {Salter}, Graeme},
        title = "{Further Defining Spectral Type ``Y'' and Exploring the Low-mass End of the Field Brown Dwarf Mass Function}",
      journal = {\apj},
         year = 2012,
        month = jul,
       volume = {753},
       number = {2},
          eid = {156},
        pages = {156},
          doi = {10.1088/0004-637X/753/2/156},
archivePrefix = {arXiv},
       eprint = {1205.2122},
 primaryClass = {astro-ph.SR},
       adsurl = {https://ui.adsabs.harvard.edu/abs/2012ApJ...753..156K}
}

@ARTICLE{kirkpatrick1991,
       author = {{Kirkpatrick}, J.~D. and {Henry}, Todd J. and {McCarthy}, Jr., Donald W.},
        title = "{A Standard Stellar Spectral Sequence in the Red/Near-Infrared: Classes K5 to M9}",
      journal = {\apjs},
         year = 1991,
        month = nov,
       volume = {77},
        pages = {417},
          doi = {10.1086/191611},
       adsurl = {https://ui.adsabs.harvard.edu/abs/1991ApJS...77..417K}
}

@ARTICLE{kirkpatrick1999,
       author = {{Kirkpatrick}, J. Davy and {Reid}, I. Neill and {Liebert}, James and {Cutri}, Roc M. and {Nelson}, Brant and {Beichman}, Charles A. and {Dahn}, Conard C. and {Monet}, David G. and {Gizis}, John E. and {Skrutskie}, Michael F.},
        title = "{Dwarfs Cooler than ``M``: The Definition of Spectral Type ``L'' Using Discoveries from the 2 Micron All-Sky Survey (2MASS)}",
      journal = {\apj},
         year = 1999,
        month = jul,
       volume = {519},
       number = {2},
        pages = {802-833},
          doi = {10.1086/307414},
       adsurl = {https://ui.adsabs.harvard.edu/abs/1999ApJ...519..802K}
}

@ARTICLE{kirkpatrick2005,
       author = {{Kirkpatrick}, J. Davy},
        title = "{New Spectral Types L and T}",
      journal = {\araa},
         year = 2005,
        month = sep,
       volume = {43},
       number = {1},
        pages = {195-245},
          doi = {10.1146/annurev.astro.42.053102.134017},
       adsurl = {https://ui.adsabs.harvard.edu/abs/2005ARA&A..43..195K}
}

@ARTICLE{kirkpatrick2010,
       author = {{Kirkpatrick}, J. Davy and {Looper}, Dagny L. and {Burgasser}, Adam J. and {Schurr}, Steven D. and {Cutri}, Roc M. and {Cushing}, Michael C. and {Cruz}, Kelle L. and {Sweet}, Anne C. and {Knapp}, Gillian R. and {Barman}, Travis S. and {Bochanski}, John J. and {Roellig}, Thomas L. and {McLean}, Ian S. and {McGovern}, Mark R. and {Rice}, Emily L.},
        title = "{Discoveries from a Near-infrared Proper Motion Survey Using Multi-epoch Two Micron All-Sky Survey Data}",
      journal = {\apjs},
         year = 2010,
        month = sep,
       volume = {190},
       number = {1},
        pages = {100-146},
          doi = {10.1088/0067-0049/190/1/100},
archivePrefix = {arXiv},
       eprint = {1008.3591},
 primaryClass = {astro-ph.SR},
       adsurl = {https://ui.adsabs.harvard.edu/abs/2010ApJS..190..100K}
}

@ARTICLE{Kothari2024,
       author = {{Kothari}, Harshil and {Cushing}, Michael C. and {Burningham}, Ben and {Beiler}, Samuel A. and {Kirkpatrick}, J. Davy and {Schneider}, Adam C. and {Mukherjee}, Sagnick and {Marley}, Mark S.},
        title = "{Probing the Heights and Depths of Y Dwarf Atmospheres: A Retrieval Analysis of the JWST Spectral Energy Distribution of WISE J035934.06─540154.6}",
      journal = {\apj},
         year = 2024,
        month = aug,
       volume = {971},
       number = {2},
          eid = {121},
        pages = {121},
          doi = {10.3847/1538-4357/ad583b},
archivePrefix = {arXiv},
       eprint = {2406.06493},
 primaryClass = {astro-ph.SR},
       adsurl = {https://ui.adsabs.harvard.edu/abs/2024ApJ...971..121K}
}

@ARTICLE{lepine2007,
       author = {{L{\'e}pine}, S{\'e}bastien and {Rich}, R. Michael and {Shara}, Michael M.},
        title = "{Revised Metallicity Classes for Low-Mass Stars: Dwarfs (dM), Subdwarfs (sdM), Extreme Subdwarfs (esdM), and Ultrasubdwarfs (usdM)}",
      journal = {\apj},
         year = 2007,
        month = nov,
       volume = {669},
       number = {2},
        pages = {1235-1247},
          doi = {10.1086/521614},
archivePrefix = {arXiv},
       eprint = {0707.2993},
 primaryClass = {astro-ph},
       adsurl = {https://ui.adsabs.harvard.edu/abs/2007ApJ...669.1235L}
}

@ARTICLE{phillips2024,
       author = {{Phillips}, Mark W. and {Liu}, Michael C. and {Zhang}, Zhoujian},
        title = "{The Carbon-to-oxygen Ratio in Cool Brown Dwarfs and Giant Exoplanets. I. The Benchmark Late-T Dwarfs GJ 570D, HD 3651B, and Ross 458C}",
      journal = {\apj},
         year = 2024,
        month = feb,
       volume = {961},
       number = {2},
          eid = {210},
        pages = {210},
          doi = {10.3847/1538-4357/ad06ba},
archivePrefix = {arXiv},
       eprint = {2312.02001},
 primaryClass = {astro-ph.SR},
       adsurl = {https://ui.adsabs.harvard.edu/abs/2024ApJ...961..210P}
}

@ARTICLE{saumon1994,
       author = {{Saumon}, D. and {Bergeron}, P. and {Lunine}, J.~I. and {Hubbard}, W.~B. and {Burrows}, A.},
        title = "{Cool Zero-Metallicity Stellar Atmospheres}",
      journal = {\apj},
         year = 1994,
        month = mar,
       volume = {424},
        pages = {333},
          doi = {10.1086/173892},
       adsurl = {https://ui.adsabs.harvard.edu/abs/1994ApJ...424..333S}
}

@ARTICLE{saumon2006,
       author = {{Saumon}, D. and {Marley}, M.~S. and {Cushing}, M.~C. and {Leggett}, S.~K. and {Roellig}, T.~L. and {Lodders}, K. and {Freedman}, R.~S.},
        title = "{Ammonia as a Tracer of Chemical Equilibrium in the T7.5 Dwarf Gliese 570D}",
      journal = {\apj},
         year = 2006,
        month = aug,
       volume = {647},
       number = {1},
        pages = {552-557},
          doi = {10.1086/505419},
archivePrefix = {arXiv},
       eprint = {astro-ph/0605563},
 primaryClass = {astro-ph},
       adsurl = {https://ui.adsabs.harvard.edu/abs/2006ApJ...647..552S}
}

@ARTICLE{zhang2017,
       author = {{Zhang}, Z.~H. and {Pinfield}, D.~J. and {G{\'a}lvez-Ortiz}, M.~C. and {Burningham}, B. and {Lodieu}, N. and {Marocco}, F. and {Burgasser}, A.~J. and {Day-Jones}, A.~C. and {Allard}, F. and {Jones}, H.~R.~A. and {Homeier}, D. and {Gomes}, J. and {Smart}, R.~L.},
        title = "{Primeval very low-mass stars and brown dwarfs - I. Six new L subdwarfs, classification and atmospheric properties}",
      journal = {\mnras},
         year = 2017,
        month = jan,
       volume = {464},
       number = {3},
        pages = {3040-3059},
          doi = {10.1093/mnras/stw2438},
archivePrefix = {arXiv},
       eprint = {1609.07181},
 primaryClass = {astro-ph.SR},
       adsurl = {https://ui.adsabs.harvard.edu/abs/2017MNRAS.464.3040Z}
}
\bibliographystyle{aasjournalv7}

\end{document}